\documentclass[fleqn,usenatbib]{mnras}

\usepackage{newtxtext,newtxmath}
\usepackage[T1]{fontenc}

\usepackage[british]{babel}
\addto\extrasbritish{%
}

\newcommand{\e}{\mathrm{e}}

\usepackage[inline]{enumitem}

\DeclareRobustCommand{\VAN}[3]{#2}
\let\VANthebibliography\thebibliography
\def\thebibliography{\DeclareRobustCommand{\VAN}[3]{##3}\VANthebibliography}

\usepackage{graphicx}	% Including figure files
\usepackage{amsmath}	% Advanced maths commands
\usepackage{mathtools}
\usepackage{hyperref}

\usepackage{caption}
\usepackage{subcaption}
\usepackage{bm}
\usepackage{ulem}

\newcommand{\sTheta}{\bm{\upTheta}} %set Theta
\graphicspath{{Figures/}}

\title[Bayesian analysis of BFCC spectrometer data - I]{A Bayesian approach to modelling spectrometer data chromaticity corrected using beam factors - I. Mathematical formalism}

\author[Sims et al.]{Peter H. Sims,$^{1,2}$\thanks{E-mail: peter.sims@mail.mcgill.ca}
Judd D. Bowman,$^{2}$
Nivedita Mahesh,$^{2}$
Steven G. Murray,$^{2}$
John P. Barrett,$^{3}$ \and
Rigel Cappallo,$^{3}$
Raul A. Monsalve,$^{2,4,5}$
Alan E. E. Rogers,$^{3}$
Titu Samson,$^{2}$
and Akshatha K. Vydula$^{2}$
\\
$^1$Department of Physics \& Trottier Space Institute, McGill University, Montreal, QC H3A 2T8, Canada \\
$^{2}$School of Earth and Space Exploration, Arizona State University, Tempe, AZ 85287, USA\\
$^{3}$MIT Haystack Observatory, Westford, MA 01886-1299, USA \\
$^{4}$Space Sciences Laboratory, University of California Berkeley,
Berkeley, CA 94720, USA \\
$^{5}$Facultad de Ingeniería, Universidad Católica de la Santísima Concepción, Alonso de Ribera 2850, Concepción, Chile
}

\date{Accepted XXX. Received YYY; in original form ZZZ}

\pubyear{2021}

\begin{document}
\label{firstpage}
\pagerange{\pageref{firstpage}--\pageref{lastpage}}
\maketitle

\begin{abstract}
Accurately accounting for spectral structure in spectrometer data induced by instrumental chromaticity on scales relevant for detection of the 21-cm signal is among the most significant challenges in global 21-cm signal analysis. In the publicly available EDGES low-band data set, this complicating structure is suppressed using beam-factor based chromaticity correction (BFCC), which works by dividing the data by a sky-map-weighted model of the spectral structure of the instrument beam. Several analyses of this data have employed models that start with the assumption that this correction is complete. However, while BFCC mitigates the impact of instrumental chromaticity on the data, given realistic assumptions regarding the spectral structure of the foregrounds, the correction is only partial. This complicates the interpretation of fits to the data with intrinsic sky models (models that assume no instrumental contribution to the spectral structure of the data). In this paper, we derive a BFCC data model from an analytic treatment of BFCC and demonstrate using simulated observations that, in contrast to using an intrinsic sky model for the data, the BFCC data model enables unbiased recovery of a simulated global 21-cm signal from beam-factor chromaticity corrected data in the limit that the data is corrected with an error-free beam-factor model.
\end{abstract}


%%%%%%%%%%%%%%%%%%%%%%%%%%%%%%%%%%%%%%%%%%%%%%%%%%
%%%%%%%%%%%%%%%%% BODY OF PAPER %%%%%%%%%%%%%%%%%%

%%%%%%%%%%%%%%%%%%%%%%%%%%%%%%%%%%%%%%%%%%%%%%%%%%
\section{Introduction}
\label{Sec:Intro}
%%%%%%%%%%%%%%%%%%%%%%%%%%%%%%%%%%%%%%%%%%%%%%%%%%

After the first stars and proto-galaxies formed at Cosmic Dawn (CD) and suffused the intergalactic medium (IGM) with photons at the Ly$\alpha$ resonance of hydrogen, their absorption and spontaneous re-emission is expected to have coupled the hydrogen spin and kinetic temperatures via the Wouthuysen--Field effect. The associated decoupling of the hydrogen spin temperature from the background radiation temperature imprints a spectral distortion in the sky-averaged radio spectrum that should be observable today in the frequency range $\nu_\mathrm{CD} \le \nu \lesssim 220~\mathrm{MHz}$, where $\nu_\mathrm{CD} = \nu_{21} / (1+z_\mathrm{CD})$ is the redshifted frequency of 21-cm hyperfine line radiation emitted at the onset of CD, $z_\mathrm{CD}$ is the corresponding redshift, and $\nu_{21} = 1420.4057~\mathrm{MHz}$ is the rest frequency of the 21-cm emission.

Multiple experiments are underway to measure the evolution of this sky-averaged `global' redshifted 21-cm signal, including: the Experiment to Detect the Global Epoch of Reionization Signature (EDGES; \citealt{2018Natur.555...67B}), the Large-aperture Experiment to Detect the Dark Ages (LEDA; e.g. \citealt{2016MNRAS.461.2847B}), the Mapper of the IGM Spin Temperature (MIST; e.g. \citealt{2022arXiv221116547S}), Probing Radio Intensity at High-Z from Marion (PRIZM; \citealt{2019JAI.....850004P}), the Radio Experiment for the Analysis of Cosmic Hydrogen (REACH; \citealt{2022NatAs.tmp..167D}) and the Shaped Antenna measurement of the background RAdio Spectrum (SARAS3; \citealt{2021arXiv210401756N}). Additionally, space- and lunar-based global 21-cm experiments, which eliminate observational challenges associated with the ionosphere and mitigate or remove those associated with human-made Radio Frequency Interference (RFI), are also planned to begin operation in the next few years. These include: Discovering Sky at the Longest wavelength (DSL; \citealt{2021RSPTA.37990566C}), the Lunar Surface Electromagnetic Experiment (LuSEE - Night; \citealt{2023arXiv230110345B}) and Probing ReionizATion of the Universe using Signal from Hydrogen (PRATUSH).

In 2018, analysing data in the 50--100 MHz range, the EDGES experiment presented evidence for a first detection of the global 21-cm signal (\citealt{2018Natur.555...67B}; hereafter B18). A best fitting flat-bottomed absorption trough was recovered centred at $78 \pm 1~\mathrm{MHz}$ ($z = 17.2 \pm 0.2$), with a width of $19^{+4}_{-2}~\mathrm{MHz}$ and with a depth of $500^{+500}_{-200}~\mathrm{mK}$, where the uncertainties correspond to 99 per cent confidence intervals, accounting for both thermal and systematic errors.

The depth of this absorption trough exceeds the $\sim 165~\mathrm{mK}$ maximum depth predicted in this redshift range by fiducial models (e.g. \citealt{2021MNRAS.506.5479R}). Assuming that the absorption parameters derived with this model are accurate and that the recovered 21-cm absorption trough is interpreted as physical, its large depth implies an increased differential brightness between the hydrogen 21-cm spin temperature and the radio background temperature than is predicted by standard models. The large absorption depth can be explained astrophysically by positing a new contribution to the radio background temperature at CD, which raises it in excess of the CMB temperature (e.g.  B18; \citealt{2018ApJ...858L..17F, 2018PhLB..785..159F, 2018ApJ...868...63E, 2019MNRAS.483.1980M, 2019MNRAS.486.1763F, 2020MNRAS.499.5993R}), or by assuming a kinetic temperature of the hydrogen gas reduced below the adiabatic cooling limit, which could occur due to interactions between cold dark matter and baryons (e.g. B18; \citealt{2018Natur.555...71B, 2018Natur.557..684M, 2018PhRvL.121a1101F, 2019PhRvD.100l3011L}). However, non-astrophysical explanations have also been put forward, with re-analyses of the EDGES data raising concerns about the original data analysis and potential presence of unmodelled systematics (e.g. \citealt{2018Natur.564E..32H, 2019ApJ...874..153B, 2019ApJ...880...26S, 2020MNRAS.492...22S, 2021MNRAS.502.4405B, 2022MNRAS.tmp.2422M}). Additionally, a recent attempt by SARAS3 to verify this absorption trough in a radiometer measurement of the spectrum of the radio sky in the 55-85 MHz band rejected with 95.3\% confidence the best-fitting profile reported in B18 (Singh et al. 2022); however, analysing the same data, \citet{2022NatAs...6.1473B} find that when defining a prior that spans models drawn from astrophysical simulations of the 21-cm signal with similar depths and central frequencies to the absorption feature reported in B18, 60\% of this physical EDGES-like parameter space is consistent with the SARAS 3 data.

The primary challenge to unbiased 21-cm signal recovery from spectrometer data is the mixing, through instrumental chromaticity, of bright, spectrally smooth foreground emission into the narrower spectral scales relevant for 21-cm signal detection. The frequency-dependent weighting of the sky by the instrument beam is a primary source of instrument chromaticity which can introduce this mixing. In the EDGES low-band data analysed in B18, a procedure called beam-factor based Chromaticity Correction (BFCC) is used to mitigate the impact of the spectral structure introduced to the data by beam chromaticity. This involves dividing the calibrated autocorrelation spectrometer data by an integrated sky-map weighted model of the instrument beam, acting as a proxy for beam chromaticity (see \autoref{Sec:ChromaticityCorrectionFormalism} for details).

If BFCC were able to completely eliminate the spectral structure introduced to the data by beam chromaticity, one would be able to recover unbiased estimates of the global 21-cm signal from a fit of accurately calibrated BFCC data with a model for the intrinsic spectral structure of the sky. The analyses of the data, thus far, have taken such a perfect correction as a starting point for the modelling of the data, with the foregrounds, after propagation through the ionosphere, being modelled by low-order or derivative-constrained polynomials.

Building on such intrinsic models, potential systematics resulting from imperfections in the BFCC correction have been fit in a data-driven manner (e.g. B18; \citealt{2018Natur.564E..32H, 2019ApJ...874..153B, 2019ApJ...880...26S, 2020MNRAS.492...22S, 2021MNRAS.502.4405B, 2021AJ....162...38M, 2022MNRAS.tmp.2422M}). However, the extent to which these models are necessary or sufficient to model these systematic effects has not been studied from a first-principles perspective. This absence of detailed characterisation of the systematic structure in the data resulting from BFCC introduces unnecessary epistemic uncertainty into astrophysical inferences from data fitted with these models. This hinders the drawing of firm conclusions regarding the 21-cm signal.

A comprehensive theoretical study of the efficacy of BFCC and of models with which one should expect to recover unbiased estimates of the global 21-cm signal from BFCC spectrometer data is required to mitigate this uncertainty. It is essential both for assessing the results of previous analyses of the EDGES low-band data and to provide a route towards obtaining a more theoretically principled model for BFCC data. The series of papers of which this is the first is designed to provide this.

In this first paper, we address the extent to which BFCC can eliminate the spectral structure introduced to the data by beam chromaticity under realistic assumptions regarding the spectral structure of the sky, in the limit that one has an error-free model for the beam-factor. We demonstrate that, in this limit, while BFCC mitigates the impact of instrumental chromaticity on the data, given realistic assumptions regarding the spectral structure of the foregrounds, the correction is only partial. However, we show that an analytic treatment of BFCC nevertheless provides a strong foundation for derivation of an accurate approximate model for BFCC data. We demonstrate using simulated observations that, in contrast to using an intrinsic model for the data, the BFCC data model derived from this treatment enables unbiased recovery of a simulated global 21-cm signal from beam-factor chromaticity corrected data.

In subsequent papers in this series we plan to: \begin{enumerate*} \item quantify the impact of uncertainties associated with one's \textit{a priori} knowledge of the instrument beam and sky on the model complexity required to describe BFCC data; \item examine the relation between 21-cm signal amplitude and the degree to which one can recover unbiased astrophysical inferences using; \item carry out Bayesian model comparison between the BFCC model and extensions to the intrinsic sky model incorporating systematic models. \end{enumerate*}

The remainder of this paper is organised as follows. In \autoref{Sec:ChromaticityCorrectionFormalism} we detail the BFCC procedure and derive the general form of BFCC data. In \autoref{Sec:ToyModel} we explore a simplified scenario in the limit that the spectral structure of the foreground emission is spatially independent\footnote{Throughout the text, we use spatially invariant and isotropic spectral structure synonymously. Both denote a situation in which the spectral structure of the emission in question does not vary as a function of celestial coordinate. This is in contrast to emission with spatially-dependent or anisotropic spectral structure in which the reverse is true.} and demonstrate that in this limit BFCC can perfectly chromaticity correct spectrometer data, yielding a closed-form\footnote{A general expression for spectrometer data can be written in terms of the integral of the instrument-response-weighted sky brightness temperature over integration time and solid angle (see \autoref{Sec:ChromaticityCorrectionFormalism}). Here, by closed-form we mean that, after beam-factor chromaticity correction one's model for the data can be written free of integral terms or infinite sums.} expression for the data model and unbiased 21-cm signal estimates. In \autoref{Sec:RealisticModels} we will show that for foregrounds with realistic spatially dependent spectral structure no such closed-form expression for the data model exists but that a perturbed version of the BFCC model derived in \autoref{Sec:ToyModel} provides a well motivated basis for describing the data. In \autoref{Sec:RealisticBFCCDemonstration}, we determine the preferred complexity of the BFCC model for EDGES low-band data using fits to simulated data. We additionally compare the preferred BFCC model to an alternate model designed to accurately describe the intrinsic spectral structure of the sky. In \autoref{Sec:Discussion}, we discuss the relative merits of the BFCC model and intrinsic model for describing BFCC EDGES low-band data. We also compare the difference in 21-cm signals recovered from simulated data with the two models to the difference between the 21-cm signal recovered in the B18 analysis and that recovered in the \citet{2018Natur.564E..32H} analysis of the publicly available EDGES low-band data with an intrinsic sky model. In \autoref{Sec:Conclusions}, we provide a summary and conclusions.

%%%%%%%%%%%%%%%%%%%%%%%%%%%%%%%%%%%%%%%%%%%%%%%%%%
\section{Beam-factor chromaticity correction}
\label{Sec:ChromaticityCorrectionFormalism}
%%%%%%%%%%%%%%%%%%%%%%%%%%%%%%%%%%%%%%%%%%%%%%%%%%

% \subsection{Mozdzen beam-factor chromaticity correction}
% \label{Sec:MozdzenBFCC}

Working in the reference frame of the antenna, a calibrated autocorrelation spectrum derived from a zenith-pointing antenna, such as EDGES, integrated over a short time interval $\upDelta t = (t_\mathrm{end} - t_\mathrm{start})$ centered on a time, $t = t_\mathrm{start} + \upDelta t / 2$, where $t_\mathrm{start}$ and $t_\mathrm{end}$ are the start and end of the observation, respectively, can be written as,
\begin{equation}
\label{Eq:Tdata}
T_\mathrm{data}(\nu, t) = \dfrac{1}{\upDelta t}\int\limits_{t_\mathrm{start}}^{t_\mathrm{end}} \int\limits_{0}^{4\pi} B(\nu, \Omega)T_\mathrm{sky}(\nu, \Omega, t^{\prime})\mathrm{d}\Omega\mathrm{d}t^{\prime} + n \ .
\end{equation}
Here, $\Omega$ is a position coordinate on the celestial sphere in the reference frame of the antenna\footnote{We use this notation for brevity and generality. In a zenith angle, $\theta$, and azimuth, $\phi$, horizontal celestial coordinate system, $\mathrm{d}\Omega = \sin(\theta)\mathrm{d}\theta\mathrm{d}\phi$, and $T(\Omega) = T(\theta, \phi)$, where $T$ is a scalar field defined on the celestial sphere.}, $T_\mathrm{sky}(\nu, \Omega, t)$ describes the time-dependent sky brightness temperature distribution above the antenna, $B(\nu, \Omega) = \frac{1}{4\pi}D(\nu, \Omega)$, describes the antenna beam, where $D(\nu, \Omega)$ is the antenna's directivity pattern, and $n$ is instrumental noise. Assuming the data is calibrated such that $T_\mathrm{data}(\nu, t)$ is an absolute temperature measurement, we have $\int_{0}^{4\pi} B(\nu, \Omega)\mathrm{d}\Omega = 1$.

For an instrument such as EDGES, incorporating a ground plane below the antenna, the dominant contribution to the integrated antenna directivity derives from the skyward hemisphere, $\Omega^{+}$ (in a zenith angle, $\theta$, and azimuth, $\phi$, horizontal celestial coordinate system, the hemisphere with $\theta \le 90$). In the limit of an infinite and perfectly electrically conducting (hereafter, PEC) ground plane the integrated antenna directivity derives exclusively from $\Omega^{+}$ (i.e. the beam is entirely skyward), and \autoref{Eq:Tdata} reduces to,
\begin{align}
\label{Eq:Tdata2}
T_\mathrm{data}(\nu, t) &= \dfrac{1}{\upDelta t}\int\limits_{t_\mathrm{start}}^{t_\mathrm{end}} \int\limits_{0}^{2\pi} \int\limits_{0}^{\pi/2} B(\nu, \Omega)T_\mathrm{sky}(\nu, \Omega, t^{\prime})\sin(\theta)\mathrm{d}\theta\mathrm{d}\phi\mathrm{d}t^{\prime} + n \\ \nonumber
&\equiv \dfrac{1}{\upDelta t}\int\limits_{t_\mathrm{start}}^{t_\mathrm{end}} \int\limits_{\Omega^{+}} B(\nu, \Omega)T_\mathrm{sky}(\nu, \Omega, t^{\prime})\mathrm{d}\Omega\mathrm{d}t^{\prime} + n \ ,
\end{align}

Absorption by the ground of some fraction of the radiation that would have been received by an antenna on an infinite PEC ground plane means the response to the sky signal of an antenna on a finite ground plane is proportionately lower. \autoref{Eq:Tdata2} provides an approximation for $T_\mathrm{data}(\nu, t)$ measured by an antenna on a finite ground plane, but neglects this small fractional loss due to the ground. This fractional loss can however be simulated and corrected with a ground loss correction procedure (e.g. \citealt{2012RaSc...47.0K06R, 2017ApJ...847...64M}). For the EDGES low-band instrument\footnote{We specifically consider the H2 configuration of the EDGES low-band instrument with a $30~\mathrm{m} \times 30~\mathrm{m}$ sawtooth ground plane, the data from which was used for the primary analysis in B18 (see B18 for alternate configurations and \autoref{Sec:PerfectBFCCSimulations} for details of the instrument simulation in this work).}, $1-\Omega^{+}/\Omega \simeq 10^{-3}$; thus, any chromatic effects introduced into the data through imperfections in ground-loss correction are expected be subdominant to those introduced through imperfect beam-factor chromaticity correction. As such, in this paper we focus on the latter and assume that ground loss correction has been performed sufficiently accurately for negligible bias to be introduced by using \autoref{Eq:Tdata2} to describe the calibrated and ground-loss corrected spectrometer data.

It can be seen from \autoref{Eq:Tdata2}, that the spectral structure of the measured data, $T_\mathrm{data}(\nu, t)$, is a function of the product of the antenna chromaticity and the spectral structure of the emission incident on the antenna. Thus, if the antenna has chromatic structure on scales relevant for detection of the 21-cm signal, then the chromaticity imparted to the foregrounds by the beam will complicate or potentially preclude unbiased estimation of the 21-cm signal even if the foreground component of the brightness distribution incident on the antenna is intrinsically spectrally smooth.

In previous work, the EDGES collaboration has taken the approach of mitigating beam chromaticity by dividing the calibrated autocorrelation spectrum by a beam chromaticity correction factor (hereafter, beam-factor) that describes the average spectral structure of the beam weighted by the brightness temperature distribution of the sky at a given reference frequency.
In general, such a correction is given by,
\begin{equation}
\label{Eq:CCgeneral}
B_\mathrm{factor,general}(\nu, t) = \frac{\int\limits_{t_\mathrm{start}}^{t_\mathrm{end}} \int\limits_{\Omega^{+}}  B^\mathrm{m}(\nu, \Omega) T_\mathrm{fg}^\mathrm{m}(\nu_\mathrm{c}, \Omega, t^{\prime}) \mathrm{d}\Omega\mathrm{d}t^{\prime}}{\int\limits_{t_\mathrm{start}}^{t_\mathrm{end}} \int\limits_{\Omega^{+}}  B^\mathrm{m}(\nu_\mathrm{c}, \Omega) T_\mathrm{fg}^\mathrm{m}(\nu_\mathrm{c}, \Omega, t^{\prime}) \mathrm{d}\Omega\mathrm{d}t^{\prime}} \ ,
\end{equation}
where, $\nu_\mathrm{c}$ is a reference frequency, which going forward we will define to be the center of the band observed by the instrument, and $B^\mathrm{m}(\nu, \Omega)$ and $T_\mathrm{fg}^\mathrm{m}(\nu_\mathrm{c}, \Omega, t^{\prime})$ are models for the frequency-dependent beam over the frequency range of the data and foreground sky brightness temperature distribution at reference frequency $\nu_\mathrm{c}$, respectively.

For a sufficiently short integration time\footnote{Single-dish integrations are time-averaged measurements. The spectrum for a given integration can be reasonably approximated by a snapshot at the center of the time step if the beam crossing time for a source is $t\approx \mathrm{\Theta_\mathrm{beam}}/\omega_\mathrm{E} \gg \tau$, with $\tau$ the integration time, $\Theta_\mathrm{beam}$ the angular scale of spatial structure in the beam and $\omega_\mathrm{E}\approx7.3\times10^{-5}~\mathrm{rad~s^{-1}}$ the maximum rotation rate of a source through the beam. To zeroth order, the EDGES beam has a characteristic width of 10s of degrees, for which this is a reasonable approximation on timescales of 10s of minutes. Scattering effects, associated with placing the antenna on a finite ground-plane on top of a realistic soil layer, introduce low-level spatial structure in the beam on smaller scales, implying a corresponding reduction in time for which a snapshot approximation is valid at the level required for 21-cm cosmology. Realistic simulations of the EDGES low-band antenna, ground plane and soil suggest a snapshot approximation is reasonable for integration times less than approximately 10 m, with statistically consistent results recovered in this limit for simulated data sets with a comparable signal-to-noise level to the publicly available data from B18.}, \autoref{Eq:CCgeneral} is well approximated by (e.g. \citealt{2017MNRAS.464.4995M, 2019MNRAS.483.4411M}),
\begin{equation}
\label{Eq:CCsnapshot}
B_\mathrm{factor}(\nu, t) = \frac{\int\limits_{\Omega^{+}}  B^\mathrm{m}(\nu, \Omega) T_\mathrm{fg}^\mathrm{m}(\nu_\mathrm{c}, \Omega, t) \mathrm{d}\Omega}{\int\limits_{\Omega^{+}}  B^\mathrm{m}(\nu_\mathrm{c}, \Omega) T_\mathrm{fg}^\mathrm{m}(\nu_\mathrm{c}, \Omega, t) \mathrm{d}\Omega} \ .
\end{equation}
A correction of this type has also been used for some applications by other global signal experiments including LEDA (e.g. \citealt{2021MNRAS.505.1575S}) and REACH (e.g. \citealt{2021MNRAS.503..344S, 2022MNRAS.509.4679A}).

In \citet{2017ApJ...847...64M}, an alternate chromaticity correction formulation incorporating additional foreground spectral information is employed. In Appendix \ref{Sec:MonsalveBFCC} we describe the dependence of the relative efficacy of the Monsalve beam-factor formulation and that of the Mozdzen beam-factor formulation described by \autoref{Eq:CCsnapshot} on the accuracy of one's foreground spectral model. As the nominal form of beam-factor chromaticity correction used in B18, in the remainder of this work we focus on BFCC using the Mozdzen beam-factor formulation.

In the short-integration `snapshot' limit considered in \autoref{Eq:CCsnapshot}, the chromaticity corrected spectrum for the observation has the form,
\begin{align}
\label{Eq:Tcorrected}
T_{\rm corrected}(\nu, t) \\ \nonumber
={}&  \left[\int\limits_{\Omega^{+}} B(\nu, \Omega) T_\mathrm{sky}(\nu, \Omega, t) \mathrm{d}\Omega + n\right] \\  \nonumber
& \times \frac{\int\limits_{\Omega^{+}} B^\mathrm{m}(\nu_\mathrm{c}, \Omega) T_\mathrm{fg}^\mathrm{m}(\nu_\mathrm{c}, \Omega, t) \mathrm{d}\Omega}{\int\limits_{\Omega^{+}} B^\mathrm{m}(\nu, \Omega) T_\mathrm{fg}^\mathrm{m}(\nu_\mathrm{c}, \Omega, t) \mathrm{d}\Omega} \ .
\end{align}
From \autoref{Eq:Tcorrected}, it can be seen that BFCC scales the instrumental noise in addition to the sky-signal. Thus, in principle, when defining a data likelihood, the covariance matrix of $T_{\rm corrected}(\nu, t)$ should be scaled by $1/B_\mathrm{factor}^{2}$. However, in practice, the EDGES low-band $B_\mathrm{factor}$ deviates from unity by a maximum of $\sim10\%$ when the Galaxy is overhead, and less than $\sim5\%$ when the Galaxy is low in the beam in the $\sim0 - 12~\mathrm{h}$ Local Sidereal Time- (LST-)range that the spectrum is averaged over in the B18 publicly released data (see \autoref{Fig:BeamFactor}). As a result, the impact of this scaling of the noise is approximately an order of magnitude smaller than the effect of data-weighting associated with RFI flagging in the same data set, which varies by approximately a factor of two between data inside vs outside the FM band above $87~\mathrm{MHz}$ (B18). With a more chromatic instrument, with a larger $B_\mathrm{factor}$ spread, the impact of correctly accounting for this re-weighting of the noise will increase. In either case, if one wishes to recover unbiased signal estimates with accurate uncertainty estimates at high precision, this re-weighting of the noise should be accounted for when defining a data likelihood for $T_{\rm corrected}(\nu, t)$.

A phenomenological model for the effect of errors on $B^\mathrm{m}$ as applied to EDGES low-band data was presented in \citet{2020MNRAS.492...22S}. Additionally, in the context of the LEDA experiment, \citet{2022MNRAS.515.1580S} showed that errors in simulated properties of the soil below the ground plane degrade the effectiveness of BFCC. In either case, accounting for residual chromaticity resulting from imperfect BFCC is found to affect 21-cm signal recovery when not accounted for in the analysis.

Here, we focus on the effectiveness of BFCC when one has an accurate model for $B^\mathrm{m}(\nu, \Omega)$ and $T_\mathrm{fg}^\mathrm{m}(\nu_\mathrm{c}, \Omega, t)$. Errors in the beam model (e.g. \citealt{2022MNRAS.515.1580S, 2022RaSc...5707558R}) and the base-map foreground model (e.g. \citealt{2022arXiv221110448P}) have potential to introduce additional spectral structure into the BFCC data, necessitating additional model complexity to accurately describe the foregrounds and prevent biased recovery of the 21-cm signal. In future work we will consider the impact of additional chromatic structure in the spectrum due to realistic deviations from the assumption of an error-free model for $T_\mathrm{fg}^\mathrm{m}(\nu_\mathrm{c}, \Omega, t)$ and $B^\mathrm{m}(\nu, \Omega)$, deriving from uncertainties in the base-map model and perturbations to the parameters of electromagnetic simulations of the beam within realistic thresholds, respectively. Additionally, we will explore the extent to which the flexible complexity of the non-21-cm component of the BFCC data model derived in this work can be used absorb this structure and reduce or eliminate bias in 21-cm signal recovery that it would otherwise affect.

%%%%%%%%%%%%%%%%%%%%%%%%%%%%%%%%%%%%%%%%%%%%%%%%%%
\section{Chromaticity correction applied to foregrounds with spatially-independent spectral structure}
\label{Sec:ToyModel}
%%%%%%%%%%%%%%%%%%%%%%%%%%%%%%%%%%%%%%%%%%%%%%%%%%

In this section, we begin by examining the impact of BFCC in a scenario in which the data is derived from observations of foregrounds with spatially-independent spectral structure. We use these `toy model foregrounds' to build intuition for the impact of BFCC in the more complex scenario that the spectral structure of the foregrounds in the data is spatially dependent, and we examine this latter scenario, in detail, in \autoref{Sec:RealisticModels}.

Starting with the toy model foregrounds scenario, one can show that a `perfect' correction for instrumental chromaticity can be achieved under the following definition (see Appendix \ref{Sec:PerfectBFCCDefinitions}):
\begin{itemize}
    \item \textit{Perfect BFCC definition} - a correction for which, in the limit of an error-free model for $B^\mathrm{m}(\nu, \Omega)$ and $T_\mathrm{fg}^\mathrm{m}(\nu_\mathrm{c}, \Omega, t)$, a closed-form solution for $T_{\rm corrected}(\nu)$ can be derived in terms of the (assumed known) intrinsic spectral structure of the sky observed by the instrument and the calculated beam-factor.
\end{itemize}
In \autoref{Sec:PerfectBFCC}, we begin by deriving the expected spectrum of BFCC data in this regime.

\subsection{Perfect BFCC}
\label{Sec:PerfectBFCC}

To illustrate the effectiveness of BFCC on data derived from observations of foregrounds with spatially-independent spectral structure it is useful to divide the sky brightness into the two components of interest for 21-cm cosmology,
\begin{equation}
\label{Eq:TskySimple}
T_\mathrm{sky}(\nu, \Omega, t) = T_\mathrm{fg}(\nu, \Omega, t) + T_{21}(\nu) \ .
\end{equation}
Here $T_{21}(\nu)$ is the global 21-cm signal, which we assume is well approximated as isotropic on the scale of the EDGES beam.
\begin{equation}
\label{Eq:Tplfg}
T_\mathrm{fg}(\nu, \Omega, t) = T_\gamma + T_{\rm plfg}(\nu, \Omega, t) \
\end{equation}
is the anisotropic, foreground sky brightness temperature distribution above the antenna at a given time, and is composed of two components, \begin{enumerate*}\item $T_\gamma$, an approximately isotropic CMB temperature and \item $T_{\rm plfg}(\nu, \Omega, t)$, an anisotropic power-law foreground brightness temperature distribution (comprised of the sum of anisotropic Galactic and approximately isotropic extragalactic foregrounds). \end{enumerate*} In this section, we assume the foregrounds have isotropic spectral structure and that the Galactic and extragalactic foreground brightness temperature distribution is described by a single spatially independent power law (in preparation for generalising our foreground spectral structure model to a realistic spatially dependent spectral index distribution in \autoref{Sec:RealisticModels}); however, the result also holds for a foreground with arbitrary spatially independent spectral structure.

With the above definitions and assuming in this section that
\begin{align}
T_{\rm plfg}(\nu, \Omega, t) =  (T_\mathrm{fg}(\nu_\mathrm{c}, \Omega, t) - T_\gamma)\left(\frac{\nu}{\nu_\mathrm{c}}\right)^{-\beta_{0}} \ ,
\end{align}
we can rewrite $T_\mathrm{sky}(\nu, \Omega, t)$ as,
\begin{align}
\label{Eq:ToySky}
T_{\rm sky}(\nu, \Omega, t) &= (T_\mathrm{fg}(\nu_\mathrm{c}, \Omega, t) - T_\gamma)\left(\frac{\nu}{\nu_\mathrm{c}}\right)^{-\beta_{0}} + T_\gamma + T_{21} \ .
\end{align}
Substituting \autoref{Eq:ToySky} into \autoref{Eq:Tcorrected}, the BFCC spectrum is given by,
\begin{align}
\label{Eq:TcorrectedC2pt1}
\MoveEqLeft[3]
T_{\rm corrected}(\nu, t) \\ \nonumber
={}& T_\mathrm{data} / B_\mathrm{factor}\\ \nonumber%
={}& \Bigg[\int\limits_{\Omega^{+}} B(\nu, \Omega) \\ \nonumber
&{}\times \left((T_\mathrm{fg}(\nu_\mathrm{c}, \Omega, t) - T_\gamma)\left(\frac{\nu}{\nu_\mathrm{c}}\right)^{-\beta_{0}} + T_\gamma + T_{21}\right) \mathrm{d}\Omega + n \Bigg] \\ \nonumber
&\times \frac{\int\limits_{\Omega^{+}} B(\nu_\mathrm{c}, \Omega) T_\mathrm{fg}(\nu_\mathrm{c}, \Omega, t) \mathrm{d}\Omega}{\int\limits_{\Omega^{+}} B(\nu, \Omega) T_\mathrm{fg}(\nu_\mathrm{c}, \Omega, t) \mathrm{d}\Omega} \\ \nonumber
={}& T_\mathrm{m_{0}}(t)\left(\frac{\nu}{\nu_\mathrm{c}}\right)^{-\beta_{0}} + \frac{(1-\left(\frac{\nu}{\nu_\mathrm{c}}\right)^{-\beta_{0}}) T_{\gamma}}{B_\mathrm{factor}(\nu, t)} \\ \nonumber
&+ \frac{T_{21}}{B_\mathrm{factor}(\nu, t)} + \frac{n}{B_\mathrm{factor}(\nu, t)} \ .
\end{align}
Here, $T_\mathrm{m_{0}}$ is the beam weighted average foreground brightness at reference frequency $\nu_\mathrm{c}$, the simplification between lines 2 and 3 relies on the spatial and spectral separability of the components of \autoref{Eq:ToySky}, and we have made use of the fact that $T_{\gamma}$ and $T_{21}$ are spatially independent to move them outside the integral. We have arranged the terms to highlight those which, in the BFCC data, are inversely scaled by the beam-factor relative to their intrinsic expectations.

We note that the purpose of BFCC is to suppress beam-chromaticity-induced spectral structure in the foreground component of the data that, if unaccounted for, will bias recovery of the 21-cm signal. The scaling of the latter three terms in \autoref{Eq:TcorrectedC2pt1} by the beam chromaticity correction factor is a side effect of BFCC rather than its intended purpose. This scaling must be accounted for when fitting the corrected data, but is not otherwise an impediment to recovery of unbiased estimates of the 21-cm signal (as will be shown in \autoref{Sec:PerfectBFCCDemonstration}).

\autoref{Eq:TcorrectedC2pt1} provides a closed-form solution for $T_{\rm corrected}(\nu)$ parametrised in terms of the intrinsic spectral structure of the sky, fulfilling \textit{perfect BFCC definition 2}, without requiring condition (iv).

For a sky described by \autoref{Eq:ToySky}, the spectrum measured by a hypothetical spectrometer with a uniform achromatic beam has the form,
\begin{multline}
\label{Eq:USToySky}
T_{\rm data,uniform}(\nu, t) = T_{\rm fg}(\nu_\mathrm{c}, t)\left(\frac{\nu}{\nu_\mathrm{c}}\right)^{-\beta_{0}} + (1-\left(\frac{\nu}{\nu_\mathrm{c}}\right)^{-\beta_{0}})T_{\gamma} \\
+ T_{21} + n \ ,
\end{multline}
where $T_{\rm fg}(\nu_\mathrm{c}, t) = \int\limits_{\Omega^{+}} T_\mathrm{fg}(\nu_\mathrm{c}, \Omega, t) \mathrm{d}\Omega$.

Comparing \autoref{Eq:TcorrectedC2pt1} and  \autoref{Eq:USToySky}, we see that:
\begin{itemize}
 \item the spectral structure of the power-law component of the foregrounds is identical. Due to the relative amplitudes of $T_{\rm fg}(\nu_\mathrm{c}, t)$ and the remaining signal components, by performing perfect chromaticity correction of $T_{\rm fg}(\nu_\mathrm{c}, t)$, BFCC has eliminated the majority of the non-sky-based chromatic structure in $T_{\rm data}(\nu, t)$.
 \item The remaining components differ by factors of $B_\mathrm{factor}$.
Thus, one should expect \autoref{Eq:USToySky} to provide a biased fit to BFCC data; however, since the largest source of unmodelled spectral structure has been eliminated, the level of bias will be significantly mitigated relative to fitting uncorrected data with the same model.
\end{itemize}

\subsection{Time-averaging}
\label{Sec:LSTaveraging}

In the limit of data derived from observations of foregrounds with spatially-independent spectral structure corrected using BFCC with an error-free beam-factor model, \autoref{Eq:TcorrectedC2pt1}, derived in the preceding section, is an accurate model for a chromaticity-corrected snapshot observation at a given time. In practice, to achieve sufficient signal-to-noise\footnote{Throughout this work, we use signal-to-noise to denote the ratio between the amplitude of the 21-cm signal and noise in the data.} to detect the cosmological 21-cm signal, one must fit either:
\begin{itemize}
\item time-dependent data, comprised of multiple spectra, where each spectrum is the average of data falling in a given time-bin (potentially over a number of sidereal days); or,
\item a single spectrum averaged in time (and potentially over a number of sidereal days).
\end{itemize}

A fit to time-dependent data enables one to leverage angular, in addition to spectral, information to distinguish between the foregrounds and 21-cm signal (e.g. \citealt{2013PhRvD..87d3002L, 2020ApJ...897..175T, 2020ApJ...897..132T, 2022arXiv221004707A}). This provides more stringent constraints on ones instrument weighted sky model and correspondingly enables the placement of stronger constraints on the cosmological signal when one has a sufficiently high fidelity model of the instrument and foregrounds.

In contrast and for the same reason, one expects averaging the data to a single spectrum to mitigate time-dependent systematics arising from imperfect sky and instrument modelling, if they are uncorrelated or weakly correlated on the time-scales being averaged over. Correspondingly, this enables a reduction in the bias of the signal estimates, at the expense of having a less well constrained estimate of the 21-cm signal. Since in this case the requirements on the precision of knowledge of the sky and instrument are lower and because this is the case applicable to modelling the publicly available EDGES low-band data, here we focus on modelling time-averaged data and, thus, on defining the appropriate model for BFCC data described by \autoref{Eq:TcorrectedC2pt1} averaged over time to form a single spectrum.

Assuming the noise in the data is uncorrelated between frequencies\footnote{This is a good approximation in the publicly-available EDGES low-band data (\citealt{2022MNRAS.tmp.2422M})}, the optimal, with respect to signal-to-noise, inverse-variance weighted average, over time, of BFCC data described by \autoref{Eq:TcorrectedC2pt1} is given by,
\begin{align}
\label{Eq:IVWeightedLSTaveragedTcorrected}
\MoveEqLeft[3]
T_{\rm corrected}(\nu) \\ \nonumber
={}&  \frac{1}{W(\nu)} \sum\limits_{i=1}^{N_{t}} \frac{1}{\sigma_{i}^{2}(\nu)}\Bigg[T_\mathrm{m_{0}}(t_{i})\left(\frac{\nu}{\nu_\mathrm{c}}\right)^{-\beta_{0}} + \frac{(1-\left(\frac{\nu}{\nu_\mathrm{c}}\right)^{-\beta_{0}}) T_{\gamma}}{B_\mathrm{factor}(\nu, t_{i})} \\ \nonumber
&+ \frac{T_{21}}{B_\mathrm{factor}(\nu, t_{i})} + \frac{n}{B_\mathrm{factor}(\nu, t_{i})}\Bigg] \ ,
\end{align}
where $N_{t}$ is the number of snapshots in the time-range being averaged over, $\sigma_{i}^{2}(\nu)$ is the expected variance of the noise on the data at frequency $\nu$, and $W(\nu) = (\sum_{i=1}^{N_{t}}(1/\sigma_{i}^{2}(\nu)))$.
Defining the time- and sky-averaged beam weighted foreground brightness at reference frequency, $\nu_\mathrm{c}$, as,
\begin{align}
\label{Eq:IVWeightedLSTaveragedTfg}
\bar{T}_\mathrm{m_{0}} = \frac{1}{W} \sum\limits_{i=1}^{N_{t}} \frac{1}{\sigma_{i}^{2}(\nu)}T_\mathrm{m_{0}}(t_{i}) \ ,
\end{align}
the time-averaged beam-factor as,
\begin{align}
\label{Eq:IVWeightedLSTaveragedBfactor}
\bar{B}_\mathrm{factor}(\nu) = \left[ \frac{1}{W} \sum\limits_{i=1}^{N_{t}} \frac{1}{\sigma_{i}^{2}(\nu)B_\mathrm{factor}(\nu, t_{i})} \right]^{-1} \ ,
\end{align}
and the time-averaged noise on the BFCC data as,
\begin{align}
\label{Eq:IVWeightedLSTaveragedNoise}
\bar{n}(\nu) = \frac{1}{W} \sum\limits_{i=1}^{N_{t}} \frac{n(\nu, t_{i})}{\sigma_{i}^{2}(\nu)B_\mathrm{factor}(\nu, t_{i})} \ ,
\end{align}
\autoref{Eq:IVWeightedLSTaveragedTcorrected} can be rewritten as,
\begin{align}
\label{Eq:IVWeightedLSTaveragedTcorrected2}
\MoveEqLeft[3]
T_{\rm corrected}(\nu) \\ \nonumber
={}&  \bar{T}_\mathrm{m_{0}}\left(\frac{\nu}{\nu_\mathrm{c}}\right)^{-\beta_{0}} + \frac{(1-\left(\frac{\nu}{\nu_\mathrm{c}}\right)^{-\beta_{0}}) T_{\gamma}}{\bar{B}_\mathrm{factor}(\nu)} + \frac{T_{21}}{\bar{B}_\mathrm{factor}(\nu)} + \bar{n} \ .
\end{align}
If we uniformly weight snapshots when averaging over time\footnote{Uniformly weighting snapshot spectra when averaging over time is optimal, with respect to signal-to-noise on the averaged data, only if the noise in each snapshot can be reasonably approximated as constant, in which case Equations \ref{Eq:IVWeightedLSTaveragedTfg}--\ref{Eq:IVWeightedLSTaveragedNoise} and Equations \ref{Eq:SimpleLSTweightedTfg}--\ref{Eq:SimpleLSTweightedNoise} are equivalent.}, Equations \ref{Eq:IVWeightedLSTaveragedTfg}--\ref{Eq:IVWeightedLSTaveragedNoise} simplify to,
\begin{subequations}
\begin{align}
\label{Eq:SimpleLSTweightedTfg}
\bar{T}_\mathrm{m_{0}} &= \frac{1}{N_{t}} \sum\limits_{i=1}^{N_{t}} T_\mathrm{m_{0}}(t_{i}) \ , \\
\label{Eq:SimpleLSTweightedBfactor}
\bar{B}_\mathrm{factor}(\nu) &= \left[ \frac{1}{N_{t}} \sum\limits_{i=1}^{N_{t}} \frac{1}{B_\mathrm{factor}(\nu, t_{i})} \right]^{-1} \ , \\
\label{Eq:SimpleLSTweightedNoise}
\bar{n}(\nu) &= \frac{1}{N_{t}} \sum\limits_{i=1}^{N_{t}} \frac{n(\nu, t_{i})}{B_\mathrm{factor}(\nu, t_{i})} \ .
\end{align}
\end{subequations}
To match the processing of the publicly available EDGES low-band data, we assume uniformly weighted averaging of the data and correspondingly apply the latter set of definitions in the analyses carried out in the remainder of this paper.

\subsection{Demonstration of perfect BFCC in simulated data with spatially-independent foreground spectra}
\label{Sec:PerfectBFCCDemonstration}

To understand the impact of BFCC on 21-cm signal recovery from spectrometer data in the simplifying limit that the foregrounds in the data have spatially-independent spectral structure, we construct simulated EDGES low-band data, in this approximate regime, and analyse it in three scenarios:
\begin{enumerate}
\item Fitting the uncorrected data, $T_{\rm data}(\nu)$, incorporating a flattened Gaussian 21-cm absorption trough, averaged over time in the same manner as described above, with a model based on the intrinsic spectral structure of the foreground and 21-cm emission in the simulations  (\autoref{Eq:USToySky}).
\item Fitting $T_{\rm corrected}(\nu)$ (the data above after BFCC has been applied) with a model based on the intrinsic spectral structure of the emission in the simulations.
\item Fitting $T_{\rm corrected}(\nu)$ using \autoref{Eq:IVWeightedLSTaveragedTcorrected2} - the correct analytic model for BFCC data derived from intrinsic emission on the sky described by  \autoref{Eq:ToySky}. This scenario illustrates unbiased recovery of the 21-cm signal when one constructs and fits the correct model for the BFCC data.
\end{enumerate}

\subsubsection{Simulated data}
\label{Sec:PerfectBFCCSimulations}

Starting with scenario (i), we construct $T_{\rm data}(\nu)$ over a $50-100~\mathrm{MHz}$ spectral band, assuming a $1~\mathrm{MHz}$ channel width and integration over a short time interval ($\upDelta t = 6~\mathrm{minutes}$) such that we can work with \autoref{Eq:Tdata} in the snapshot limit (see \autoref{Sec:ChromaticityCorrectionFormalism}),
\begin{equation}
\label{Eq:TdataSimSimple}
T_\mathrm{data}(\nu, t) = \int\limits_{\Omega^{+}} B(\nu, \Omega)T_\mathrm{sky}(\nu, \Omega, t)~\mathrm{d}\Omega + n \ .
\end{equation}
For these simulations, $n$ is zero-mean Gaussian random noise and to illustrate the relatively subtle differences between the fits of models (ii) and (iii) in this simplified foreground scenario, we inject noise such that the resultant noise in the BFCC data after time-averaging is Gaussian and white, with an RMS amplitude of $1~\mathrm{mK}$; finally, $T_\mathrm{sky}(\nu, \Omega, t) = T_\mathrm{fg}(\nu, \Omega, t) + T_{21}(\nu)$, with,
\begin{multline}
\label{Eq:TFgSimple}
T_\mathrm{fg}(\nu, \Omega, t) = (T_\mathrm{fg}(408~\mathrm{MHz}, \Omega, t) - T_\gamma)\left(\frac{\nu}{408~\mathrm{MHz}}\right)^{-\beta_{0}}  + T_\gamma \ .
\end{multline}
Here, $T_\mathrm{fg}(408~\mathrm{MHz}, \Omega, t)$ is given by the Haslam 408 MHz all-sky map (\citealt{1981A&A...100..209H, 1982A&AS...47....1H}) reprocessed by \citet{2015MNRAS.451.4311R}, $T_\gamma = 2.725~\mathrm{K}$ is the CMB temperature and $T_{21}$ is a flattened Gaussian model for the 21-cm signal of the form,
\begin{equation}
\label{Eq:FlattenedGaussian}
T_\mathrm{21}(\nu) = -A\left(\frac{1-\e^{-\tau \e^{B_{21}}}}{1-\e^{-\tau}}\right) \ ,
\end{equation}
where,
\begin{equation}
\label{Eq:FlattenedGaussianB}
B_{21} = \frac{4(\nu-\nu_0)^2}{w^2}\log\left[-\frac{1}{\tau}\log\left(\frac{1+\e^{-\tau}}{2}\right)\right] \ ,
\end{equation}
and $A=100~\mathrm{mK}$, $\nu_0=75~\mathrm{MHz}$, $w=10~\mathrm{MHz}$ and $\tau=4$ describe the amplitude, central frequency, width and flattening of the absorption trough included in the simulated data, respectively.

For our beam model, $B(\nu, \Omega)$, we use a {\sc{FEKO}} EM simulation of the EDGES low-band blade dipole  antenna, with a $30~\mathrm{m} \times 30~\mathrm{m}$ sawtooth ground plane on top of soil with a conductivity of $\sigma_\mathrm{c} = 0.02~\mathrm{S~m^{-1}}$ and relative permittivity of $\epsilon_\mathrm{r} = 3.5$, consistent with the soil properties reported by \citet{2015ITAP...63.5433S} for the Murchison Radio-astronomy Observatory, where EDGES is located. For further details regarding this beam model, see \citet{2021AJ....162...38M}.

\begin{figure}
	\centerline{
	\includegraphics[width=0.5\textwidth]{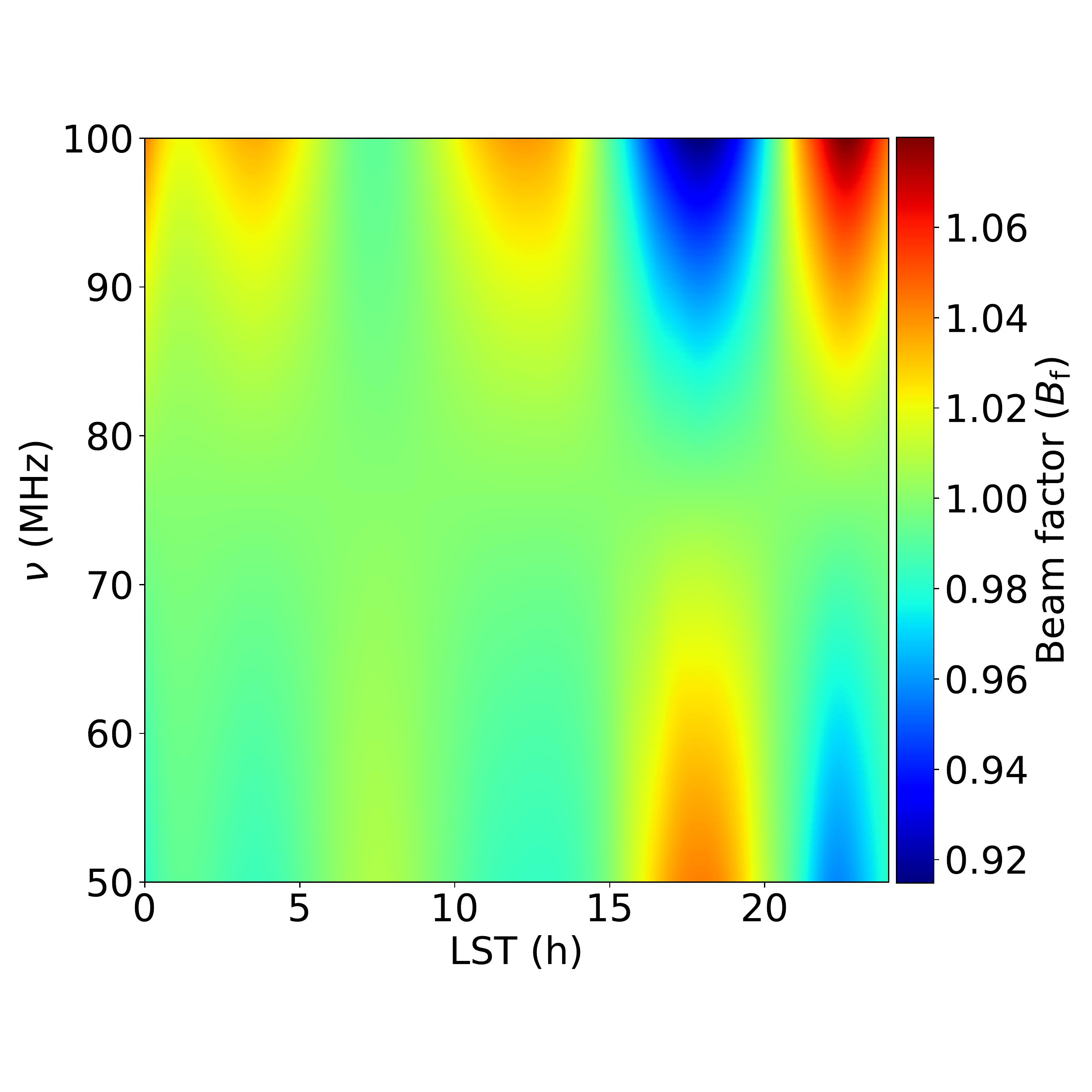}
	}
\caption{Simulated beam chromatic correction factor $B_\mathrm{factor}(\nu, t)$ (\autoref{Eq:CCsnapshot}) derived using a model of the EDGES blade antenna $B^\mathrm{m}(\nu, \Omega)$ and for the time-dependent foreground sky above the antenna at $\nu_\mathrm{c}=75~\mathrm{MHz}$.}
\label{Fig:BeamFactor}
\end{figure}

We construct our time-dependent beam-factor model, $B_\mathrm{factor}(\nu, t)$, using \autoref{Eq:CCsnapshot}, with $B^\mathrm{m}(\nu, \Omega) \equiv B(\nu, \Omega)$ and $T_\mathrm{fg}^\mathrm{m}(\nu_\mathrm{c}, \Omega, t) \equiv T_\mathrm{fg}(\nu_\mathrm{c}, \Omega, t)$. Here, $B^\mathrm{m}(\nu, \Omega)$ and $T^\mathrm{m}_\mathrm{fg}(\nu_\mathrm{c}, \Omega, t)$ are, respectively, the EM simulation of the EDGES low-band beam, and the time-dependent foreground sky above the antenna used for construction of the simulated data described above (\autoref{Eq:TFgSimple}), evaluated at $\nu_\mathrm{c}=75~\mathrm{MHz}$. The resulting beam-factor model is shown in \autoref{Fig:BeamFactor}. Using our simulated data, $T_\mathrm{data}(\nu, t)$, and beam-factor, $B_\mathrm{factor}(\nu, t)$, we  derive the BFCC data, $T_{\rm corrected}(\nu, t)$, using \autoref{Eq:Tcorrected}.

We calculate the corresponding time-averaged $T_\mathrm{data}(\nu)$ (used in analysis scenario (i)) and $T_{\rm corrected}(\nu)$ (BFCC data used in analysis scenarios (ii) and (iii)) by averaging $T_\mathrm{data}(\nu, t)$ and $T_{\rm corrected}(\nu, t)$ over the 120 simulated snapshot spectra derived at 6 minute intervals in the LST range $0 \le LST < 12~\mathrm{h}$. This range is selected to match the LST window of the publicly available EDGES low-band data, when the Galactic plane is relatively low in the beam.

\subsubsection{Instrumentally induced chromaticity}
\label{Sec:InstrumentallyInducedChromaticity}

We define $P_{B_{\mathrm{f}}}(\eta, t)$, the power spectrum of the mean-subtracted, windowed beam-factor, as:
\begin{equation}
\label{Eq:GainAmplitudePS}
\left\langle \widetilde{B_\mathrm{factor}^\prime}(\eta, t)\widetilde{B_\mathrm{factor}^\prime}^{*}(\eta^\prime, t) \right\rangle(\Delta\nu)^{2} \equiv \delta_\mathrm{K}(\eta-\eta^\prime)P_{B_{\mathrm{f}}}(\eta, t) \ .
\end{equation}
This provides a simple metric to describe the level of instrumental chromaticity imparted by the beam to the time-dependent spectrum measured by the instrument. Here, $B_\mathrm{factor}^\prime = W \delta B_\mathrm{factor}$, with $W$ a Blackmann-Harris window function to prevent spectral leakage of power on scales in excess of the bandwidth of the data into the spectral scales of interest and $\delta B_\mathrm{factor}(\nu, t) = B_\mathrm{factor}(\nu, t) - B_\mathrm{factor}(t)$, with $B_\mathrm{factor}(t)$ the average of $B_\mathrm{factor}(\nu, t)$ over frequency; $\eta$ measures inverse spectral scale, $\widetilde{B_\mathrm{factor}^\prime}(\eta, t)$ is the Fourier transform of the mean-subtracted, windowed beam-factor with respect to frequency, $\Delta\nu = 1~\mathrm{MHz}$ is the channel width and $\delta_\mathrm{K}$ is the Kronecker delta function.

\begin{figure}
	\centerline{
        \includegraphics[width=0.5\textwidth]{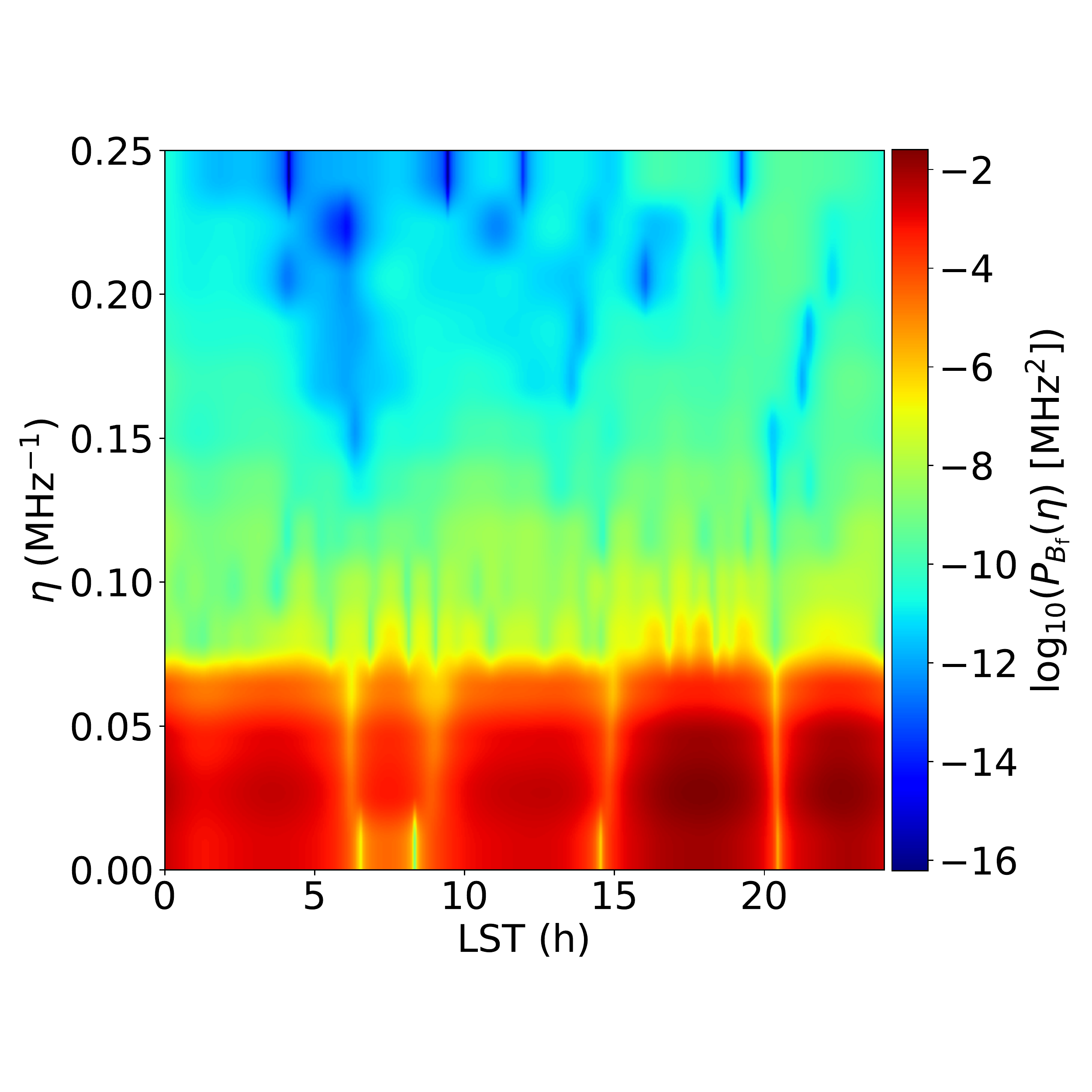}
        }
\caption{Power spectrum, $P_{B_{\mathrm{f}}}(\eta, t)$, of the mean-subtracted, windowed beam-factor, $B_\mathrm{factor}^\prime(\nu, t)$, shown in \autoref{Fig:BeamFactor}. $P_{B_{\mathrm{f}}}(\eta, t)$ provides a simple metric of the level of instrumental chromaticity imparted by the beam to the time-dependent spectrum measured by the instrument. Since the beam-factor is real and, thus, its Fourier transform is symmetric, we display the power spectrum only for positive $\eta$.
}
\label{Fig:BeamFactorPS}
\end{figure}

A plot of $P_{B_{\mathrm{f}}}(\eta, t)$ for the EDGES low-band beam is shown in \autoref{Fig:BeamFactorPS}. The $10^{5} - 10^{6}:1$ dynamic range between the order $10^{3}$--$10^{4}~\mathrm{K}$ foregrounds in the spectral range of interest and the order of magnitude $10~\mathrm{mK}$ noise level in the publicly available EDGES low-band data correspond to a requirement that, for unbiased 21-cm signal inference in the absence of BFCC, $P_{B_{\mathrm{f}}}(\eta, t) \ll 10^{-10}~\mathrm{MHz^{2}}$ on spectral scales relevant for recovery of the 21-cm signal. From \autoref{Fig:BeamFactorPS} it can be seen that there is no LST range where this condition is met for $\eta < 0.15~\mathrm{MHz^{-1}}$ and only in limited LST windows for $\eta \gtrsim 0.15~\mathrm{MHz^{-1}}$.

\begin{figure}
	\centerline{
        \includegraphics[width=0.5\textwidth]{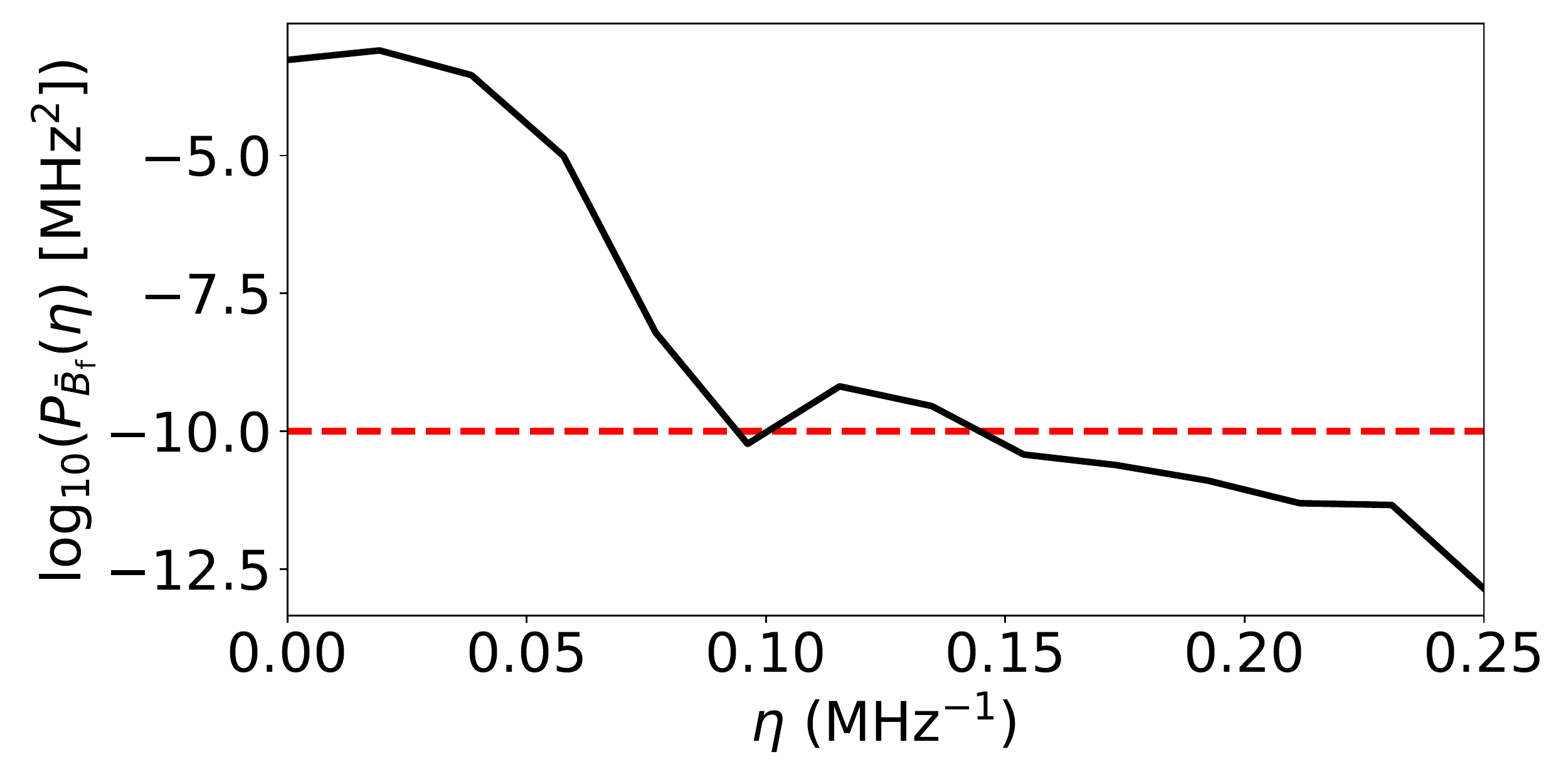}
        }
    \caption{Power spectrum, $P_{\bar{B}_{\mathrm{f}}}(\eta)$, of $\bar{B}_\mathrm{factor}(\nu)$, the mean-subtracted, windowed, simulated beam-factor, time-averaged over the LST range $0 \le LST < 12~\mathrm{h}$, matching the LST window of the publicly available EDGES low-band data. The dashed red line displays the power in fluctuations in $\bar{B}_\mathrm{factor}(\nu)$ above which coupling between flat-spectrum foregrounds and the beam will induce spectral structure in excess of the noise level in the publicly available EDGES low-band data.
        }
\label{Fig:LSTaveragedBeamFactorPS}
\end{figure}

\autoref{Fig:LSTaveragedBeamFactorPS} shows $P_{\bar{B}_{\mathrm{f}}}(\eta)$, the power spectrum of $\bar{B}_\mathrm{factor}(\nu)$, the mean-subtracted, windowed, simulated beam-factor, time-averaged over the LST range $0 \le LST < 12~\mathrm{h}$, matching the LST window of the publicly available EDGES low-band data. The reduction in power, particularly on large spectral scales (low-$\eta$) results from averaging down of beam-factor fluctuations uncorrelated on $12~\mathrm{h}$ time-scales. Nevertheless, power in $\bar{B}_\mathrm{factor}(\nu)$ remains sufficiently high on the vast majority of spectral scales ($\eta \lesssim 0.2~\mathrm{MHz^{-1}}$, or frequency scales larger than $5~\mathrm{MHz}$) to bias recovery of the 21-cm signal unless one employs a high complexity foreground model capable of absorbing this structure or accurately chromaticity corrects the data prior to estimating the 21-cm signal.

\subsubsection{Bayes' Theorem}
\label{Sec:BayesTheorem}

Bayesian inference provides a consistent approach to estimate a set of parameters, $\sTheta$, from a model, $M$, given a set of data $\bm{D}$ and, through the use of the Bayesian evidence, $\mathrm{Pr}(\bm{D}\vert M_{i})\equiv\mathcal{Z}_{i}$, and model priors, $\mathrm{Pr}(M_{i})$, to estimate from a set of models, the ones that are the most probable given the data. Bayes' theorem states that,
\begin{equation}
\label{Eq:BayesEqn}
\mathrm{Pr}(\upTheta\vert\bm{D},M) = \dfrac{\mathrm{Pr}(\bm{D}\vert\upTheta,M)\ \mathrm{Pr}(\upTheta\vert M)}{\mathrm{\mathrm{Pr}}(\bm{D}\vert M)} \ .
\end{equation}
Here, $\mathrm{Pr}(\upTheta\vert\bm{D},M)$ is the posterior probability distribution of the parameters, $\mathrm{Pr}(\bm{D}\vert\upTheta,M)$ is the likelihood, and $\mathrm{Pr}(\upTheta\vert M)$ is the prior probability distribution of the parameters.

The Bayesian model evidence (BME; the factor required to normalise the posterior over the parameters), is given by,
\begin{equation}
\label{Eq:Evidence}
\mathcal{Z}=\int\mathrm{Pr}(\bm{D}\vert\upTheta,M)\ \mathrm{Pr}(\upTheta\vert M)\mathrm{d}^{p}\upTheta \ ,
\end{equation}
where $p$ is the dimensionality of the parameter space.

Given samples from the posterior distribution of the parameters, $\mathrm{Pr}(\upTheta\vert\bm{D},M)$, one can estimate $\mathrm{Pr}(y\vert\nu,\bm{D},M)$, the predictive posterior of a function $y = f(\nu,\upTheta)$ by calculating the corresponding set of samples from $\mathrm{Pr}(y\vert\nu,\bm{D},M)$.
When analysing data in \autoref{Sec:DataModelSimple}, and later in \autoref{Sec:Results}, we sample from the posteriors on the model parameters given the data using nested sampling as implemented by the \textsc{polychord} algorithm \citep{2015MNRAS.453.4384H, 2015MNRAS.450L..61H}, and derive contour plots of functional posterior probability distributions using the \textsc{fgivenx} software package (\citealt{2018JOSS....3..849H}).

\subsubsection{Data likelihood}
\label{Sec:InstrumentallyInducedChromaticity}

In each of our three analysis scenarios, we assume zero-mean uncorrelated Gaussian random noise, $n$, on our data, with covariance matrix $\mathbfss{N}$ given by,
\begin{equation}
N_{ij} = \left< n_in_j^*\right> = \delta_{ij}\sigma_{j}^{2} \ ,
\end{equation}
where $\left< .. \right>$ represents the expectation value and $\sigma_{j}$ is the RMS value of the noise term for spectral channel $j$. In the frequency range of interest, the intrinsic noise in EDGES low-band data is expected to be dominated by power law sky noise, with a smaller $\sim200~\mathrm{K}$ contribution from the receiver (see B18, Extended Data Figure 5). However, the reduced data volume resulting from RFI flagging due to FM radio ($\nu \gtrsim 85~\mathrm{MHz}$) increases the effective noise level in the upper end of the band, such that the effective noise is roughly flat. In this paper, for simplicity, we simulate the noise as flat across the band and model it as such when fitting the data.

Writing our data vectorised over frequency as $\mathbf{d}$, a vectorised model for the data constructed from a set of parameters $\Theta$ as $\mathbf{m(\Theta)}$, and the residuals between the data and model as $\mathbf{r} = \mathbf{d} - \mathbf{m(\Theta)}$, we can write a Gaussian likelihood for $\mathbf{r}$ as,
\begin{equation}
\label{Eq:BasicVisLike}
\mathrm{Pr}(\mathbf{d} | \mathbf{\Theta}) = \frac{1}{\sqrt{(2\pi)^{N_{\mathrm{chan}}})\mathrm{det}(\mathbfss{N})}} \exp\left[-\frac{1}{2}\mathbf{r(\Theta)}^T\mathbfss{N}^{-1}\mathbf{r(\Theta)}\right].
\end{equation}
When fitting the data in analysis scenario (i) $\mathbf{d} = \mathrm{vec}(T_{\rm data}(\nu))$ and in scenarios (ii) and (iii) $\mathbf{d} = \mathrm{vec}(T_{\rm corrected}(\nu))$; here, for a function $X$, we define $\mathrm{vec}(.)$ such that for a frequency-dependent measurement $X$, $\mathrm{vec}(X(\nu)) = [X_{0}, X_{1}, \cdots, X_{N_{\mathrm{chan}}}]^{T}$, where $X_{i}$ is the value of $X$ in frequency channel $i$ and $N_{\mathrm{chan}}$ is the number of channels in the data set. We define our models for the data in the three analysis scenarios and their fits to the data in \autoref{Sec:DataModelSimple}.

\subsubsection{Data models and fits}
\label{Sec:DataModelSimple}

We construct our data model for scenarios (i) and (ii) as the time-average of \autoref{Eq:USToySky} (the spectrum obtained by observing a sky brightness temperature distribution given by \autoref{Eq:ToySky} with an achromatic beam),
\begin{equation}
\label{Eq:TskySimSimpleIntrinsicModel}
T_\mathrm{sky}^\mathrm{m}(\nu) = \bar{T}_\mathrm{m_{0}}\left(\frac{\nu}{\nu_\mathrm{c}}\right)^{-\beta_{0}} + (1-\left(\frac{\nu}{\nu_\mathrm{c}}\right)^{-\beta_{0}})T_{\gamma} + T_{21} \ .
\end{equation}

Here, $\nu_\mathrm{c}=75~\mathrm{MHz}$ is a reference frequency which we set to the center of the spectral band of the data, $\bar{T}_\mathrm{m_{0}}$ is a free parameter with an expected value equal to the average, over the time-range of the data, of the observed CMB-subtracted foreground brightness temperature at frequency $\nu_\mathrm{c}$, $\beta_{0}=2.5$ is the foreground temperature spectral index, which, here, we assume is known, $T_\gamma \simeq 2.725~\mathrm{K}$ is the CMB temperature, and $T_{21}$ is a model for the flattened Gaussian absorption profile described by \autoref{Eq:FlattenedGaussian}, with free parameters $A$, $\nu_0$, $w$, and $\tau$ as described below \autoref{Eq:FlattenedGaussianB}. For scenario (iii), we use \autoref{Eq:IVWeightedLSTaveragedTcorrected2} (minus the noise) as our data model and fit as free parameters $\bar{T}_\mathrm{m_{0}}$ and the four flattened Gaussian absorption profile parameters.

We note that Equations \ref{Eq:IVWeightedLSTaveragedTcorrected2} and \ref{Eq:TskySimSimpleIntrinsicModel} have similar parametrisations but the latter lacks explicit accounting for the beam-factor scaling of the non-power-law foregrounds components of the data present in \autoref{Eq:TskySimSimpleIntrinsicModel}. The (shared) priors on the parameters of the two models are listed in \autoref{Tab:DataModelPriors}.

\begin{table}
\caption{Priors on the parameters of the 21-cm signal model component fit in \autoref{Sec:DataModelSimple}.}
\centerline{
\begin{tabular}{l l l }
\hline
Parameter & Model component & Prior     \\
\hline
$\bar{T}_\mathrm{m_{0}}$ & foreground     & $U(10^{3},10^{4})~\mathrm{K}$ \\
$A$                      & 21-cm signal   & $U(0,1000)~\mathrm{mK}$ \\
$\nu_0$                  & 21-cm signal   & $U(55,95)~\mathrm{MHz}$ \\
$w$                      & 21-cm signal   & $U(5,30)~\mathrm{MHz}$ \\
$\tau$                   & 21-cm signal   & $U(0,20)$  \\
\hline
\end{tabular}
}
\label{Tab:DataModelPriors}
\end{table}

\begin{figure*}
	\centerline{
	\begin{subfigure}[t]{0.5\textwidth}
	\caption{}
	\includegraphics[width=\textwidth]{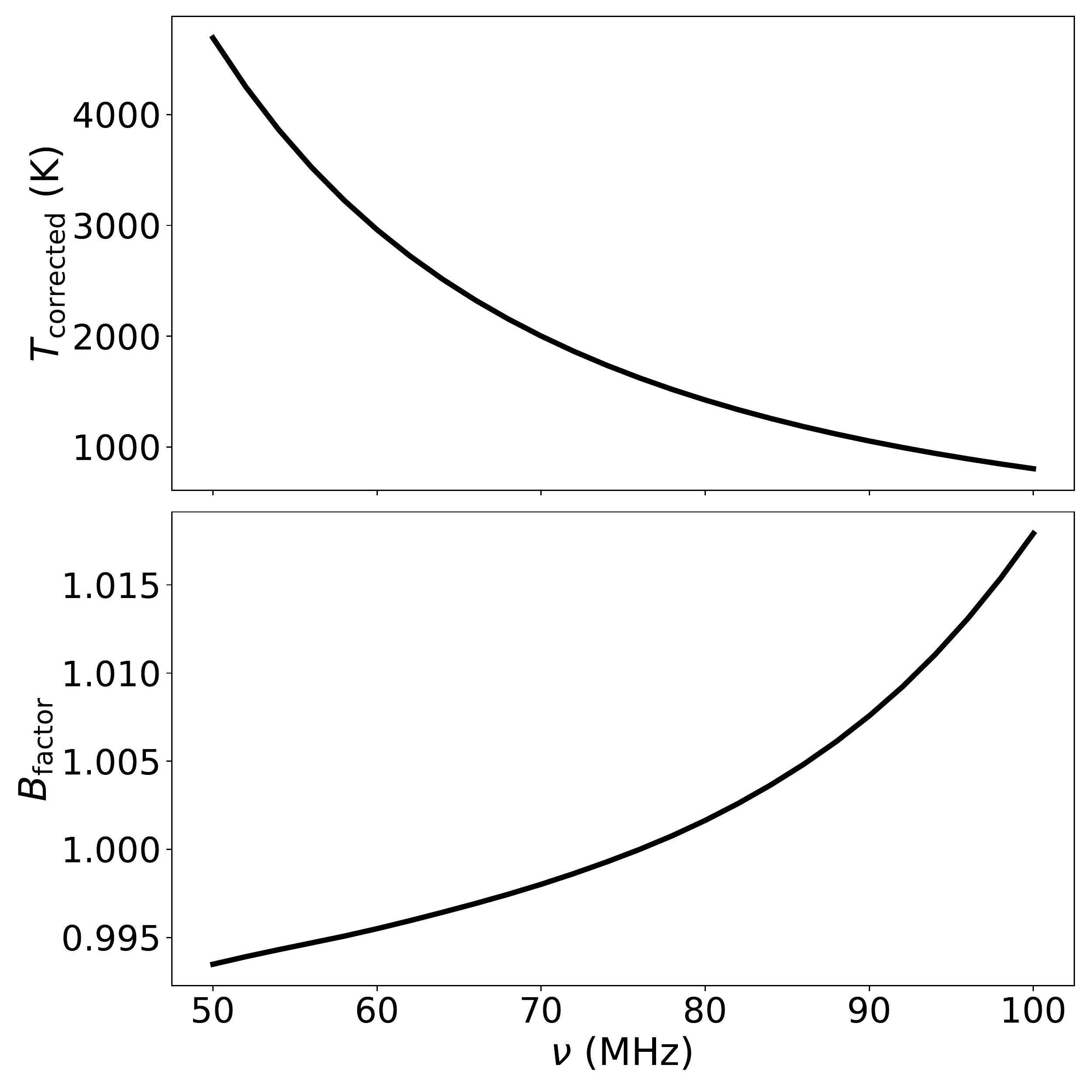}
	\end{subfigure}
	\begin{subfigure}[t]{0.5\textwidth}
	\caption{}
	\includegraphics[width=\textwidth]{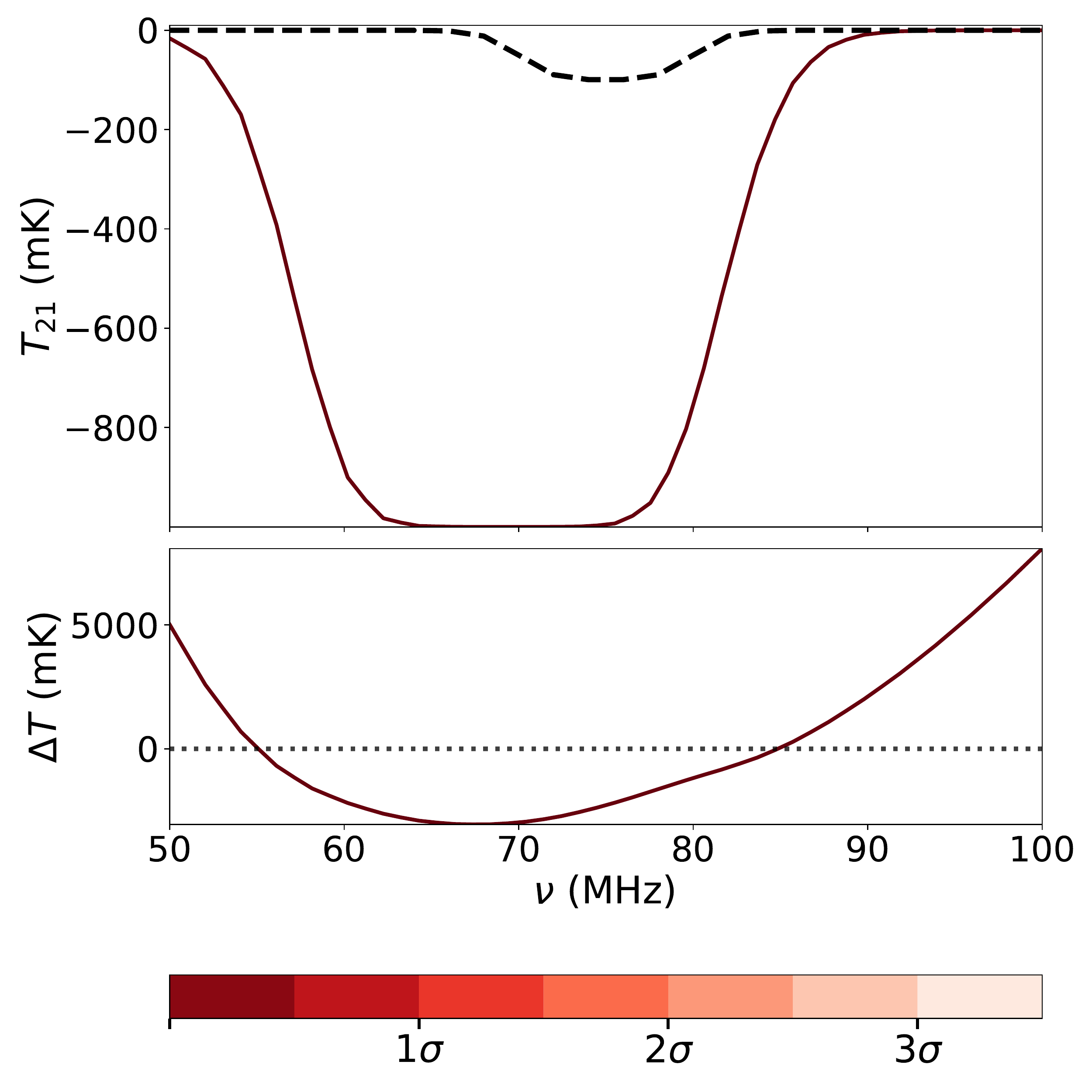}
	\end{subfigure}
	}
	\centerline{
	\begin{subfigure}[t]{0.5\textwidth}
	\caption{}
	\includegraphics[width=\textwidth]{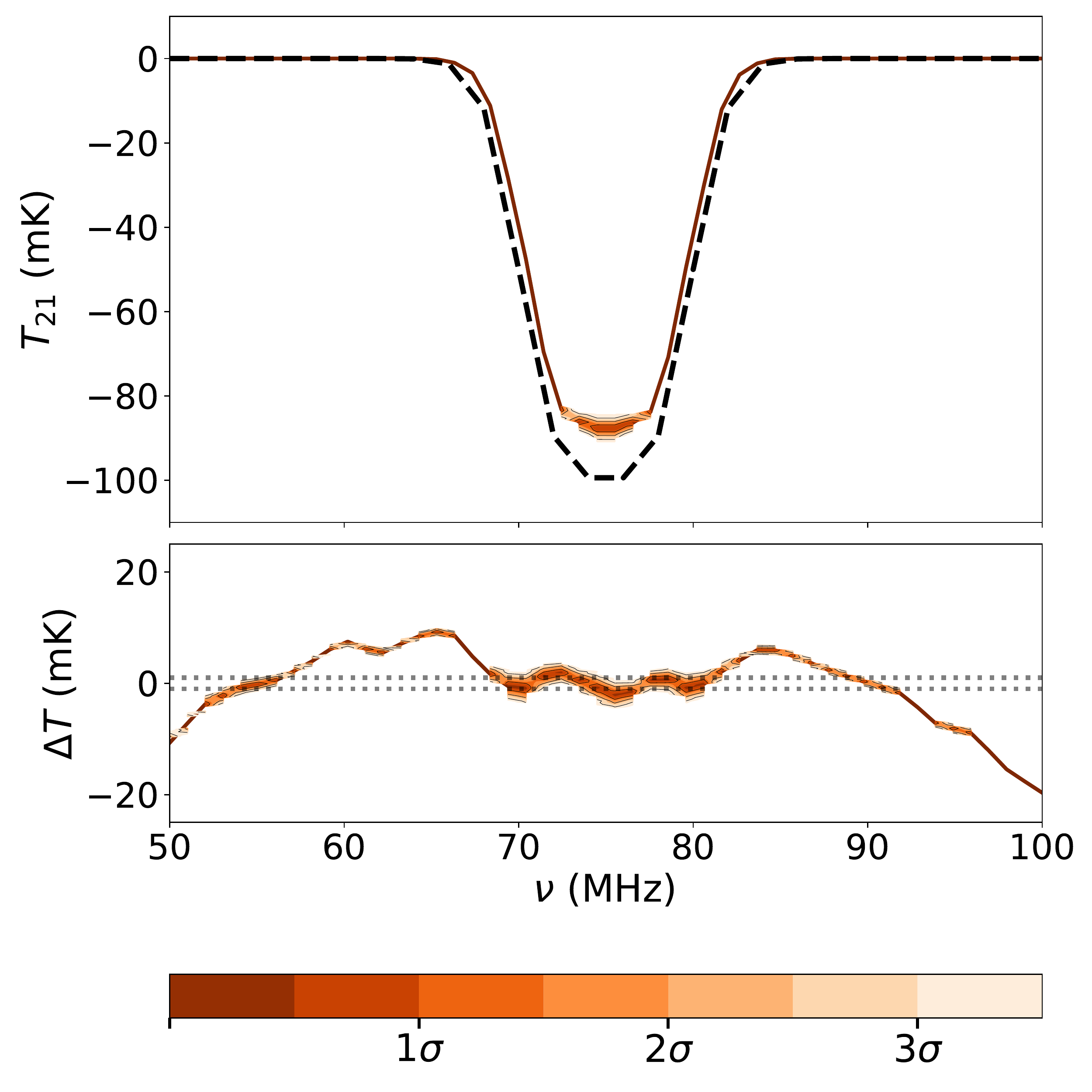}
	\end{subfigure}
	\begin{subfigure}[t]{0.5\textwidth}
	\caption{}
	\includegraphics[width=\textwidth]{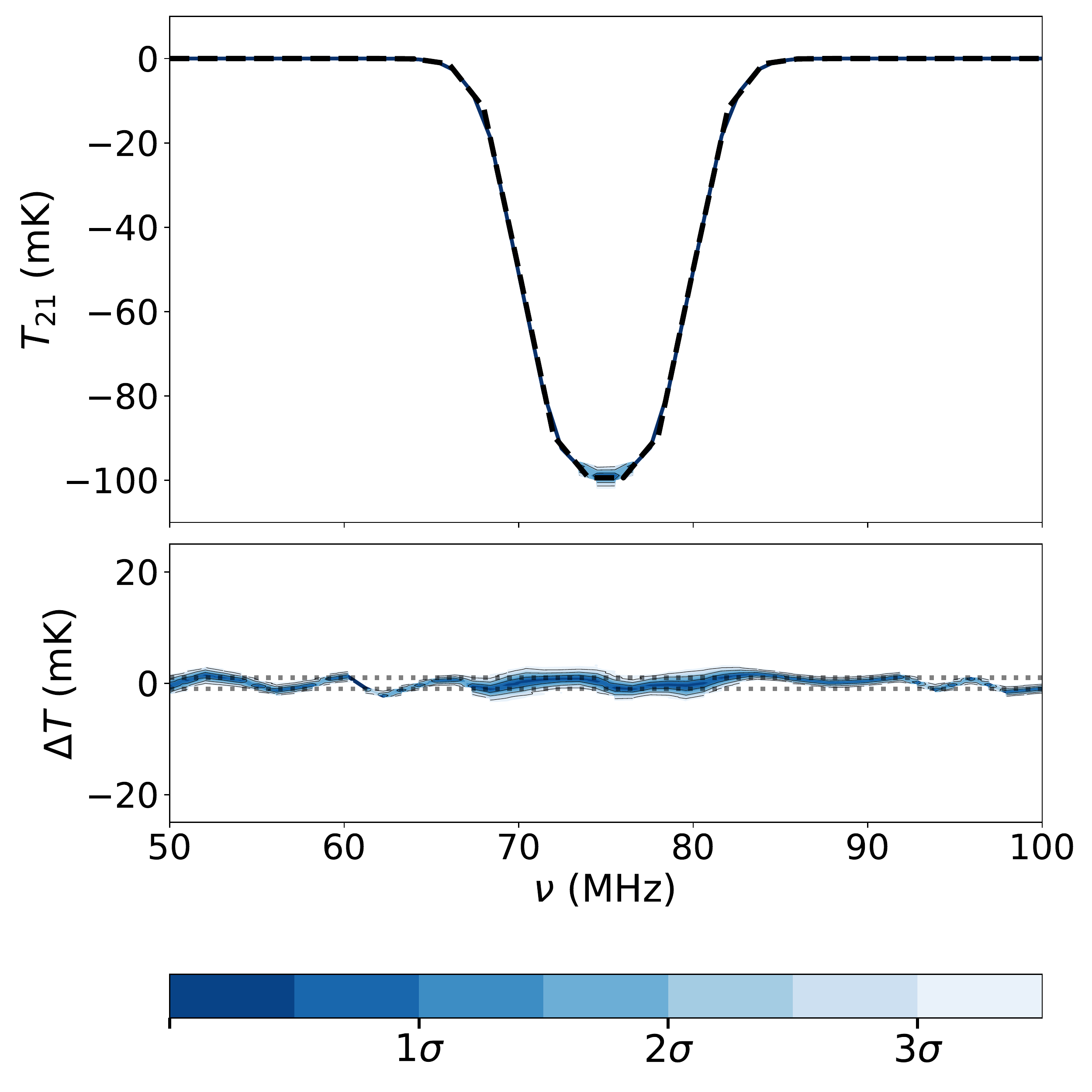}
	\end{subfigure}
	}
\caption{
Subplot (a) shows the time-averaged, BFCC corrected data ($T_{\rm corrected}(\nu)$; top) and beam-factor ($\bar{B}_\mathrm{factor}(\nu)$; bottom). On the scale shown $T_{\rm corrected}(\nu)$ and $T_{\rm data}(\nu)$ are visually indistinguishable, so we plot only the former. Subplots (b)--(d) show results for scenarios (i) -- fitting the uncorrected data, $T_{\rm data}(\nu)$, with a model based on the intrinsic spectral structure of the foreground and 21-cm emission in the simulations, (ii) -- fitting $T_{\rm corrected}(\nu)$ with a model based on the intrinsic spectral structure of the emission in the simulations and (iii) -- fitting $T_{\rm corrected}(\nu)$ using the correct analytic model for BFCC data, respectively. In each, the black dashed line in the top subfigure shows the input 21-cm signal, $T_{21}$, and they are overlaid with coloured iso-probability contours of the recovered functional posteriors for the signal. The bottom subplots show the recovered functional posteriors on the residuals $\upDelta T = (d-m)$, where $d$ and $m$ are the data analysed and model fit in the relevant analysis scenario, respectively - the uncorrected data fit with the intrinsic sky model in scenario (i), the BFCC corrected data fit with the intrinsic sky model in scenario (ii), and the  BFCC corrected data fit with the analytic BFCC model in scenario (iii). The dotted lines display the $1$-$\sigma$ noise level in each case.
}
\label{Fig:SpatiallyInvariantForegrounds}
\end{figure*}

Functional posterior distributions resulting from the fits in scenarios (i)--(iii) are shown in \autoref{Fig:SpatiallyInvariantForegrounds}. The time-averaged BFCC corrected data, $T_{\rm corrected}(\nu)$, and beam-factor, $\bar{B}_\mathrm{factor}(\nu)$, are displayed in the top and bottom subpanels of subfigure (a). $T_{\rm corrected}(\nu)$ and $T_{\rm data}(\nu)$ would be visually indistinguishable on the scale shown, so we show only the former. Subplots (b)--(d) show the results for scenarios (i)--(iii), respectively. The means and $1$-$\sigma$ uncertainties of the 21-cm signal parameter estimates associated with subplots (b)--(d) (as well as the input 21-cm parameter values in the simulated data) are listed in \autoref{Tab:DataModelSimpleResultsTable}.

\begin{table}
\caption{Input values of the parameters of the 21-cm signal in \autoref{Sec:DataModelSimple} and the recovered means and $1$-$\sigma$ uncertainties of the parameters as a function of the analysis scenario considered.}
\centerline{
\begin{tabular}{l l l l }
\hline
Parameter & Scenario & Mean & $1$-$\sigma$ uncertainty     \\
\hline
$A$     & input value & $100~\mathrm{mK}$ & - \\
        & (i)         & $1000~\mathrm{mK}$ & $4\times 10^{-5}~\mathrm{mK}$ \\
        & (ii)         & $88~\mathrm{mK}$ & $1~\mathrm{mK}$ \\
        & (iii)         & $100~\mathrm{mK}$ & $1~\mathrm{mK}$ \\
$\nu_0$ & input value & $75~\mathrm{MHz}$ & - \\
        & (i)         & $69.386~\mathrm{MHz}$ & $0.001~\mathrm{MHz}$ \\
        & (ii)         & $75.04~\mathrm{MHz}$ & $0.02~\mathrm{MHz}$ \\
        & (iii)         & $74.99~\mathrm{MHz}$ & $0.02~\mathrm{MHz}$ \\
$w$     & input value & $10~\mathrm{MHz}$ & - \\
        & (i)         & $25.0~\mathrm{MHz}$ & $10.0~\mathrm{MHz}$ \\
        & (ii)         & $9.7~\mathrm{MHz}$ & $0.1~\mathrm{MHz}$ \\
        & (iii)         & $10.1~\mathrm{MHz}$ & $0.1~\mathrm{MHz}$ \\
$\tau$  & input value & $4$ & - \\
        & (i)         & $10.0$ & $8\times 10^{-6}$ \\
        & (ii)         & $6.5$ & $0.6$ \\
        & (iii)         & $4.4$ & $0.3$ \\
\hline
\end{tabular}
}
\label{Tab:DataModelSimpleResultsTable}
\end{table}

\paragraph{The results for scenario (i), of fitting the uncorrected data, with a model based on the intrinsic spectral structure of the emission in the simulations,}
\label{Sec:Scenario1Results}
shown in subplot (b), demonstrate that the incoherent averaging over time of instrumentally induced foreground systematics alone does not reduce their amplitude sufficiently for recovery of unbiased estimates of the underlying 21-cm signal in the data in the simplified foreground scenario considered in this section. The fractional bias,
\begin{equation}
\label{Eq:FractionalBias}
\frac{\uptheta_\mathrm{input} - \overline{\uptheta}_\mathrm{recovered}}{\uptheta_\mathrm{input}} \ ,
\end{equation}
where $\uptheta_\mathrm{input}$ and $\overline{\uptheta}_\mathrm{recovered}$ are the input value and mean of the posterior distribution of the parameter $\uptheta$, respectively, exceeds 100\% for all of the 21-cm signal parameter estimates barring the central frequency which has a fractional bias of $\sim7\%$. Additionally, the amplitude and flattening parameter are both prior-limited such that more biased values would have been recovered if their priors had been extended. Furthermore, the inaccuracy of the foreground component of the model leads to spurious high precision constraints on the signal parameters as the signal component of the model attempts to fit high significance foreground systematics. This results in recovered parameter estimates that are highly inconsistent with the true values of the 21-cm signal in the data ($(\uptheta_\mathrm{input} - \overline{\uptheta}_\mathrm{recovered})/\sigma_{\uptheta} \gg 1$, where $\sigma_{\uptheta}$ is the $1$-$\sigma$ uncertainty on $\uptheta$). Finally, the recovered functional posteriors on the residuals $\upDelta T = (d-m)$, where $d$ and $m$ are the data analysed and model fit, respectively, demonstrate that in this scenario, the inadequacy of the intrinsic foreground model for describing the non-chromaticity corrected data combined with our prior on the signal amplitude preventing the fitting of foreground systematics with very deep ($A>1~\mathrm{K}$) amplitude 21-cm signal models, means that significant ($>1~\mathrm{K}$) residuals remain for the best fitting model and the data.

\paragraph{The results for scenario (ii), of fitting the corrected data with a model based on the intrinsic spectral structure of the emission in the simulations,}
\label{Sec:Scenario2Results}
shown in subplot (c), illustrate that BFCC is successful at significantly mitigating bias in recovered estimates of the 21-cm signal, relative to scenario (i), when fitting the same data model, despite that model neglecting the residual non-intrinsic chromaticity in the BFCC data and the impact of BFCC on the 21-cm signal component of the data. However, this mitigation is partial and recovered estimates of the 21-cm signal remain biased in this analysis (albeit at a lower level).

The fractional biases on the recovered parameters are now reduced to order-10\% on the amplitude and flattening factor, and they are smaller still on the position and width. Additionally, the overall fit to the data shown in the bottom panel is greatly improved, with the $>1~\mathrm{K}$ residual features present when analysing the non-chromaticity corrected data in scenario (i) reduced by nearly three orders of magnitude to $\sim10~\mathrm{mK}$ when analysing the BFCC data with the same model. However, this residual structure still exceeds the $1~\mathrm{mK}$ RMS noise level in the data by an order of magnitude. This illustrates the limitations of this model in the high signal-to-noise regime, even if the foregrounds are simulated in the limit of having spatially independent spectral structure. Furthermore, despite the significantly improved fractional errors, which are useful only as a rough measure of the performance of the analysis, the level of statistical consistency of the recovered estimates and the true underlying signal is the more fundamentally important metric for the purposes of unbiased astrophysical and cosmological inference. In this respect, we find that even with the simplified foreground spectral structure in the simulated data analysed in this section, in the high signal-to-noise regime
significant statistical inconsistency between the input and recovered amplitude and flattening of the 21-cm signal is still apparent. In particular, while the mean recovered estimates of the central frequency and width of the absorption feature are in agreement with the inputs to within $~2$- and $~3$-$\sigma$, respectively, the recovered estimates of the mean amplitude and flattening of the 21-cm signal are inconsistent with the input values of these parameters in the simulated data at $~12$- and $~4$-$\sigma$, respectively. These offsets will translate to comparably statistically significant biases in any inferences regarding the astrophysics of CD derived from the signal; making them more problematic than one may otherwise infer from the moderate fractional errors.

This problem would be ameliorated in the lower signal-to-noise regime applicable to current data sets; however, as will be shown in \autoref{Sec:RealisticModels}, this bias is further compounded when modelling foregrounds with more realistic spectral structure and, as a result, remains statistically significant even at more moderate signal-to-noise levels.

\paragraph{The results for scenario (iii), of fitting the corrected data with the correct analytic model,}
\label{Sec:Scenario3Results}
shown in subplot (d), illustrate that the 21-cm signal is perfectly recovered, free from bias, even in the high signal-to-noise regime, when the foregrounds in the data have simple spatially-independent spectral structure. In particular, all 21-cm signal parameters are recovered consistent with the input parameters to within $~1.5-\sigma$ or better and the functional posteriors on the residuals between the full model and the data are fully consistent with the $1~\mathrm{mK}$ RMS noise level in the data.

\subsubsection{Realistic foreground spectral structure}
\label{Sec:SimpleForegroundsDiscussion}

Despite the excellent performance of the analytic model for the BFCC data demonstrated in scenario (iii), its derivation relied on the separability of the spatial and spectral structure of the foreground component of the data. When the foregrounds have spatially dependent spectral structure, this is not possible. However, by subdividing the spatially dependent spectral structure of the foregrounds into spatially dependent spectral perturbations on top of a spatially isotropic background (i.e. small spectral index perturbations to an isotropic power law spectrum), a similar approach can be taken to derive a closed-form model that accurately approximates the data in this more realistic scenario. We explore this in more detail in the next section.

%%%%%%%%%%%%%%%%%%%%%%%%%%%%%%%%%%%%%%%%%%%%%%%%%%
\section{A general model for autocorrelation spectra chromaticity corrected using beam-factors}
\label{Sec:RealisticModels}
%%%%%%%%%%%%%%%%%%%%%%%%%%%%%%%%%%%%%%%%%%%%%%%%%%

In this section, we will show that, when the spectral structure of the foregrounds in the data is spatially dependent, one can model the data to arbitrary precision with a closed-form expression\footnote{While there is no closed-form expression for the data, an exact solution can be obtained with a full forward model of the data (e.g. \citealt{2014MNRAS.437.1056V, 2021MNRAS.506.2041A})} building on the analytic BFCC data model derived in \autoref{Sec:ToyModel}, preventing foreground bias in recovered estimates of the 21-cm signal. However, unlike in the case considered in \autoref{Sec:ToyModel}, here, modelling higher signal-to-noise data requires an increasingly complex foreground model that is increasingly correlated with models for the 21-cm signal, reducing the significance with which the signal can be recovered from the data. Nevertheless, we find, using Bayesian model comparison to select for a preferred set of models for the data, that the highest evidence BFCC data model enables unbiased recovery of the 21-cm signal at the signal-to-noise level of the publicly available EDGES low-band data.

\subsection{Foreground emission with spatially-dependent spectral structure}
\label{Sec:SpatiallyDependentSpectralStructure}

As in \autoref{Sec:PerfectBFCC}, we divide the sky brightness into two components, $T_\mathrm{sky}(\nu, \Omega, t) = T_\mathrm{fg}(\nu, \Omega, t) + T_{21}(\nu)$, where $T_{21}(\nu)$ is the global 21-cm signal and $T_\mathrm{fg}(\nu, \Omega, t) = T_\gamma + T_{\rm plfg}(\nu, \Omega, t)$ is the anisotropic, time-dependent, foreground sky brightness temperature distribution above the antenna and is composed of two components: \begin{enumerate*}\item an isotropic CMB temperature, $T_\gamma$, and \item an anisotropic spectral power law foreground brightness temperature distribution (comprised of the sum of anisotropic Galactic and isotropic extragalactic foregrounds), $T_{\rm plfg}(\nu, \Omega, t)$. \end{enumerate*}

\subsubsection{Spatially varying spectral structure}
\label{Sec:SpatiallyVarying}

Here, we assume that the spectrum of $T_{\rm plfg}$ is a spatially varying power law with a temperature spectral index distribution that can be modelled as a Gaussian random field (GRF) characterised by a mean temperature spectral index over the sky, $\beta_{0}$, and a variance, $\sigma_{\beta}^{2}$ (e.g. \citealt{2016MNRAS.462.3069S, 2019MNRAS.488.2904S}). We also assume that $\sigma_{\beta} \ll \beta_{0}$, which at the resolution of EDGES is an excellent approximation (e.g. \citealt{2017MNRAS.464.4995M, 2019MNRAS.483.4411M}). Thus, $\beta$ varies across the sky and, in our topocentric coordinate system, with $\Omega$ and time. However, at any given $\Omega$ and time, $\beta$ is a random draw from the GRF and thus has no functional dependence on those parameters (we discuss deviations from this assumption in \autoref{Sec:SpatialCorrelation}).

By Taylor expanding $T_{\rm fg}$ about $\beta=\beta_{0}$, along a given line of site and frequency, one can separate the spectral structure into a dominant component that can be perfectly chromaticity corrected, as in \autoref{Sec:ToyModel}, and smaller, spatially dependent spectral perturbations, which can be fit with a parametric model as detailed below. With the above definitions and labelling the temperature spectral index in a given direction, $\Omega$, and time, $t$, as, $\beta_{\Omega,t} = \beta_{0} + \upDelta\beta_{\Omega,t}$, we can rewrite $T_\mathrm{sky}(\nu, \Omega, t)$ as,
\begin{equation}
\label{Eq:TskySpatiallyDependentFg}
T_\mathrm{sky}(\nu, \Omega, t, \beta_{\Omega,t}) = T_\mathrm{fg}(\nu, \Omega, t, \beta_{\Omega,t}) + T_{21}
\ ,
\end{equation}
with,
\begin{equation}
\label{Eq:TskySpatiallyDependentFg2}
T_\mathrm{fg}(\nu, \Omega, t, \beta_{\Omega,t}) = (T_\mathrm{fg}(\nu_\mathrm{c}, \Omega, t) - T_{\gamma})\left(\frac{\nu}{\nu_\mathrm{c}}\right)^{-\beta_{\Omega,t}} \\
+  T_{\gamma}
\ .
\end{equation}
Taylor expanding $T_\mathrm{fg}(\nu, \Omega, t, \beta_{\Omega,t})$ about $\beta_{0}$ (for fixed $\nu$, $\Omega$, and $t$), we have,
\begin{align}
\label{Eq:FgTaylorExpansion1}
T_\mathrm{fg}(\nu, \Omega, t, \beta_{\Omega,t}) \\ \nonumber
={}& \sum_{m=0} ^ {\infty} \frac {T_\mathrm{fg}^{(m)}(\nu, \Omega, t, \beta_{0})}{m!} (\Delta \beta_{\Omega,t})^{m}  \\ \nonumber
={}& T_\gamma + T_\mathrm{plfg}(\nu_\mathrm{c}, \Omega, t)\left(\frac{\nu}{\nu_\mathrm{c}}\right)^{-\beta_{0}} \\ \nonumber
+{}& \sum_{m=1} ^ {\infty} \frac {T_\mathrm{fg}^{(m)}(\nu, \Omega, t, \beta_{0})}{m!} (\Delta \beta_{\Omega,t})^{m} \ .
\end{align}
Here, $!$ is the factorial operator, $T_\mathrm{fg}^{(m)}(\nu, \Omega, t, \beta_{0})$ denotes the $m$th derivative of $T_\mathrm{fg}$ with respect to $\beta$, evaluated at $\beta = \beta_{0}$, and $\Delta \beta_{\Omega,t} = \beta_{\Omega,t}-\beta_{0}$.

In the limit that $\Delta \beta_{\Omega,t}$ tends to zero everywhere on the sky, \autoref{Eq:FgTaylorExpansion1} simplifies to \autoref{Eq:Tplfg}, and we recover the spatially independent spectral structure foreground model used in \autoref{Sec:ToyModel}. For non-zero $\Delta \beta_{\Omega,t}$, \autoref{Eq:FgTaylorExpansion1} allows us to express foregrounds with spatially dependent spectral structure as the sum of two parts that are impacted differently by BFCC:
\begin{enumerate}
\item the foregrounds with spatially independent spectral structure considered in \autoref{Sec:ToyModel}, and
\item a smaller, spatially dependent spectral perturbation given by the third term in the final line of \autoref{Eq:FgTaylorExpansion1}.
\end{enumerate}
Part (i) has separable spatial and spectral structure and is perfectly chromaticity correctable using BFCC. Part (ii), in contrast, does not have separable spatial and spectral structure and thus the instrumentally induced chromaticity in the contribution to the spectrometer data  of this component will be imperfectly corrected by BFCC.

To see this, we can rewrite \autoref{Eq:Tcorrected} using \autoref{Eq:FgTaylorExpansion1} as,
\begin{align}
\label{Eq:TcorrectedC4pt4}
T_{\rm corrected}&(\nu, t) \\ \nonumber
&= \Bigg[ \int B(\nu, \Omega) \Bigg(T_\gamma + (T_\mathrm{fg}(\nu_\mathrm{c}, \Omega, t) - T_\gamma)\left(\frac{\nu}{\nu_\mathrm{c}}\right)^{-\beta_{0}} \\ \nonumber
& + T_{21} + \sum_{m=1} ^ {\infty} \frac {T_\mathrm{fg}^{(n)}(\nu, \Omega, t, \beta_{0})}{m!} (\Delta \beta_{\Omega,t})^{m} \Bigg) \mathrm{d}\Omega + n \Bigg] \\ \nonumber
&\times \frac{\int B(\nu_\mathrm{c}, \Omega) T_\mathrm{fg}(\nu_\mathrm{c}, \Omega, t) \mathrm{d}\Omega}{\int B(\nu, \Omega) T_\mathrm{fg}(\nu_\mathrm{c}, \Omega, t) \mathrm{d}\Omega} \\ \nonumber
&= T_\mathrm{m_{0}}(\nu_\mathrm{c}, t)\left(\frac{\nu}{\nu_\mathrm{c}}\right)^{-\beta_{0}} + \frac{(1-\left(\frac{\nu}{\nu_\mathrm{c}}\right)^{-\beta_{0}}) T_{\gamma}}{B_\mathrm{factor}} + \frac{T_{21}}{B_\mathrm{factor}} \\ \nonumber
&+ \frac{1}{B_\mathrm{factor}} \int B(\nu, \Omega)  \sum_{m=1} ^ {\infty} \frac {T_\mathrm{fg}^{(n)}(\nu, \Omega, t, \beta_{0})}{m!} (\Delta \beta_{\Omega,t})^{m} \mathrm{d}\Omega \\ \nonumber
&+ \frac{n}{B_\mathrm{factor}} \ .
\end{align}

\subsubsection{An accurate low expansion-order approximation}
\label{Sec:SpatialCorrelation}

The penultimate term in \autoref{Eq:TcorrectedC4pt4} marks a departure from the simple closed-form description of $T_{\rm corrected}(\nu)$, free from direction dependent integrals found in \autoref{Sec:ToyModel}.
Nevertheless, progress can be made with respect to deriving a compact finite-term approximation for the data by analysing the expansion orders for a given choice of beam model.

We start by rewriting this term as, $T_\mathrm{pert}/B_\mathrm{factor}$, where we define the perturbation spectrum, resulting from the beam-weighted spatially dependent fluctuations of the spectral index distribution about its mean, as,
\begin{multline}
\label{Eq:Tpert1}
T_\mathrm{pert} = \sum_{m=1}^{\infty} \Bigg[\frac{1}{m!}\left(\ln\left(\frac{\nu}{\nu_\mathrm{c}}\right)\right)^{m} \int B(\nu, \Omega) T_\mathrm{plfg}(\nu_\mathrm{c}, \Omega, t)\left(\frac{\nu}{\nu_\mathrm{c}}\right)^{-\beta_{0}} \\
\times (\Delta \beta_{\Omega,t})^{m} \mathrm{d}\Omega \Bigg] \ .
\end{multline}
The power law component of the spectrum is common to all expansion orders; thus, we can write,
\begin{equation}
\label{Eq:TpertModel1}
\frac{T_\mathrm{pert}}{B_\mathrm{factor}} = \left(\frac{\nu}{\nu_\mathrm{c}}\right)^{-\beta_{0}} f_\mathrm{pert}(\nu, t) \ ,
\end{equation}
where,
\begin{align}
\label{Eq:fpert2}
f_\mathrm{pert}(\nu, t) = \sum_{m=1}^{\infty} \frac{1}{m!}\left(\ln\left(\frac{\nu}{\nu_\mathrm{c}}\right)\right)^{m} a_{m}(\nu, t) \ ,
\end{align}
with,
\begin{align}
\label{Eq:TaylorIntegral3}
a_{m}(\nu, t) = \frac{1}{B_\mathrm{factor}} \int B(\nu, \Omega, t) T_\mathrm{plfg}(\nu_\mathrm{c}, \Omega, t) \Delta \beta(\Omega)^{m} \mathrm{d}\Omega \ .
\end{align}

To derive an accurate finite-term closed-form approximation for $f_\mathrm{pert}(\nu, t)$, we first make use of the fact that for an arbitrary choice of $\nu$, the prefactor to $a_{m}(\nu, t)$,
\begin{equation}
\label{Eq:TaylorPrefactor}
\frac{1}{m!}\left(\ln\left(\frac{\nu}{\nu_\mathrm{c}}\right)\right)^{m} \ , \nonumber
\end{equation}
tends to zero for large $m$ and thus $f_\mathrm{pert}(\nu, t)$ is dominated by relatively low-order terms of the summation in \autoref{Eq:fpert2}. Additionally, for ratios of observation frequency and chromaticity correction reference frequency in the range $1/\mathrm{e} < \nu/\nu_\mathrm{c} < \mathrm{e}$, where $\mathrm{e}$ is Euler's number, the absolute value of the logarithm alone, as well as the full prefactor, exponentially tends to zero with increasing $m$ and both are less than unity for all $m > 1$. EDGES low data, with $50 < \nu < 100~\mathrm{MHz}$ and $\nu_\mathrm{c} = 75~\mathrm{MHz}$, falls within this band and thus the error associated with a finite-term approximation of the summation in \autoref{Eq:fpert2} decreases rapidly\footnote{Despite the rapid reduction with increasing $m$ of modelling error associated with a finite-term approximation of the summation in \autoref{Eq:fpert2}, the huge dynamic range between the foregrounds and the order of magnitude $10~\mathrm{mK}$ noise level associated with the publicly available EDGES low data means a simple first-order approximation is insufficient to model the foregrounds without leaving statistically significant foreground systematics. In simulated EDGES low data, depending on the intensity of the foregrounds in the field observed, we find an expansion including terms up to $m\sim5-6$ is necessary to recover residuals at a sub-$1~\mathrm{mK}$ level.} with increasing $m$.

In addition to the prefactor downweighting higher order terms in the summation, for a spatially uncorrelated GRF description of $\beta_{\Omega,t}$, $a_{m}(\nu, t)$ has an expectation value of zero for odd values of $m$ (regardless of the specific form of the beam). A spatially uncorrelated GRF is a reasonable description for spectra along lines of sight for which the emission is dominated by independent unresolved objects, such as extragalactic point sources. However, in practice, there will also be spatial correlations in the spectral structure of $T_\mathrm{plfg}$ due to emission from astronomical objects dispersed over larger solid angles or physically-connected emission regions such as the observed steepening of the synchrotron spectrum of the Galaxy with increasing distance from the Galactic plane. The large-scale gradient in the SI distribution away from the Galactic plane shifts the expected amplitudes of $a_{m}(\nu, t)$ for odd $m$ to non-zero values. However, these amplitudes remain suppressed relative to even expansion order coefficient due to the cancellation of $\Delta \beta(\Omega)^{m}$ on scales smaller than the characteristic size of the beam.

\subsubsection{Maximally smooth polynomial model}
\label{Sec:MSPolynomialModel}

In principle, this knowledge of the expected trend in relative amplitude of $a_{m}(\nu, t)$ with increasing $m$ enables one to place more informative priors on fitted values of $a_{m}$ via constraints on the hyperparameters of a function designed to fit the spectral structure of $a_{m}$ in the data.

Due to the oscillation of $a_{m}$ between non-convexity and convexity for odd and even $m$, respectively, there is no guarantee of a monotonically decreasing contribution to $f_\mathrm{pert}(\nu, t)$ from subsequent terms in the expansion, despite what one may expect from examining the prefactor alone. However, we do expect a monotonically decreasing contribution to $f_\mathrm{pert}(\nu, t)$ from subsequent terms in the odd-$m$ and even-$m$ terms of the expansion, respectively.  One way this information could be incorporated into an analysis of the data would be to describe each $a_{m}(\nu, t)$ with a maximally smooth (MS) polynomial (e.g. \citealt{2015ApJ...810....3S, 2017ApJ...840...33S}) or derivative-constrained function (e.g. \citealt{2021MNRAS.502.4405B}) and place priors enforcing a constrained and decreasing upper limit on the amplitude coefficient of the MS polynomials as a function of increasing $a_{m_\mathrm{even}}$ and $a_{m_\mathrm{odd}}$, where $m_\mathrm{even} = (2m)$ and $m_\mathrm{odd} = (2m+1)$. In practice, setting these priors would require estimation of the expected distribution of $a_{m}$ values. These estimates could be derived from simulated data using realistic models for the beam and $\beta(\Omega)$ derived from low-frequency sky surveys. However, even with realistic estimates for $a_{m}$, this approach results in a model for $f_\mathrm{pert}(\nu, t)$ parametrised by a large number of MS polynomial coefficients that are computationally expensive to fit for.

Testing with EDGES low simulations we find this requires simultaneously fitting $N_\mathrm{T} \sim 6$ MS polynomials, where $N_\mathrm{T}$ is the Taylor expansion order required to fit $T_\mathrm{pert}$ to a sub-$1~\mathrm{mK}$ RMS-residual threshold and where, individually, each MS polynomial is of between 2nd and $\sim10$th order, resulting in an $\sim 40$ parameter fit for $f_\mathrm{pert}(\nu, t)$.

\subsubsection{Log-polynomial model}
\label{Sec:LogPolynomialModel}

The arithmetic increase, with Taylor expansion order, of the number of parameters necessary to model an amplitude constrained sequence of $a_{m}$ can be improved on by fitting the $a_{m}$ with coefficient-constrained log-polynomial models,
\begin{align}
\label{Eq:LogPolynomial}
a^\mathrm{model}_{m}(\nu, t) =  \sum_{\alpha=1}^{N_{m}} p_{m,\alpha}\ln\left(\frac{\nu}{\nu_\mathrm{c}}\right)^{\alpha} \ .
\end{align}
In this case, since the product of log-polynomials leaves the functional form of the model unchanged and \autoref{Eq:TpertModel1} is linear with respect to $a_{m}$, at the cost of a less constrained prior on the amplitudes of individual $a_{m}$, the $\sum_{i=1}^{N_\mathrm{T}} N_{i}$ coefficients of our MS polynomial models, that would be necessary if we were to fit each $a_{m}$ individually, can be condensed into $\sim\max_{m}(m + N_{m}) \ll \sum_{m} N_\mathrm{T} N_{m}$ coefficients\footnote{$\max_{m}(m + N_{m})$ is an upper limit on the necessary number of coefficients because interference between the spectra corresponding to the different expansion orders can decrease the number of coefficients required to fit the sum over expansion orders.} of a composite log-polynomial, significantly reducing the dimensionality of our parameter space. Here, $(m + N_{m})$ is the log-polynomial order required to fit the $m$th component of $f_\mathrm{pert}$ to a prescribed precision and the operator $\max_{m}(.)$ denotes the maximum over Taylor expansion order $m$.

Following this approach, we write our final model for $T_\mathrm{pert}/B_\mathrm{factor}$ as,
\begin{align}
\label{Eq:PLLogPolynomial}
m_\mathrm{pert}(\nu, t) = T_\mathrm{m_{0}}(\nu_\mathrm{c}, t)\left(\frac{\nu}{\nu_\mathrm{c}}\right)^{-\beta_{0}} \sum_{\alpha=1}^{N_\mathrm{pert}} p_{\alpha}(t)\ln\left(\frac{\nu}{\nu_\mathrm{c}}\right)^{\alpha} \ .
\end{align}
Here, we have included $T_\mathrm{m_{0}}(\nu_\mathrm{c}, t)$ as a prefactor and specify $p_{\alpha}(t)$ as fractional perturbations about the mean brightness temperature at the reference frequency $\nu=\nu_\mathrm{c}$, and we define $N_\mathrm{pert}$ as the log-polynomial order necessary to fit $T_\mathrm{pert}/B_\mathrm{factor}$ to within the level of the noise in the data. In practice, we will use Bayesian model selection to determine the preferred $N_\mathrm{pert}$ for describing the data (see \autoref{Sec:BayesianInference} for details).

\subsubsection{BFCC data model}
\label{Sec:BFCCDataModel}

Since, as a linear operation, time-averaging of $m_\mathrm{pert}(\nu, t)$ leaves the functional form unchanged, our model for the contribution of $T_\mathrm{pert}/B_\mathrm{factor}$ to time-averaged data can be trivially derived from \autoref{Eq:PLLogPolynomial}  as,
\begin{align}
\label{Eq:PLLogPolynomialLSTAverage}
m_\mathrm{pert}(\nu) = \bar{T}_\mathrm{m_{0}}\left(\frac{\nu}{\nu_\mathrm{c}}\right)^{-\beta_{0}} \sum_{\alpha=1}^{N_\mathrm{pert}} p_{\alpha}\ln\left(\frac{\nu}{\nu_\mathrm{c}}\right)^{\alpha} \ .
\end{align}

Time-averaging the remaining terms in \autoref{Eq:TcorrectedC4pt4} and substituting for $T_\mathrm{pert}/B_\mathrm{factor}$ using \autoref{Eq:PLLogPolynomialLSTAverage}, our model for BFCC data including foregrounds with spatially dependent spectral structure becomes,
\begin{align}
\label{Eq:TcorrectedModel2}
T_{\rm corrected}^\mathrm{model}&(\nu) \\ \nonumber
=& \bar{T}_\mathrm{m_{0}}\left(\frac{\nu}{\nu_\mathrm{c}}\right)^{-\beta_{0}}(1 + \sum_{\alpha=1}^{N_\mathrm{pert}} p_{\alpha}\ln\left(\frac{\nu}{\nu_\mathrm{c}}\right)^{\alpha}) + \frac{(1-\left(\frac{\nu}{\nu_\mathrm{c}}\right)^{-\beta_{0}}) T_{\gamma}}{\bar{B}_\mathrm{factor}(\nu)} \\ \nonumber
&+ \frac{T_{21}}{\bar{B}_\mathrm{factor}(\nu)} \ .
\end{align}

\subsection{Ionospheric effects}
\label{Sec:IonosphericEffects}

\autoref{Eq:TskySpatiallyDependentFg} is a description of the intrinsic brightness temperature of the emission on the celestial sphere in the radio frequency range of interest for measurement of redshifted 21-cm emission from CD and the EoR, prior to propagation of the emission through the Earth's ionosphere. However, this emission is refracted and absorbed by the ionosphere in a frequency-dependent manner (e.g. \citealt{2014MNRAS.437.1056V, 2021MNRAS.503..344S}), prior to its measurement by an antenna on Earth. Furthermore, the ionosphere radiates thermal emission, which contributes to the total emission arriving at our instrument (e.g. \citealt{2015RaSc...50..130R}).

Of these three ionospheric radiative transfer effects, here we will focus on the latter two. The former, ionospheric refraction, acts as a lens that shifts the apparent positions of sources on the celestial sphere in a zenith angle-dependent manner. This can be modelled as a transfer function that can be absorbed as a component of our effective instrument beam model (e.g. \citealt{2014MNRAS.437.1056V}).

In detail, the ionosphere is non-uniform and non-stationary. To account for this, and in particular for varying ionospheric refraction during turbulent ionospheric conditions, the transfer function, and correspondingly the effective instrument beam, can be estimated in a time-dependent manner (e.g. \citealt{2022MNRAS.515.4565S}). The cadence with which the transfer function must be estimated can be reduced, and the ionospheric radiative transfer effects can be better approximated by a uniform and stationary ionospheric model, by restricting one's analysis to periods of low solar activity and nighttime data when the ionosphere has greater temporal stability and uniformity. This approximation will further be improved upon by restricting one's analysis to data that has been averaged over a number of sidereal days such that systematic structure associated with uncorrelated ionospheric fluctuations about a mean ionospheric model is averaged down. If this procedure is carried out, we do not expect ionospheric refraction to otherwise impact our conclusions, and we do not consider it further here.

Assuming the above data cuts and a uniform and stationary approximation for the ionosphere, we can account for ionospheric absorption and emission by writing the intrinsic spectral structure of the emission incident on the antenna as,
\begin{multline}
\label{Eq:TskySDFgStationaryIonosphere}
T_\mathrm{sky}(\nu, \Omega, t, \beta_{\Omega,t}) = \Big[(T_\mathrm{fg}(\nu, \Omega, t, \beta_{\Omega,t}) + T_{21} \Big] \e^{-\tau_\mathrm{ion}(\nu)} \\
+ T_{\mathrm{e}}(1-\e^{-\tau_\mathrm{ion}(\nu)})
\ .
\end{multline}
Here, the first term on the RHS of the expression is the fraction of astrophysical emission passing through the ionosphere and the second is the unabsorbed component of thermal emission by electrons in the ionosphere;  $T_{\mathrm{e}}$ and $\tau_\mathrm{ion}$  are the temperature of electrons and opacity of the ionosphere, respectively, and, $T_\mathrm{fg}(\nu, \Omega, t, \beta_{\Omega,t})$ is as defined in \autoref{Eq:TskySpatiallyDependentFg2}.

Assuming the ionospheric opacity scales as $\nu^{-2}$ (e.g. \citealt{2015RaSc...50..130R}), we can write,
\begin{align}
\label{Eq:TauIon}
\tau_\mathrm{ion} = \tau_0(\nu/\nu_\mathrm{c})^{-2} \ .
\end{align}
Here, $\nu_\mathrm{c}=75~\mathrm{MHz}$ is the reference frequency for $\tau_0$ and for the simulations considered in this section.

Accounting for the ionospheric effects described above, the assumption of stationarity and uniformity in our ionospheric model allows us to generalise our BFCC model for the data, \autoref{Eq:TcorrectedModel2}, to incorporate them, as,
\begin{multline}
\label{Eq:BFCCdataModel}
T_\mathrm{BFCC}^\mathrm{model}(\nu)
= \Bigg[\bar{T}_\mathrm{m_{0}}\left(\frac{\nu}{\nu_\mathrm{c}}\right)^{-\beta_{0}}(1 + \sum_{\alpha=1}^{N_\mathrm{pert}} p_{\alpha}\ln\left(\frac{\nu}{\nu_\mathrm{c}}\right)^{\alpha}) + \frac{(1-\left(\frac{\nu}{\nu_\mathrm{c}}\right)^{-\beta_{0}}) T_{\gamma}}{\bar{B}_\mathrm{factor}(\nu)} \\
+ \frac{T_{21}}{\bar{B}_\mathrm{factor}(\nu)}\Bigg] \e^{-\tau_\mathrm{ion}(\nu)} + \frac{T_{\mathrm{e}}}{\bar{B}_\mathrm{factor}(\nu)}(1-\e^{-\tau_\mathrm{ion}(\nu)}) \ .
\end{multline}

\subsection{Can residual instrumental chromaticity be mitigated with a revised form of the BFCC formalism?}
\label{Sec:MitigationWithMonsalveBFCC}

\autoref{Eq:BFCCdataModel} is an accurate model for autocorrelation spectra chromaticity corrected using an error-free beam-factor model formulated as described in \autoref{Eq:CCsnapshot}. In Appendix \ref{Sec:MitigationWithMonsalveBFCC1} we describe how, if one has sufficiently accurate knowledge of the spatially dependent spectral structure of the foregrounds, the number of terms, $N_\mathrm{pert}$, required to model $T_\mathrm{pert}/B_\mathrm{factor}$ can be reduced using the alternate beam-factor formulation described in \citet{2017ApJ...847...64M}. Additionally, in Appendix \ref{Sec:ParametricBFCC}, we describe an approach to mitigating the impact of uncertainties in the sky and beam models used to construct the beam-factor.

%%%%%%%%%%%%%%%%%%%%%%%%%%%%%%%%%%%%%%%%%%%%%%%%%%
\section{Demonstration on simulated BFCC data with spectrally complex foregrounds}
\label{Sec:RealisticBFCCDemonstration}
%%%%%%%%%%%%%%%%%%%%%%%%%%%%%%%%%%%%%%%%%%%%%%%%%%

To understand the impact of the choice of data model on our ability to recover unbiased estimates of the 21-cm signal from BFCC EDGES low-band spectrometer data deriving from observations of more realistic sky-emission, we construct simulations incorporating the following sky-model components:
\begin{itemize}
\item foregrounds with realistic spatially dependent spectral structure,
\item spectrally-dependent absorption by the ionosphere,
\item ionospheric emission,
\item a flattened Gaussian 21-cm absorption profile.
\end{itemize}
We analyse the simulated data using the BFCC model derived in \autoref{Sec:RealisticModels} (\autoref{Eq:BFCCdataModel}) as well as a data model describing the intrinsic spectral structure of this more complex sky model but omitting the impact of imperfect correction for instrumental chromaticity. In both analyses, we impose priors on the parameters of the flattened Gaussian 21-cm absorption profile as listed in \autoref{Tab:DataModelPriors}.

We describe the construction of the simulated data in \autoref{Sec:RealisitcBFCCSimulations}, the physically-motivated priors we impose on the BFCC model when fitting the data in \autoref{Sec:BFCCpriors}, the comparison intrinsic spectral structure data model in \autoref{Sec:ComparisonBFCCDataModels}, the Bayesian inference framework in which we analyse the data in \autoref{Sec:BayesianInference}, and the results of the analyses in \autoref{Sec:Results}.

\subsection{Simulated data}
\label{Sec:RealisitcBFCCSimulations}

\begin{figure*}
	\centerline{
	\includegraphics[width=0.5\textwidth]{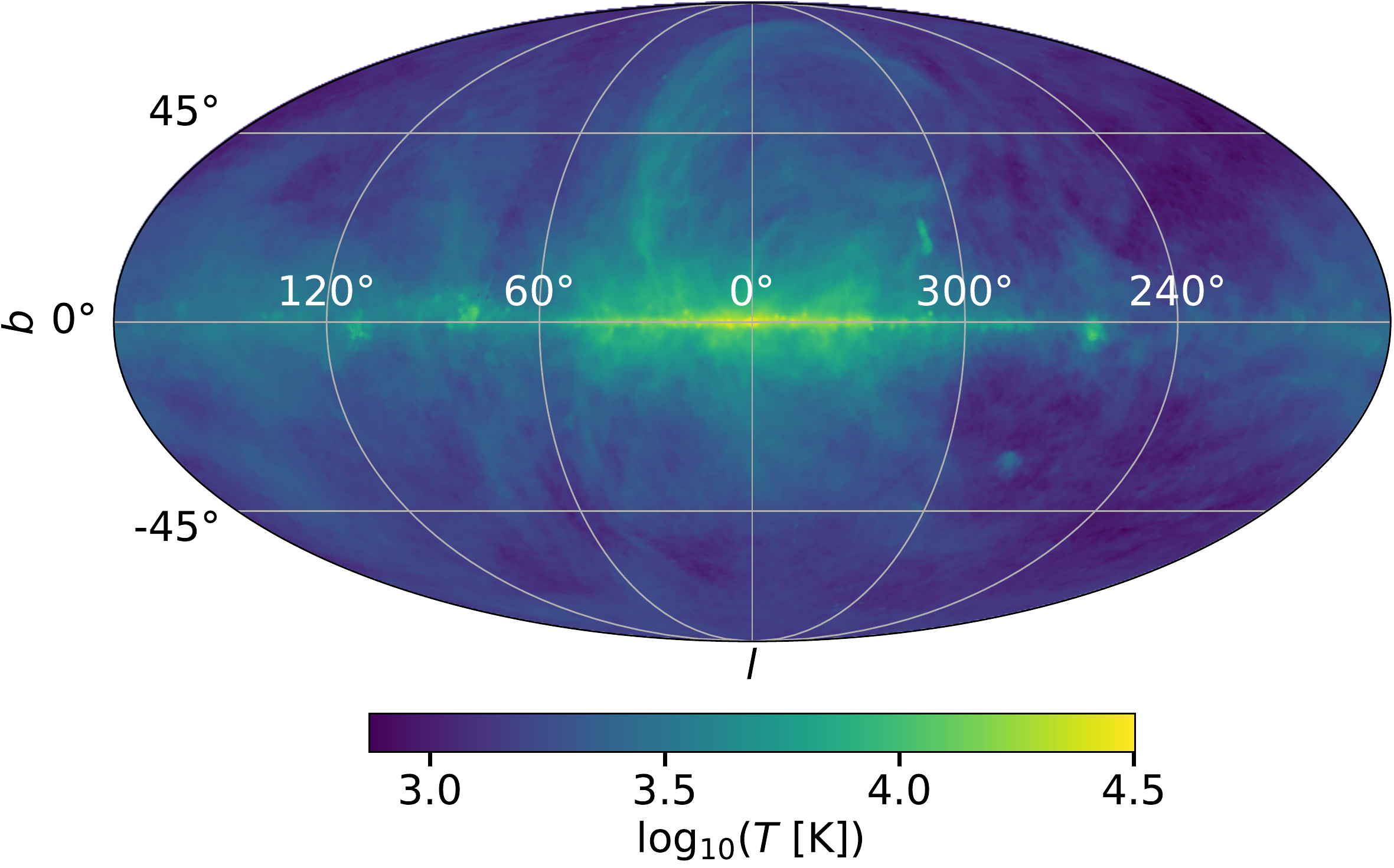}
	\includegraphics[width=0.5\textwidth]{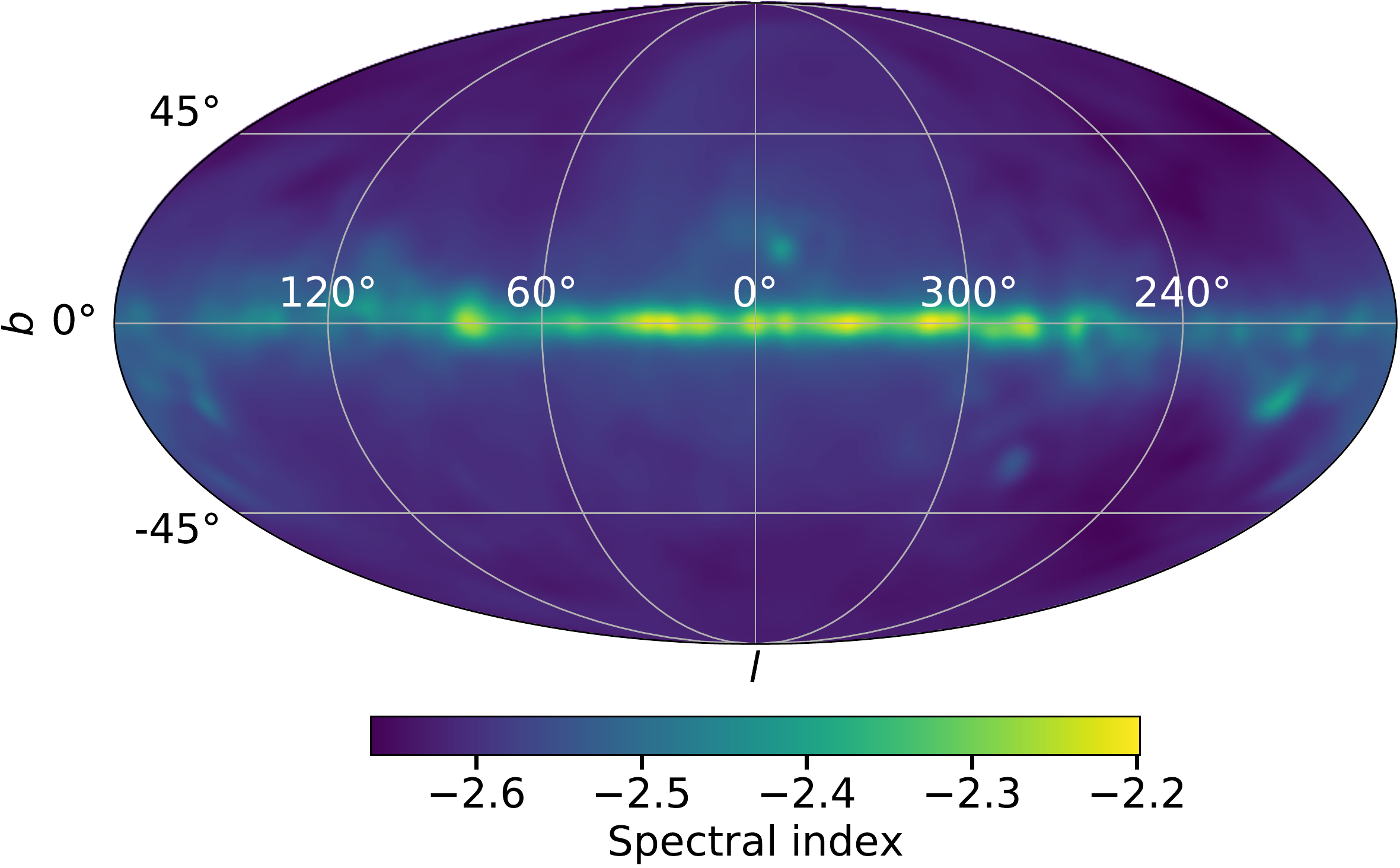}
	}
\caption{[Left] Intrinsic foreground brightness temperature distribution model, evaluated at the center of our simulated spectral band, $T_\mathrm{fg}(75~\mathrm{MHz}, l, b)$, and [right] spatially-dependent foreground spectral index distribution $\beta(l, b)$ used when constructing simulated observational data.}
\label{Fig:ForegroundAndSImodel}
\end{figure*}

We construct simulations following the procedure described in \autoref{Sec:PerfectBFCCSimulations} with modifications as detailed below. We construct $T_{\rm data}(\nu)$ over a $50-100~\mathrm{MHz}$ spectral band, assuming a $1~\mathrm{MHz}$ channel width and integration over a short time interval, $\upDelta t = 6~\mathrm{minutes}$, such that we can work with \autoref{Eq:Tdata} in the snapshot limit,
\begin{equation}
\label{Eq:TdataSimRealistic}
T_\mathrm{data}(\nu, t) = \int\limits_{\Omega^{+}} B(\nu, \Omega)T_\mathrm{sky}(\nu, \Omega, t, \beta_{\Omega,t})~\mathrm{d}\Omega + n \ .
\end{equation}
For the simulations considered here, we add noise to the data at a level such that the resultant noise in the BFCC data, after time-averaging, is Gaussian and white, with an RMS amplitude of $20~\mathrm{mK}$ that is comparable to estimates of the noise in the publicly available EDGES low-band data (e.g. \citealt{2019ApJ...880...26S}). Since the BFCC data model that we will fit to the simulations constructed in this section are approximate, and thus the Bayesian evidence maximising model complexity for describing it will be signal-to-noise dependent, this noise level is chosen to simplify comparison between the preferred model complexity for describing the simulated data here and that required to describe the publicly available EDGES low data in upcoming work (Sims et al. in prep.). We define $T_\mathrm{sky}(\nu, \Omega, t, \beta_{\Omega,t})$ via \autoref{Eq:TskySDFgStationaryIonosphere}. For the 21-cm signal we use a flattened Gaussian absorption profile described by \autoref{Eq:FlattenedGaussian}. We simulate an absorption profile with parameters: $A=500~\mathrm{mK}$, $\nu_0=78~\mathrm{MHz}$, $w=19~\mathrm{MHz}$ and $\tau=8$, consistent\footnote{In B18 a best fitting flat-bottomed absorption trough was recovered centred at $\nu_0 = 78 \pm 1~\mathrm{MHz}$, with a width of $w = 19^{+4}_{-2}~\mathrm{MHz}$, a flattening factor of $\tau = 7^{+5}_{-3}$ and with a depth of $500^{+500}_{-200}~\mathrm{mK}$, where the uncertainties correspond to $99\%$ confidence intervals, accounting for both thermal and systematic errors.} with the best-fit parameters recovered in B18, within their estimated uncertainties.

We construct the intrinsic foreground component of our simulations via,
\begin{multline}
\label{Eq:TFgSDSI}
T_\mathrm{fg}(\nu, \Omega, t, \beta_{\Omega,t}) = \\
(T_\mathrm{fg}(408~\mathrm{MHz}, \Omega, t) - T_\gamma)\left(\frac{\nu}{408~\mathrm{MHz}}\right)^{-\beta_{\Omega,t}} + T_\gamma \ .
\end{multline}
Here, $T_\mathrm{fg}(408~\mathrm{MHz}, \Omega, t)$ and  $T_\gamma=2.725~\mathrm{K}$ are the Haslam all-sky map and CMB temperature, respectively. We assume $T_{\mathrm{e}}=450~\mathrm{K}$ and $\tau_0=0.014$ at $\nu_\mathrm{c}=75~\mathrm{MHz}$, consistent with nighttime ionospheric electron temperatures and opacity, at the location of EDGES, inferred in \citet{2015RaSc...50..130R}. We calculate $\beta(l, b)$, from which we derive $\beta_{\Omega,t}$, as the spectral index distribution between the sky brightness temperature distribution at 408 MHz and 45 MHz, encompassing our $50-100~\mathrm{MHz}$ spectral band of interest, as\footnote{For the spectral index sign convention we assume $T_\mathrm{plfg}(\nu) =  T_\mathrm{plfg}(\nu_\mathrm{c})(\nu/\nu_\mathrm{c})^{-\beta(l, b)}$},
\begin{equation}
\label{Eq:SImap}
\beta(l, b) = \frac{\log{\left(\frac{T_{45}(l, b)-T_{\gamma}}{T_{408}(l, b)-T_{\gamma}}\right)}}{\log{\left(\frac{45}{408}\right)}}.
\end{equation}
Here, $l$ and $b$ are Galactic longitude and latitude, respectively. For $T_{45}(l, b)$ and $T_{408}(l, b)$ we use the global sky model (GSM; \citealt{2017MNRAS.464.3486Z}) evaluated at 408 MHz and 45 MHz, respectively, and in both cases we smooth the resulting maps to a common resolution of 5 degrees to remove structure on scales below the angular resolution of the 45 MHz data used in the derivation of the GSM model. Our resulting model for the intrinsic foreground brightness temperature, evaluated at the center of our simulated spectral band, $T_\mathrm{fg}(75~\mathrm{MHz}, l, b)$, and $\beta(l, b)$, are shown in \autoref{Fig:ForegroundAndSImodel}.

\begin{figure}
	\centerline{
	\includegraphics[width=0.5\textwidth]{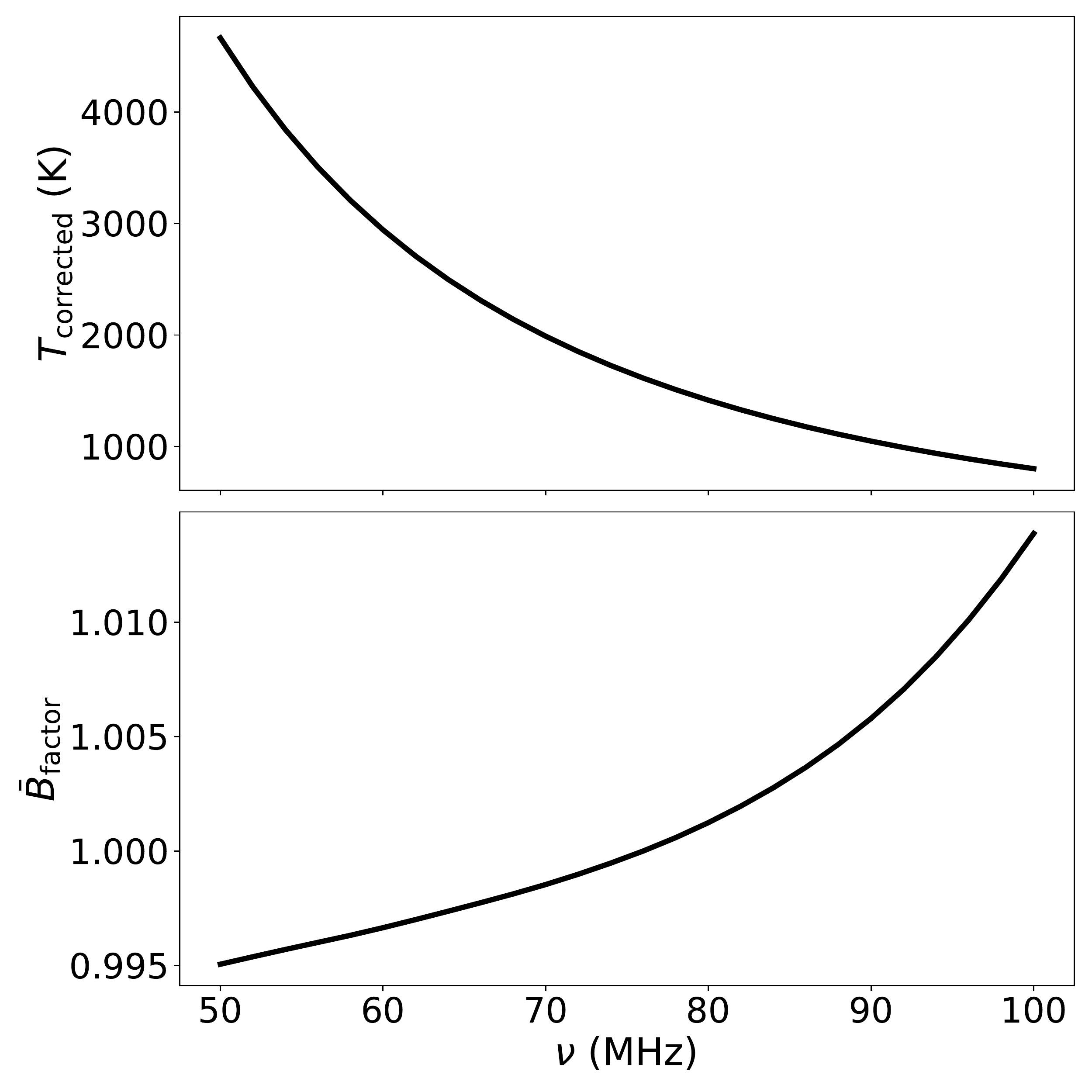}
	}
\caption{Simulated time-averaged spectrum and beam-factor resulting from time-averaging simulated BFCC EDGES low-band data and beam-factors, respectively, over 120 simulated snapshot spectra derived at 6 minute intervals in the LST range $0 \le LST < 12~\mathrm{h}$, matching the LST window of the publicly available EDGES low-band data.}
\label{Fig:SimulatedDataAndBeamFactor}
\end{figure}

For our beam model, $B(\nu, \Omega)$, we use the {\sc{FEKO}} EM simulation of the EDGES low-band blade dipole  antenna with a $30~\mathrm{m} \times 30~\mathrm{m}$ sawtooth ground plane on top of soil with properties described in \autoref{Sec:PerfectBFCCSimulations}. We construct our time-dependent beam-factor model, $B_\mathrm{factor}(\nu, t)$, and BFCC data, $T_{\rm corrected}(\nu, t)$, in the same manner described in \autoref{Sec:PerfectBFCCSimulations}, with $B^\mathrm{m}(\nu, \Omega) \equiv B(\nu, \Omega)$ and $T_\mathrm{fg}^\mathrm{m}(\nu_\mathrm{c}, \Omega, t) \equiv T_\mathrm{fg}(\nu_\mathrm{c}, \Omega, t)$. Here, $T^\mathrm{m}_\mathrm{fg}(\nu_\mathrm{c}, \Omega, t)$ is given by \autoref{Eq:TFgSDSI}, evaluated at $\nu_\mathrm{c}=75~\mathrm{MHz}$. We calculate the corresponding time-averaged BFCC data, $T_{\rm corrected}(\nu)$, by averaging $T_{\rm corrected}(\nu, t)$ over the 120 simulated snapshot spectra derived at 6 minute intervals in the LST range $0 \le LST < 12~\mathrm{h}$, matching the LST window of the publicly available EDGES low-band data, when the Galactic plane is relatively low in the beam. The resulting time-averaged data, $T_{\rm corrected}(\nu)$, and beam-factor, $\bar{B}_\mathrm{factor}(\nu)$, are shown in \autoref{Fig:SimulatedDataAndBeamFactor}.

\subsection{BFCC priors}
\label{Sec:BFCCpriors}

Considering the parameters of \autoref{Eq:BFCCdataModel}, one can define physical priors for $\bar{T}_\mathrm{m_{0}}$, $\beta_{0}$, $T_{\mathrm{e}}$ and  $\tau_\mathrm{ion}$ based on existing observations. The EDGES beam weighted sky brightness temperature at $\nu_\mathrm{c} = 75~\mathrm{MHz}$ varies with time between $\sim1500~\mathrm{K}$ and $\sim5000~\mathrm{K}$ and the temperature spectral index varies between approximately 2.45 and 2.6 (e.g. \citealt{2019MNRAS.483.4411M}). Typical nighttime attenuation through the ionosphere at $150~\mathrm{MHz}$ is 0.015 dB (which, via \autoref{Eq:TauIon}, yields $\tau_0 \simeq 0.014$), with the average magnitude of perturbations in $\tau_\mathrm{ion}$, as measured over 16 days of observations from 18 April to 6 May 2014, at night and over $24~\mathrm{h}$ in time, of order $\Delta\tau_\mathrm{ion}=0.001$ and $0.01$, respectively. Nighttime average electron temperatures inferred by EDGES in the same timeframe were of order a few hundred kelvin (e.g. \citealt{2015RaSc...50..130R}). The parameters of $m_\mathrm{pert}(\nu)$ correspond to the temperatures of individual perturbation spectral model vectors at reference frequency $\nu_\mathrm{c} = 75~\mathrm{MHz}$. The fraction of the antenna temperature described by $m_\mathrm{pert}$ is expected to be small relative to $\bar{T}_\mathrm{m_{0}}$ (see \autoref{Sec:SpatiallyVarying}). For the simulated observations considered in \autoref{Sec:RealisitcBFCCSimulations}, we find that conservatively limiting individual perturbation model vectors to $10\%$ absolute fractional perturbations provides sufficient flexibility for \autoref{Eq:BFCCdataModel} to accurately model simulated foreground-only BFCC data, without adding a significant degree of superfluous flexibility. We incorporate this information when fitting \autoref{Eq:BFCCdataModel} to the simulated data in \autoref{Sec:Results}, in a conservative manner, using broad\footnote{In principle, one could consider contracting the priors on $\bar{T}_\mathrm{m_{0}}$ and $\beta_{0}$ and $\tau_\mathrm{ion}$; however, we find that such a contraction does not lead to a qualitative change in our conclusions and this conservative choice of priors benefits from being more general - enforcing physicality while being sufficiently uninformative to be applicable also to smaller temporal-subsets of the data and instruments with narrower beams for which the variation of these quantities will be larger.} physical priors on the parameters of the model as listed in \autoref{Tab:DataModelBFCCPriors}.

\begin{table}
\caption{Priors on the parameters of the BFCC foreground model defined in \autoref{Sec:RealisticModels} and fit in \autoref{Sec:Results}.
}
\centerline{
\begin{tabular}{l l l }
\hline
Parameter & Model component & Prior     \\
\hline
$\bar{T}_\mathrm{m_{0}}$     & foreground & $U(1000,6000)~\mathrm{K}$ \\
$\beta_{0}$     & foreground & $U(2.0,3.0)$ \\
$p_{\alpha}$     & foreground & $U(-0.1,0.1)$ \\
$T_{\mathrm{e}}$     & ionosphere & $U(100,800)~\mathrm{K}$ \\
$\tau_0$     & ionosphere & $U(0.005,0.025)$ \\
\hline
\end{tabular}
}
\label{Tab:DataModelBFCCPriors}
\end{table}

\subsection{Intrinsic data model}
\label{Sec:ComparisonBFCCDataModels}

In \autoref{Sec:Results}, we will analyse the simulated BFCC data constructed in \autoref{Sec:RealisitcBFCCSimulations} using the analytic BFCC model derived in \autoref{Sec:RealisticModels} (\autoref{Eq:BFCCdataModel}) and we will compare it to a data model describing the intrinsic spectral structure of the sky but omitting the impact of imperfect correction for instrumental chromaticity. For both models, we use the flattened Gaussian signal model used in B18 as a model for the global 21-cm absorption trough. For the foreground component of the latter model, we use the physically motivated parametrisation of the foreground component of the sky signal after propagation through the ionosphere given in B18. We provide a first-principles derivation of this model in \autoref{Sec:IntrinsicForegroundModelDerivation}. Here, we quote the form of the model given in B18 and used in \autoref{Sec:Results}:
\begin{multline}
\label{Eq:B18IntrinsicForegroundModel}
T_\mathrm{Intrinsic,fg}^\mathrm{model}(\nu) = b_{0}\left(\frac{\nu}{\nu_\mathrm{c}} \right)^{-2.5 + b_{1} + b_{2}\log\left(\frac{\nu}{\nu_\mathrm{c}} \right)} \mathrm{e}^{-b_{3}\left(\frac{\nu}{\nu_\mathrm{c}} \right)^{-2}} + b_{4}\left(\frac{\nu}{\nu_\mathrm{c}} \right)^{-2}
\ .
\end{multline}
Here, $b_{i}$ with $i \in [0,\cdots,4]$ are foreground and ionospheric parameters to be determined in the fit of the model to the data. The factor of -2.5 in the first exponent is the typical power-law spectral index of the foreground, $b_{0}$ is an overall foreground scale factor, $b_{1}$ allows for a correction relative to the typical spectral index of the foreground, for the power-law index in the field being analysed, and $b_{2}$ models the contribution to the foreground spectral structure from spatial variation in the spectral index distribution. The amplitude of ionospheric effects is described by $b_{3}$ and $b_{4}$, which model the strength of ionospheric absorption of the foreground and emission from hot electrons in the ionosphere, respectively (see \autoref{Sec:IntrinsicForegroundModelDerivation} for the detailed relations between the parameters of \autoref{Eq:B18IntrinsicForegroundModel} and physical parameters from which they derive.).

Going forward, we refer to this model (and its foreground component) as the 'Intrinsic' (foreground) model, and use the notation $T_\mathrm{intrinsic}^\mathrm{model}$ to describe the sum of $T_\mathrm{intrinsic, fg}^\mathrm{model}$ and 21-cm signal model components. When fitting this model to the simulated data in \autoref{Sec:Results}, we impose physically motivated priors on the parameters of $T_\mathrm{intrinsic}^\mathrm{model}$. The priors we use are listed in \autoref{Tab:DataModelIntrinsicPriors}, and they are set such that they are equivalent to the priors on $\bar{T}_\mathrm{m_{0}}$, $\beta_{0}$, $T_{\mathrm{e}}$ and $\tau_\mathrm{ion}$ in the BFCC model listed in \autoref{Tab:DataModelBFCCPriors}.

\begin{table}
\caption{Priors on the parameters of the Intrinsic foreground model defined in \autoref{Sec:ComparisonBFCCDataModels} and fit in \autoref{Sec:Results}.
}
\centerline{
\begin{tabular}{l l l }
\hline
Parameter & Model component & Prior     \\
\hline
$b_{0}$     & foreground & $U(1000,6000)~\mathrm{K}$ \\
$b_{1}$     & foreground & $U(-0.5,0.5)$ \\
$b_{2}$     & foreground & $U(0,0.2)$ \\
$b_{3}$     & ionosphere & $U(0.005,0.025)$  \\
$b_{4}$     & ionosphere & $U(0.5,20.0)~\mathrm{K}$ \\
\hline
\end{tabular}
}
\label{Tab:DataModelIntrinsicPriors}
\end{table}

\subsection{Bayesian model comparison}
\label{Sec:BayesianInference}

In \autoref{Sec:Results}, we will analyse the realistic simulated BFCC corrected EDGES data derived in \autoref{Sec:RealisitcBFCCSimulations} using the closed-form model for BFCC data derived in \autoref{Sec:RealisticModels} and the Intrinsic data model described in \autoref{Sec:ComparisonBFCCDataModels}. Since the number of components of the $m_\mathrm{pert}(\nu)$ component of the BFCC model required to adequately describe the component of the foregrounds imperfectly corrected for chromatic effects by BFCC is field- and signal-to-noise-dependent and \textit{a priori} unknown, in order to optimally use the BFCC model, one requires a means to determine the preferred number of components of $m_\mathrm{pert}(\nu)$ to fit for when analysing a given data set. Comparison of the posterior odds in favour of models for the data with differing numbers of components, within a Bayesian framework, provides a statistically robust means to determine the preferred number of model components. We outline below the quantities necessary for the calculation of the posterior odds between two models.

\subsubsection{Model comparison}
\label{Sec:ModelComparison}

Bayesian inference addresses model comparison between two possible models for a data set, $M_{i}$ and $M_{j}$, via consideration of $\mathcal{R}_{ij}$, the posterior odds in favour of $M_{i}$ over $M_{j}$,
\begin{equation}
\label{Eq:RgivenData}
\mathcal{R}_{ij} = \dfrac{\mathrm{Pr}(M_{i}\vert\bm{D})}{\mathrm{Pr}(M_{j}\vert\bm{D})} = \dfrac{\mathrm{Pr}(\bm{D}\vert M_{i})\mathrm{Pr}(M_{i})}{\mathrm{Pr}(\bm{D}\vert M_{j})\mathrm{Pr}(M_{j})} = \mathcal{B}_{ij}\dfrac{\mathrm{Pr}(M_{i})}{\mathrm{Pr}(M_{j})} \ .
\end{equation}
Here, the ratio of posterior and prior odds in favour of $M_{i}$ over $M_{j}$ is called the `Bayes Factor', $\mathcal{B}_{ij}$, $\mathrm{Pr}(\bm{D}\vert M_{i}) \equiv \mathcal{Z}_{i}$ and $\mathrm{Pr}(\bm{D}\vert M_{j}) \equiv \mathcal{Z}_{j}$ are the BME of $M_{i}$ and $M_{j}$, respectively, and $\mathrm{Pr}(M_{i})/\mathrm{Pr}(M_{j})$ is the prior odds in favour of the $M_{i}$ over $M_{j}$, set before any conclusions have been drawn from the data set.

The Bayes factor is a summary of the evidence provided by the data for one model, as opposed to another. Different classification schemes exist for interpreting the significance that is implied by a given Bayes factor. In this paper, when comparing models for the data, we follow \citet{KandR}, who, in the limit that the different models are \textit{a priori} equally likely, consider $\ln({\mathcal{B}_{ij}}) = \ln(\mathcal{Z}_{i}) -\ln(\mathcal{Z}_{j}) \ge  3$, corresponding to a ratio of the marginal probability of the data given $M_{i}$ relative to $M_{j}$ of better than $\sim20$, to constitute strong evidence in favour of $M_{i}$ over $M_{j}$ and $1 \ge \ln({\mathcal{B}_{ij}}) \ge 3$ to constitute positive evidence.

In this work, when analysing the data, we estimate model evidences using nested sampling as implemented by the \textsc{polychord} algorithm \citep{2015MNRAS.453.4384H, 2015MNRAS.450L..61H}. Additionally, we assume \textit{a priori} that the different models we consider are equally likely, in which case, $R=B$.

\subsubsection{Model priors}
\label{Sec:ModelPriors}

For the data analysis and models considered here, we find that using the Bayes factor to compare between models is sufficient to arrive at reasonable conclusions regarding the preferred model for the data. However, in general, for models with correlated components, in addition to the model as a whole being able to fit the data, to recover unbiased parameter estimates with a model requires that the model components accurately describe the components of the data they are designed to model. If this is \textit{a priori} poorly quantified, one can encounter situations in which models can accurately fit the data despite being comprised of one or more components that are inaccurate descriptions of the components of the data they are intended to model. Such a situation will occur when the error associated with the poor modelling of particular components can be absorbed by the others, leading to good fits to the data but with biased component models and biased estimates of the corresponding model parameters. If the extent to which a subset of the components of the model can be expected to describe the components of the data that they are designed to model is better understood than the remaining components, informative priors on the better understood components can be used to inform the prior on models that contain them. In the context of 21-cm cosmology, this is true of the foreground component of the data, which is far more stringently constrained by existing observations than the 21-cm signal. In upcoming work, in the context of a wider comparison of foreground models and data sets, we will demonstrate that, when one requires both that the model as a whole is an accurate fit to the data \textit{and} the model components are accurate models for the respective components of the data they describe, model priors can be essential for deriving robust conclusions with respect to preferred models for the data; however, this is most important when the true amplitude of the 21-cm signal absorption trough is smaller than the 21-cm absorption trough reported in B18 and considered here (Sims et al. in prep.).

\subsection{Results}
\label{Sec:Results}

In this section, we analyse the simulated data described in \autoref{Sec:RealisitcBFCCSimulations} with the BFCC model derived in \autoref{Sec:RealisticModels} and the Intrinsic model described in \autoref{Sec:ComparisonBFCCDataModels}. In total, we analyse 9 models including the Intrinsic model and 8 BFCC models with numbers of foreground terms of between $N=3$ and $10$. When discussing summary statistics we use the median posterior solutions and uncertainties defined by the 68\% credibility interval about the median. These estimates match the mean and standard deviation in the limit that the posterior distribution is Gaussian and more informatively characterise the distributions when they are not. We use $1$-$\sigma$ as a shorthand when referring to these uncertainty estimates and $N$-$\sigma$ to refer to deviations from the median parameter estimates $N$-times larger.

\subsubsection{BFCC model complexity}
\label{Sec:BFCCmodelComplexity}

\begin{figure}
	\centerline{
	\includegraphics[width=0.5\textwidth]{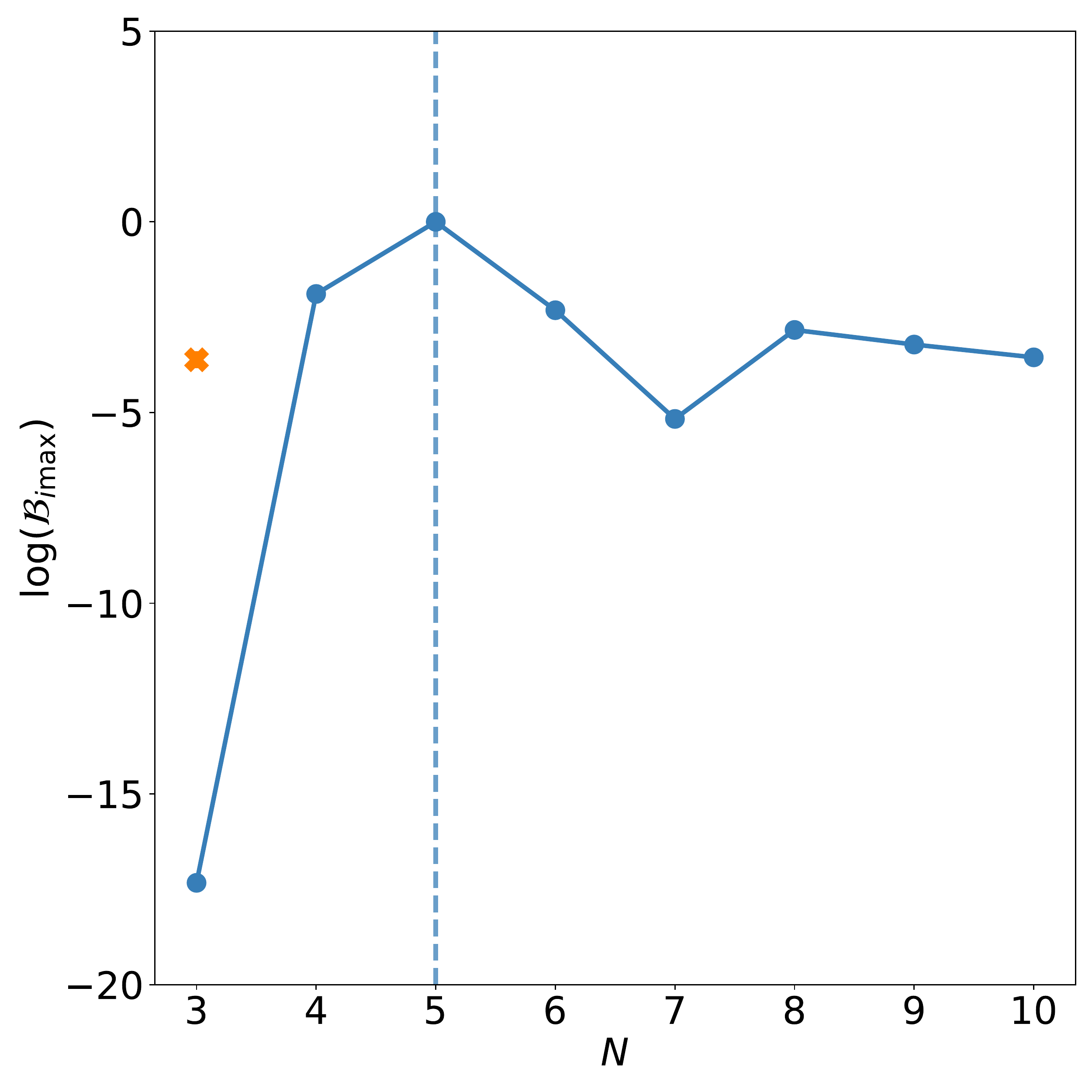}
	}
	\centerline{
	\includegraphics[width=0.5\textwidth]{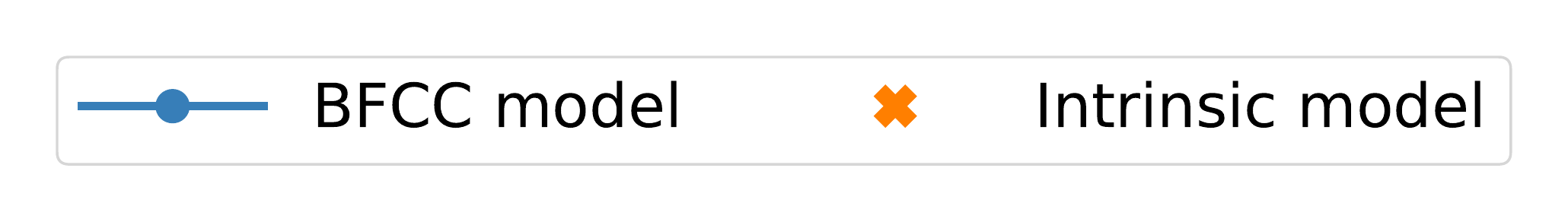}
	}\caption{
        Bayes-factors ($\mathcal{B}_{i\mathrm{max}}$) of model $M_{i}$ over $M_\mathrm{max}$ for data described by \autoref{Eq:TdataSimRealistic} and incorporating a flattened Gaussian 21-cm absorption trough with an amplitude $A=500~\mathrm{mK}$. Here, $M_\mathrm{max}$ is the highest evidence model for the data.
        For the BFCC model, the number of foreground terms included in the model that yields the maximum evidence total model for the data is indicated by the blue dashed line.
}
\label{Fig:BwithAeq500mK}
\end{figure}

\autoref{Fig:BwithAeq500mK} shows the Bayes-factors ($\mathcal{B}_{i\mathrm{max}}$) of the 9 models considered in this section relative to the highest evidence model for the simulated EDGES low-band BFCC data, derived from the sky simulations including foregrounds with realistic spatially dependent spectral structure described in \autoref{Sec:RealisitcBFCCSimulations}. The highest evidence model for the data, $M_\mathrm{max}$, is the BFCC model with $N=5$ foreground terms. There is positive evidence in favour of $M_\mathrm{max}$ over the two next most probable BFCC models with $N=4$ and $6$, respectively, and strong evidence in favour of $M_\mathrm{max}$ over the remaining BFCC models, as well as the Intrinsic model. Additionally, we find that with all models tested, there is strong evidence in favour of including the 21-cm signal component of the models. This implies that there is statistically significant level of structure in the data that is not described by the foreground model but which can be described with the 21-cm signal component of the full model. However, this does not necessarily imply a 21-cm signal has been detected. Rather, it implies either a detection of a signal in the data or a detection of systematic structure fit by the 21-cm model (e.g. \citealt{2020MNRAS.492...22S}). In practice, separation of these two scenarios requires additional evidence in favour of one scenario or the other, such as LST-independence of the detected signature, as expected if the structure results from an isotropic cosmological signal (e.g. see B18).

When assessing the model posteriors in the next section we use the highest Bayesian evidence BFCC model (using $N=5$ foreground terms), and compare the results recovered with it to those recovered with the Intrinsic model. In principle one could extend the analysis to use a weighted combination of multiple models using Bayesian model averaging and weighting the models by the posterior odds in their favour (equal to the Bayes factors relative to a reference model in the limit that the models are \textit{a priori} equally likely). This would not qualitatively change the conclusions arrived at here; however, we explore this approach further, in the context of lower amplitude input 21-cm signals, in upcoming work.

\subsubsection{21-cm signal recovery}
\label{Sec:21cmSignalRecovery}

\begin{figure*}
	\centerline{
	\includegraphics[width=0.5\textwidth]{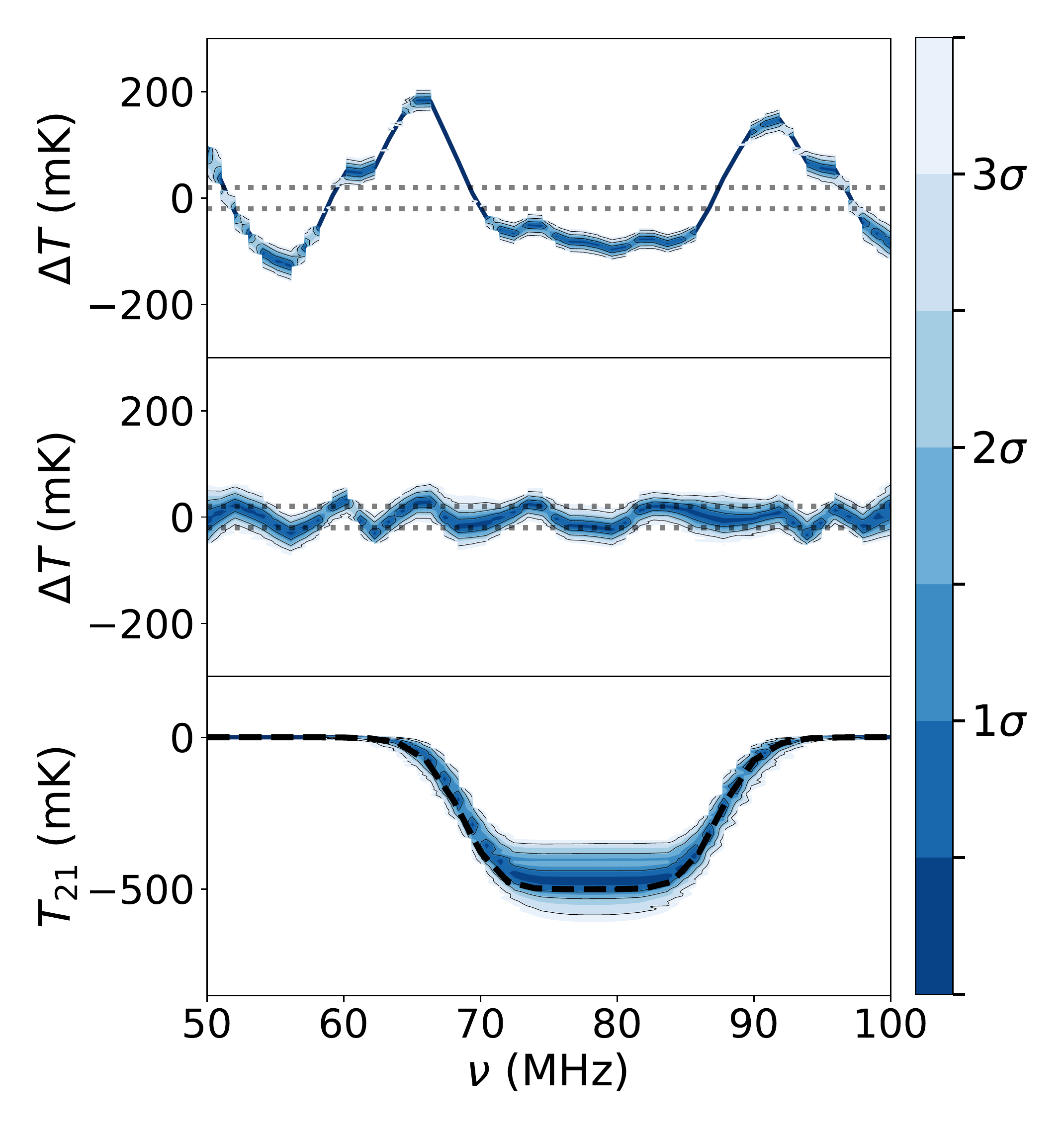}
	\includegraphics[width=0.5\textwidth]{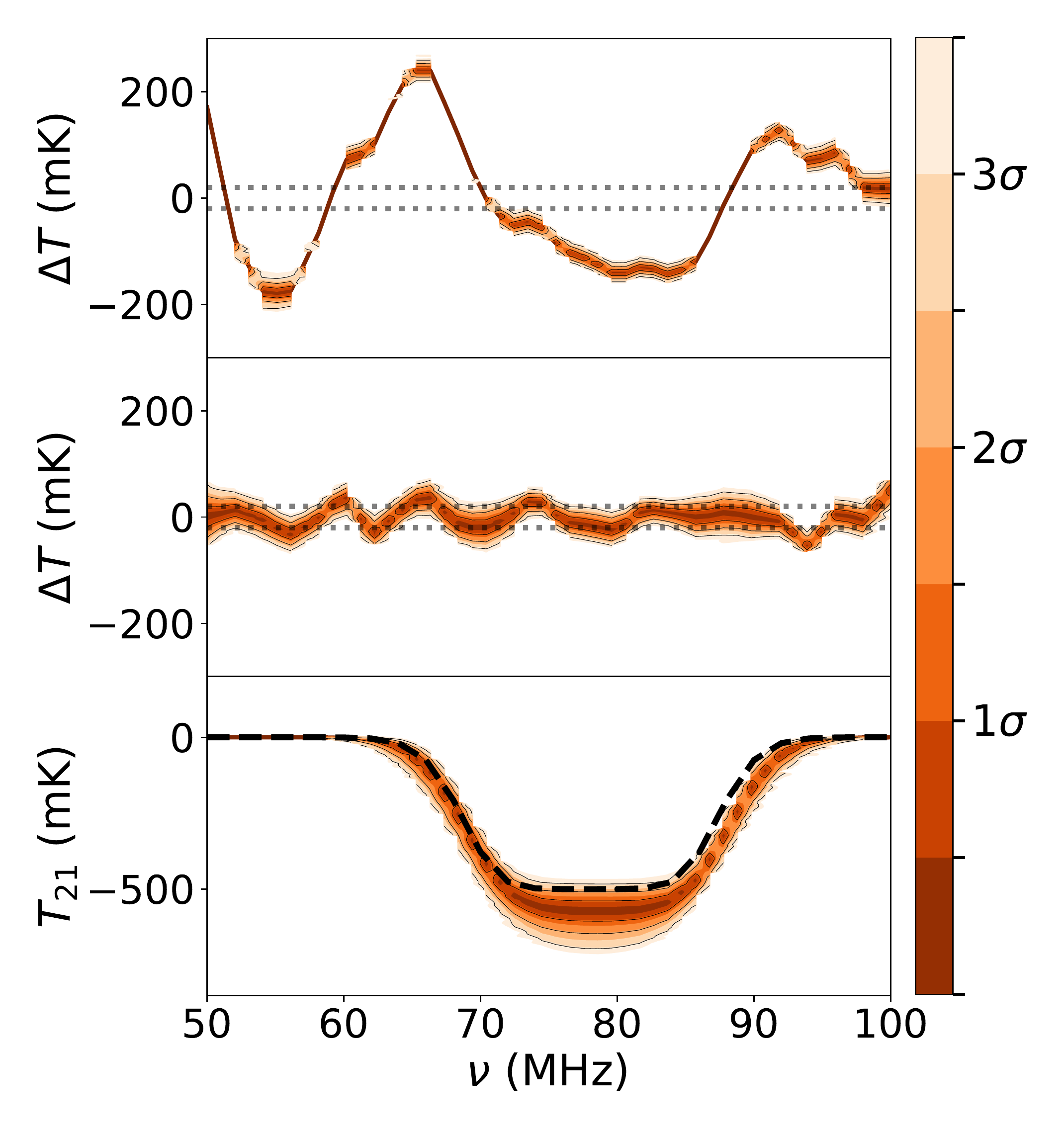}
	}
	\centerline{
	\includegraphics[width=0.6\textwidth]{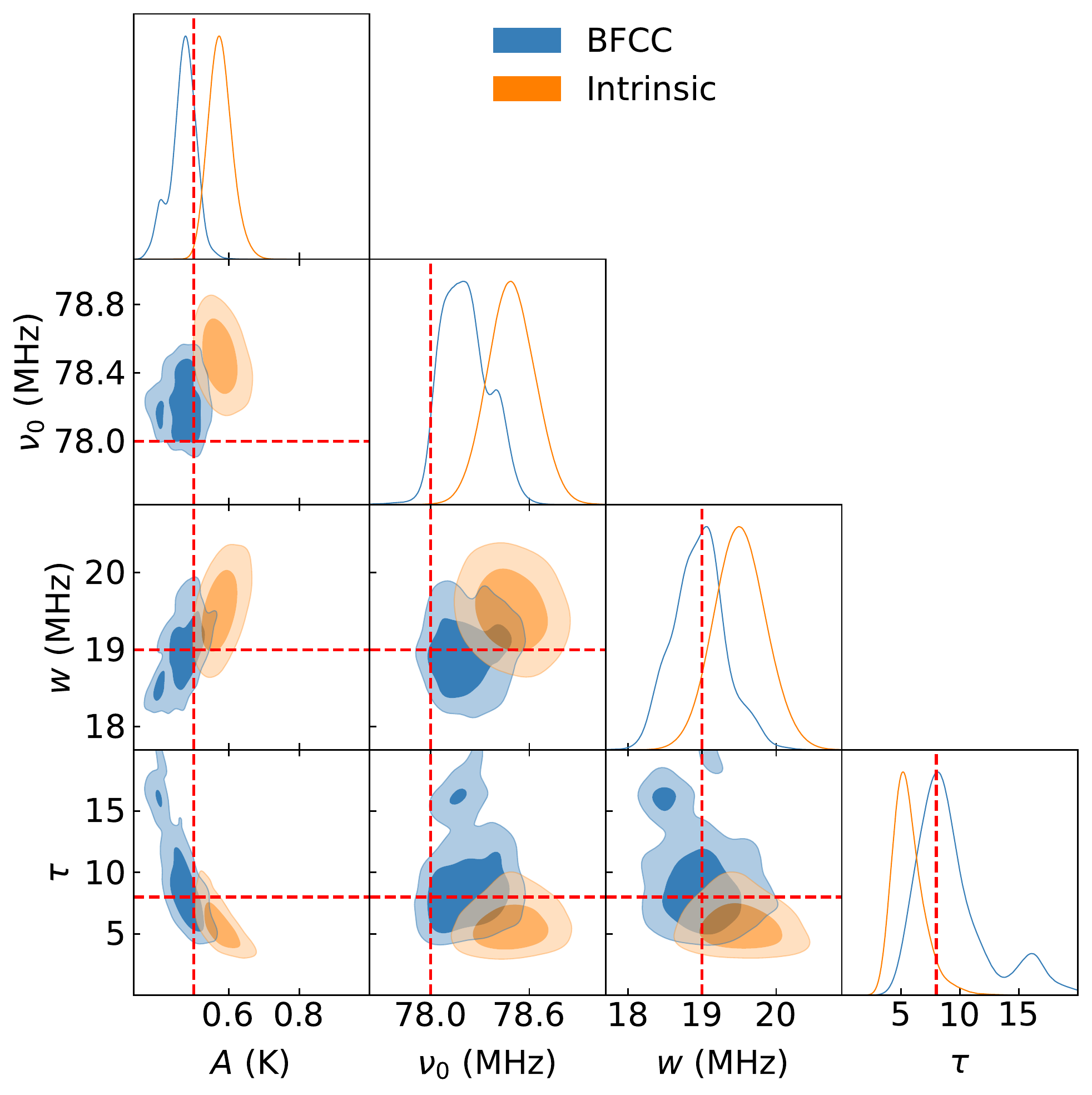}
	}
\caption{
    Signal recovery plots for the BFCC with $N=5$ foreground terms model (top left) and Intrinsic model (top right). In both cases, subplots show the functional posterior probability distributions of the residuals in a fit of the foreground-and-ionosphere-model component of the model to the data (top), the residuals in a fit of the full models to the data (middle) and the recovered 21-cm signal in the fit of the full models to the data (bottom). The dotted lines in the top and middle subplots denote the noise level in the simulated data. The dashed black line shows the input 21-cm signal in the simulated data.  The bottom plot shows one- and two-dimensional marginal posterior probability distributions of the 21-cm signal parameters in the two full models. The dashed red lines show the true values of the flattened Gaussian absorption trough parameters in the simulated data.
    }
\label{Fig:SignalRecovery500mK}
\end{figure*}

The top, middle and bottom subplots, of the top row of \autoref{Fig:SignalRecovery500mK} show functional posterior probability distributions of:
\begin{itemize}
    \item the residuals in a fit of the foreground-and-ionosphere-model components of the models to the data,
    \item the residuals in a fit of the full models to the data, and
    \item the recovered 21-cm absorption trough in the fit of the full models to the data, respectively,
\end{itemize}
for the $N=5$ foreground term BFCC model (left) and Intrinsic model (right).

The RMS residual of the maximum a posteriori models derived in fits of the foreground-and-ionosphere-model components of the models to the data are equal to $92$ and $122~\mathrm{mK}$, for the BFCC and Intrinsic models respectively, relative to the $20~\mathrm{mK}$ RMS expectation value of the noise on the data. In both cases, the significant excess RMS over the expectation for the noise is demonstrative that the spectrometer data including the simulated 21-cm signal can not be well described by only the foreground-and-ionosphere-model components of either model. In contrast, the RMS residual of the maximum a posteriori models derived in fits of the full models to the data are equal to $21$ and $25~\mathrm{mK}$, for the BFCC and Intrinsic models respectively.

Kolmogorov-Smirnov tests for consistency between the residuals, in the fits of these models to the data and the $20~\mathrm{mK}$ standard deviation Gaussian distribution, which describes the expected noise on the data, yield $p$-values less than $10^{-5}$ for both the BFCC and Intrinsic models in the former case and greater than $0.7$ for both models in the latter cases. Thus, from this comparison, one arrives at the same conclusion as from the comparison of the Bayesian evidences in \autoref{Sec:BFCCmodelComplexity}, that including a 21-cm signal component yields preferred models for the data over excluding it.

Looking at the functional posteriors on the recovered 21-cm absorption trough in the data, shown in the bottom subplots, for the BFCC model the recovered signal is consistent at $\sim 1$-$\sigma$ with the underlying signal across the full $50 - 100~\mathrm{MHz}$ spectral band of the data. In contrast, the 21-cm absorption trough recovered with the Intrinsic model is consistent with the underlying signal only at $\sim 3$-$\sigma$ in the $\sim 75 - 85~\mathrm{MHz}$ spectral range and inconsistent at $>3$-$\sigma$ in the $\sim 85 - 90~\mathrm{MHz}$ spectral range.

The corresponding consistency and bias in the recovered 21-cm absorption trough with the BFCC and Intrinsic models, respectively, can be seen in the posteriors for the parameters\footnote{The marginalised posterior probability distributions plot is generated using adaptive kernel density estimates (with corrections for boundary conditions and smoothing biases) with the \textsc{getdist} software package (\citealt{2019arXiv191013970L, 2019ascl.soft10018L}).} of the signal shown in bottom plot of \autoref{Fig:SignalRecovery500mK}. The 21-cm signal parameters recovered with the BFCC model are consistent with the input parameters of the 21-cm absorption trough in the simulated data at $1$-$\sigma$ for all parameters other than $\nu_{0}$, which is consistent at $1.5$-$\sigma$ (see \autoref{Tab:DataModelRealisticResultsTable}). In contrast, a signal biased moderately high in both amplitude and central frequency ($2.3$- and $3.5$-$\sigma$ biases, respectively) relative to the underlying absorption trough in the simulated data is recovered with the Intrinsic model.

\begin{table*}
    \caption{
    Summary of 21-cm parameter inference for the BFCC and Intrinsic models. The 50th quantile posterior parameter estimates and uncertainties corresponding to the 16th and 84th quantiles of the posterior distributions are quoted. The input parameters of the 21-cm absorption trough in the analysed data are: $A=500~\mathrm{mK}$, $\nu_0=78~\mathrm{MHz}$, $w=19~\mathrm{MHz}$ and $\tau=8$. Consistency (or not) of the recovered 21-cm signal with the true signal in the data is noted in the comments.
    }
    \begingroup
    \renewcommand{\arraystretch}{1.5} % Default value: 1
    \centerline{
    \begin{tabular}{l l c c c c l }
    \hline
    Model & $A\ (\mathrm{mK})$ & $\nu_0\ (\mathrm{MHz})$ & $w\ (\mathrm{MHz})$ & $\tau$ & Comment   \\
    \hline
       BFCC
       & $0.47^{+0.03}_{-0.04}$
       & $78.20^{+0.18}_{-0.14}$
       & $18.97^{+0.32}_{-0.37}$
       & $8.56^{+3.44}_{-1.99}$
       & Consistent with input \\
       Intrinsic
       & $0.57^{+0.03}_{-0.03}$
       & $78.49^{+0.15}_{-0.14}$
       & $19.51^{+0.35}_{-0.34}$
       & $5.44^{+1.48}_{-1.05}$
       & Biased \\
    \hline
    \end{tabular}
    }
    \label{Tab:DataModelRealisticResultsTable}
    \endgroup
    \end{table*}

\subsubsection{Ionospheric effects}
\label{Sec:IonosphericEffects}

The median posterior values and 68\% credible regions (centered on the medians) for the beam-weighted and time-averaged foreground temperature and mean spectral index, as well as for the electron temperature and opacity of the ionosphere at $\nu=75~\mathrm{MHz}$ ($\bar{T}_\mathrm{m_{0}}$, $\beta_0$, $T_\mathrm{e}$ and $\tau_0$, respectively) in both the Intrinsic and BFCC model are listed in \autoref{Tab:DataModelRealisticFGIonoResultsTable}, along with the underlying values of the parameters in the simulated data. The underlying value of $\bar{T}_\mathrm{m_{0}}$ quoted in the caption of \autoref{Tab:DataModelRealisticFGIonoResultsTable} is calculated as the time-averaged and beam-weighted temperature of the power-law foregrounds at reference frequency $\nu=\nu_\mathrm{c}$, where $\nu_\mathrm{c}=75~\mathrm{MHz}$ and the power-law foreground temperature is given by ($T_\mathrm{fg}(\nu, \Omega, t, \beta_{\Omega,t}) - T_\gamma$) with $T_\mathrm{fg}(\nu, \Omega, t, \beta_{\Omega,t})$ given by \autoref{Eq:TFgSDSI}. The underlying value of $\beta_0$ is calculated as the time-averaged and apparent sky temperature weighted spectral index in the simulated data. Here, the apparent sky temperature weighting is given by the normalised product of the beam\footnote{The effective mean spectral index of the region of sky observed by the instrument, calculated in this manner, evolves with frequency due to the change in the apparent sky temperature weighting of the underlying spectral index distribution. In practice, the variation with frequency is small ($\sim2$ parts in $1000$); therefore, for simplicity, in \autoref{Tab:DataModelRealisticFGIonoResultsTable} we quote the time-averaged and apparent sky temperature weighted spectral index averaged over the band of the data.} and foreground sky temperature.

The BFCC model recovers solutions for $\bar{T}_\mathrm{m_{0}}$, $\beta_0$ and $T_\mathrm{e}$ consistent, to within 1-$\sigma$, with the underlying values of the parameters in the simulated data. The Intrinsic model recovers solutions for $\beta_0$ and $T_\mathrm{e}$ consistent to within $\sim1$-$\sigma$ with the underlying values of the parameters in the simulated data\footnote{Nominally, the Intrinsic model recovers $\bar{T}_\mathrm{m_{0}}$ consistent with the underlying value in the data to within 1.5-$\sigma$. However, if we subtract, from $\bar{T}_\mathrm{m_{0}}$, the CMB temperature, which is not explicitly modelled as an independent foreground component in the Intrinsic model fit here, this parameter would also agree with the underlying value in the data to within its $1$-$\sigma$ uncertainty.}. In contrast, the median posterior values for $\tau_0$ recovered with the BFCC and Intrinsic models are biased at $2.4$- and $7.4$-$\sigma$, respectively. In the case of the BFCC model, the difference between the median posterior value for $\tau_0$ and the underlying value in the simulation corresponds to differential ionospheric absorption across the band in the model of $\sim1\%$ relative to the data. This differential absorption scales the 21-cm signal model, but this has negligible direct impact at the signal-to-nose level considered here. However, by indirectly providing an additional degree of freedom to the approximate $m_\mathrm{pert}$ model for the contribution of the imperfectly BFCC corrected component of foreground emission resulting from the interaction between the chromatic structure of the beam and the component of foreground emission corresponding to spatially dependent spectral variations about the mean of the foreground spectral index distribution, the mild bias in $\tau_0$ enables the composite foreground-and-ionosphere model to better fit the foreground-and-ionosphere component of the data. As shown in the preceding section, this results in a sufficiently accurate model for the composite foreground-and-ionosphere component of the data to enable unbiased recovery of the 21-cm signal at the $1$-$\sigma$ level.

In contrast, in the case of the Intrinsic model, even with biased $\tau_0$, the foreground-and-ionosphere model has insufficient degrees of freedom to describe the data and the maximum a posteriori foreground-and-ionosphere model remains an inaccurate model for the foreground-and-ionosphere component of the data. Additionally, in combination with this, the systematic error resulting from this inaccuracy summed with the underlying 21-cm signal in the data is reasonably well described by a biased fit of the 21-cm absorption trough component of the Intrinsic model. This, in turn, leads to a reasonably accurate fit of the full Intrinsic model to the data but only with biased recovery of the 21-cm signal.

\begin{table}
\caption{
Summary of common foreground and ionosphere parameters between the BFCC and Intrinsic models. The 50th quantile posterior parameter estimates and uncertainties corresponding to the 16th and 84th quantiles of the posterior distributions are quoted. The underlying values of the parameters in the analysed data are: $\bar{T}_\mathrm{m_{0}}=1627.8~\mathrm{K}$, $\beta_0=2.583$, $T_\mathrm{e}=450~\mathrm{K}$ and $100\tau_0=1.4$.
}
\begingroup
\renewcommand{\arraystretch}{1.5} % Default value: 1
\centerline{
\begin{tabular}{l l c c c c l }
\hline
Model & $ \bar{T}_\mathrm{m_{0}}\ (\mathrm{K})$ & $\beta_0$ & $T_\mathrm{e}\ (\mathrm{K})$ & $100\tau_0$ \\
\hline
BFCC
& $1624.8^{+3.2}_{-2.4}$
& $2.562^{+0.074}_{-0.061}$
& $302^{+193}_{-127}$
& $0.832^{+0.241}_{-0.146}$
\\
Intrinsic
& $1634.0^{+4.2}_{-4.1}$
& $2.582^{+0.002}_{-0.002}$
& $460^{+234}_{-242}$
& $1.704^{+0.040}_{-0.042}$
\\
\hline
        \end{tabular}
}
\label{Tab:DataModelRealisticFGIonoResultsTable}
\endgroup
\end{table}

%%%%%%%%%%%%%%%%%%%%%%%%%%%%%%%%%%%%%%%%%%%%%%%%%%
\section{Discussion}
\label{Sec:Discussion}
%%%%%%%%%%%%%%%%%%%%%%%%%%%%%%%%%%%%%%%%%%%%%%%%%%

The primary goal of global 21-cm signal experiments is to obtain unbiased measurements of, or constraints on, the redshifted 21-cm signal. Our results with respect to this goal have been described in \autoref{Sec:21cmSignalRecovery}, with the BFCC model, in contrast to the Intrinsic model, found to enable unbiased recovery of the 21-cm absorption trough in the simulated BFCC EDGES low-band data.

Beyond deriving a data model capable of realizing this primary goal, all other things being equal, employing physical models for the emission with physical prior ranges on the model parameters (e.g. \citealt{2018Natur.564E..32H, 2021MNRAS.506.2041A}) has the advantage of providing improved interpretability over more general alternative models, as well as reducing the level of correlation between the model components relative to an equivalent model with less constrained priors. However, the more constrained the non-21-cm component of the model, in general the less capable that model will be of absorbing low-level systematic effects in the data, necessitating explicit characterisation and modelling of such effects, if present in the data.

\subsection{Efficacy of the Intrinsic model}
\label{Sec:EfficacyOfTheIntrinsicModel}

The foreground and ionospheric parameters of the Intrinsic model correspond to the amplitudes of physically interpretable parameters, and our priors on those parameters (see \autoref{Tab:DataModelIntrinsicPriors}) ensure that their posteriors are within physically plausible ranges. However, they do not ensure that the posterior distributions of the parameters are unbiased relative to their underlying values in the data (see \autoref{Sec:Results}).

The bias in the 21-cm signal estimates recovered with the Intrinsic model results from the following two conditions being present in the analysis of the data:
\begin{enumerate}
    \item the non-21-cm component of the Intrinsic model is unable to accurately model the corresponding component of the data,
    \item a significant fraction of the resulting systematic structure introduced by the inaccuracy of the non-21-cm component of the Intrinsic model, summed with the underlying 21-cm absorption trough in the simulated data, is modellable with a flattened Gaussian absorption trough that is biased relative to the cosmological signal in the data (i.e. the sum of foreground systematic and signal can be well fit by the signal model).
\end{enumerate}
We consider these properties to make the Intrinsic model a \textit{poor data model}, for the purposes of modelling BFCC EDGES low-band data.

\subsection{Efficacy of the BFCC model}
\label{Sec:EfficacyOfTheBFCCModel}

The physical parameters of the Intrinsic model comprise a subset of the BFCC model parameters. With the BFCC model, as with the intrinsic model, we employ priors on those parameters (see \autoref{Tab:DataModelBFCCPriors}) that ensure their posteriors are within physically plausible ranges. The BFCC model additionally accounts for the inverse beam-factor scaling of all emission components of BFCC data and includes the $m_\mathrm{pert}$ component designed to model structure in $T_{\rm corrected}(\nu, t)$ that is imperfectly corrected by the beam-factor-based chromaticity correction of the data. If $m_\mathrm{pert}$ were a sufficiently accurate standalone description of the component of the data it is designed to model, one would expect to recover accurate 21-cm signal parameters, and not only physically plausible but also unbiased values for the physically interpretable foreground and ionospheric parameters relative to those in the simulated data. We term a model fulfilling this condition a \textit{perfect data model}\footnote{In practice, such a model is perfect insofar as it is unbiased, and models with $m_\mathrm{pert}$ components of varying complexity, and correlations with the other model components, can in principle fall in this category. In this case, models with minimal correlation between $m_\mathrm{pert}$ and other model components will enable more precise parameter constraints and, thus, are preferred. However, since such models will naturally have less flexibility for absorbing systematics they necessitate meticulous care that such systematic structure is not present in any new data being analysed.} of BFCC EDGES low-band data.

In contrast, an intermediate scenario can also occur if, in combination, the foreground-plus-ionosphere model is an accurate model for the foreground-plus-ionosphere model component of the data without either being perfect data models for their respective components of the data. In such a situation one would still expect to recover unbiased estimates of the 21-cm signal but to recover biased estimates for some or all of the foreground-plus-ionosphere-specific model parameters. Such a model achieves the primary goal of 21-cm signal experiments but falls short of the condition for a \textit{perfect data model}, defined above. Despite this, our primary analysis goal is recovery of unbiased estimates of the 21-cm signal, and such a model nevertheless constitutes an \textit{excellent data model} for this purpose.

The variant of the BFCC model (\autoref{Eq:BFCCdataModel}) tested in \autoref{Sec:Results} enables reliable 21-cm signal inference from realistic simulated BFCC EDGES low-band data in addition to recovering unbiased recovery of $\bar{T}_\mathrm{m_{0}}$, $\beta_0$, $T_\mathrm{e}$. In this sense, it approaches a \textit{perfect data model} of BFCC EDGES low-band data, as defined above. However, further refinement of the model for $T_\mathrm{pert}$, the imperfectly BFCC corrected component of foreground emission resulting from the interaction between the chromatic structure of the beam and the component of foreground emission corresponding to spatially dependent spectral variations about the mean of the foreground spectral index distribution, would be necessary to eliminate bias in the recovered estimates of $\tau_0$. Thus, the variant of the BFCC model using the power law-damped log-polynomial parametrisation for $m_\mathrm{pert}$ tested here falls under our classification for an \textit{excellent data model} of BFCC EDGES low-band data. Recovery of $\tau_0$ is not of principle importance in the application of the BFCC model developed in this paper, we therefore leave further consideration of models with additional optimisation for recovery of this parameter to future work.

\subsection{Comparison to Intrinsic model results in \citet{2018Natur.564E..32H}}
\label{Sec:ComparisonH18}

In \citet{2018Natur.564E..32H} (hereafter, H18), a number of models are fit to the publicly available BFCC EDGES low-band data. Of these, the one of interest for comparison here is the analysis using a model equivalent to the Intrinsic model considered here, with restrictions to the fit, comparable to the priors on the model parameters imposed in this paper, to ensure recovery of physically plausible posteriors on the foreground and ionospheric parameters of the model. In this fit, the amplitude and central frequency of the recovered signal are increased relative to the solution recovered with a model  substituting foreground-plus-ionosphere component of the Intrinsic model with the less constrained and more flexible variant of that model employed in B18. Comparing the 21-cm absorption trough recovered with the BFCC and Intrinsic model, we find qualitatively the same behaviour here as that found in the above comparison, with the amplitude and central frequency of the recovered 21-cm absorption trough biased high (here, at $2.3$- and $3.5$-$\sigma$ significance, respectively) relative to the 21-cm signal posteriors recovered with the BFCC model.

While the foreground models fit in B18 are not equivalent to the BFCC model considered here, they share the trait of being more flexible foreground models relative to the Intrinsic model. As a result, it is of interest to compare the difference between the signal recovered in the B18 analysis and that recovered in the analysis of the same data with the Intrinsic model in H18 to the difference in the signals recovered with the BFCC model and Intrinsic model here\footnote{In upcoming work, focussed on model priors, we find that qualitatively the same result holds with respect to the 21-cm signal recovered with the linearised form of the Intrinsic model fit with uninformative priors on the effective foreground and ionosphere parameters, as used in B18.}. We leave to upcoming work analysis of the publicly available data from B18 with the BFCC model, enabling a more direct comparison.

Despite the qualitative agreement in the parameters and direction of bias with the intrinsic model, in the two cases, the results of the H18 analysis of the publicly available BFCC EDGES low-band data and the analysis of the simulated BFCC EDGES low-band data considered here differ quantitatively in two ways of interest:
\begin{enumerate}
    \item While the bias in the amplitude and central frequency of the absorption trough recovered with the Intrinsic model are statistically significant, the absolute values of the offsets are smaller than the offsets between those model parameters for the 21-cm signals recovered when fitting the Intrinsic model in H18 relative to the 21-cm signal found in B18. In particular, here, the median posterior value of the absorption trough recovered with the Intrinsic model is $\sim100~\mathrm{mK}$ larger than both that recovered with the BFCC model and the amplitude of the true signal in the simulated data, while the central frequency is approximately $300~\mathrm{kHz}$ high than in the BFCC model. In contrast, it is shown in H18 that the absorption trough recovered with the Intrinsic model is approximately $1.5~\mathrm{K}$ larger in amplitude and with an order $10~\mathrm{MHz}$ higher in central frequency than the absorption trough recovered with a more flexible foreground model in B18.
    \item While the 21-cm signal recovered using the Intrinsic model to fit the simulated BFCC EDGES low-band data in this work is biased, the overall fit of the intrinsic model to the data leaves residuals with an RMS amplitude of $\sim25~\mathrm{mK}$, comparable to the $20~\mathrm{mK}$ expectation value of the noise in the simulated data. In contrast, H18 find that the fit of the Intrinsic model to the publicly available EDGES low-band data yields residuals with an RMS of $121~\mathrm{mK}$, significantly larger than the $24~\mathrm{mK}$ RMS residuals found when jointly fitting the data with a 21-cm absorption trough and the more flexible linearised form of the Intrinsic foreground-and-ionosphere model with uninformative priors, applied in B18.
\end{enumerate}

Both of the above points are suggestive of additional structure present in the publicly available BFCC EDGES low-band data relative to the simulated BFCC EDGES low-band data in this work. In this paper we assumed no uncertainty on the base-map and beam model used in beam-factor chromaticity correction of the data. In practice both of these assumptions will be violated at some level. Correspondingly, imperfections in the beam and base-map models used to derive the beam-factors which were applied to the publicly available BFCC data are a candidate for this additional structure. We plan to explore this possibility further, as well as assess the efficacy of the BFCC model for accounting for such structure and mitigating its impact on 21-cm signal recovery in simulations and in the analysis of EDGES data in upcoming work.

%%%%%%%%%%%%%%%%%%%%%%%%%%%%%%%%%%%%%%%%%%%%%%%%%%
\section{Summary and conclusions}
\label{Sec:Conclusions}
%%%%%%%%%%%%%%%%%%%%%%%%%%%%%%%%%%%%%%%%%%%%%%%%%%

Accurately accounting for spectral structure in spectrometer data induced by instrumental chromaticity on scales relevant for detection of the 21-cm signal is among the most significant challenges in global 21-cm signal analysis. In the publicly available EDGES low-band data set, this complicating structure is suppressed using beam-factor based chromaticity correction (BFCC), which works by dividing the data by a sky-map-weighted model of the spectral structure of the instrument beam.

BFCC perfectly corrects for chromatic effects\footnote{Here, by perfectly correct we mean that in the limit of an error-free model for the instrument beam, $B^\mathrm{m}$, and  $T_\mathrm{fg}^\mathrm{m}$, a closed-form solution for $T_{\rm corrected}(\nu)$ can be derived in terms of the (assumed known) intrinsic spectral structure of the sky observed by the instrument and the calculated beam-factor.} only in the hypothetical scenario that the foregrounds contributing to the measured spectrometer data have spatially independent spectral structure. In this case, we have shown that the correction is perfect in the sense that, in the limit that one has an error-free model for the beam and of the sky at a given reference frequency from which a beam-factor model is constructed, one can write down a closed-form solution for BFCC data in terms of the intrinsic spectral structure of the sky observed by the instrument and the calculated beam-factor (\autoref{Eq:TcorrectedC2pt1}). However, even in this simplified scenario, it is not the case that BFCC produces data that is proportional to the autocorrelation spectrum that would be measured if the spectrometer had an achromatic beam, $T_\mathrm{sky}(\nu, t)$.

For realistic foreground spectral structure dominated by Galactic diffuse synchrotron emission and synchrotron emission from extragalactic sources, with a smaller contribution from Galactic free-free emission, the intrinsic foreground spectrum is spatially-dependent. In this case, we have shown that there is not an exact closed-form expression for the resulting BFCC data. However, by writing the spectral structure of the foregrounds in terms of spatially dependent spectral perturbations on top of a spatially isotropic background (corresponding to small spatially-dependent spectral index perturbations to an isotropic power law spectrum), we derive a physically-motivated approximate model for the foregrounds comprised of two components:
\begin{enumerate}
    \item the dominant spatially isotropic power law component of the foregrounds, and
    \item the subdominant contribution of the spatially dependent spectral perturbations, about a spectrally isotropic background.
\end{enumerate}

 We have shown that in the limit that one has an error-free beam-factor model, the first component can be perfectly corrected for instrumental chromaticity via BFCC. However, the second component is not exactly describable with a finite-term closed-form expression. Nevertheless, we show that it can be accurately approximated with a moderate number of terms and that the optimal complexity for this component of the model when describing a given data set can be ascertained via Bayesian model comparison. Furthermore, we show that recovery of unbiased estimates of the underlying 21-cm signal in the data is possible when using the maximum Bayesian evidence BFCC models to analyse realistic simulated observations of the BFCC EDGES low-band data incorporating:
\begin{itemize}
\item foregrounds with realistic spatially dependent spectral structure,
\item spectrally-dependent absorption by the ionosphere,
\item ionospheric emission,
\item a flattened Gaussian 21-cm absorption profile with parameters consistent with those found in B18.
\end{itemize}
In contrast, even in the limit of an error-free beam-factor model, fitting BFCC data with an Intrinsic model, which neglects the residual effects of beam chromaticity on BFCC data, recovers biased estimates of the underlying 21-cm signal in the data.

Comparing the difference between the 21-cm signal recovered with the Intrinsic model and the BFCC model from simulated BFCC EDGES low-band data using a realistic foreground model and error-free beam-factor model to the difference between the 21-cm signal recovered with the Intrinsic model component of the H18 analysis of the publicly available EDGES low-band data and the signal recovered in B18 with a more flexible foreground model, we find qualitatively the same behaviour, with the amplitude and central frequency of the recovered 21-cm absorption trough biased high. However, the size of the difference in these two quantities is substantially larger in the latter case, and the overall fit of the model to the data is worse. Both of the above points are suggestive of additional structure present in the publicly available BFCC EDGES low-band data relative to the simulated BFCC EDGES low-band data in this work.

An important caveat to the simulated results derived here is that errors in the beam-factor model have the potential to reduce the efficacy of BFCC, introducing an additional source of the systematic structure in observational data, relative to the simulated data analysed here. Such systematic structure is expected to contribute to the above difference.
In future work in this series, to answer this question we plan to explore the impact of realistic deviations from the assumption of an error-free model for the beam-factor used in BFCC, due to uncertainties in the base-map sky model and beam model. Additionally, we will assess the efficacy of the BFCC model at accounting for such structure and for mitigating its impact on 21-cm signal recovery in simulations and in the analysis of EDGES data.

%%%%%%%%%%%%%%%%%%%%%%%%%%%%%%%%%%%%%%%%%%%%%%%%%%
\section*{Acknowledgements}
%%%%%%%%%%%%%%%%%%%%%%%%%%%%%%%%%%%%%%%%%%%%%%%%%%

This work was supported by the NSF through research awards for EDGES (AST-1609450, AST-1813850, and AST-1908933). PHS was supported in part by a Trottier Space Institute fellowship and funding from the Canada 150 Research Chairs Program and would like to thank Jonathan Sievers and Irina Stefan for valuable discussions related to the work in this manuscript. NM was supported by the Future Investigators in NASA Earth and Space Science and Technology (FINESST) cooperative agreement 80NSSC19K1413. EDGES is located at the Murchison Radio-astronomy Observatory. We acknowledge the Wajarri Yamatji people as the traditional owners of the Observatory site. We thank CSIRO for providing site infrastructure and support.

%%%%%%%%%%%%%%%%%%%%%%%%%%%%%%%%%%%%%%%%%%%%%%%%%%
\section*{Data Availability}
%%%%%%%%%%%%%%%%%%%%%%%%%%%%%%%%%%%%%%%%%%%%%%%%%%

The data from this study will be shared on reasonable request to the corresponding author.
Software used in this work to generate beam-factors, given an electromagnetic simulation of the beam, is publicly available at \url{https://github.com/edges-collab}.



\bibliographystyle{mnras}

%%%%%%%%%%%%%%%%%%%%%%%%%%%%%%%%%%%%%%%%%%%%%%%%%%

%%%%%%%%%%%%%%%%% APPENDICES %%%%%%%%%%%%%%%%%%%%%

\appendix

% %%%%%%%%%%%%%%%%%%%%%%%%%%%%%%%%%%%%%%%%%%%%%%%%%%
% \section{Some extra material}
% %%%%%%%%%%%%%%%%%%%%%%%%%%%%%%%%%%%%%%%%%%%%%%%%%%

% %%%%%%%%%%%%%%%%%%%%%%%%%%%%%%%%%%%%%%%%%%%%%%%%%%
\section{Monsalve beam-factor chromaticity correction}
\label{Sec:MonsalveBFCC}
% %%%%%%%%%%%%%%%%%%%%%%%%%%%%%%%%%%%%%%%%%%%%%%%%%%

In \citet{2017ApJ...847...64M}, an alternate chromaticity correction formulation is employed with a similar form to \autoref{Eq:CCsnapshot} but with a model of the foreground brightness temperature distribution across the observing band replacing the foreground sky brightness temperature distribution at reference frequency $\nu_\mathrm{c}$,
\begin{equation}
    \label{Eq:CCsnapshot2}
    B_\mathrm{factor}^{\prime}(\nu, t) = \frac{\int\limits_{\Omega^{+}}  B^\mathrm{m}(\nu, \Omega) T_\mathrm{fg}^\mathrm{m}(\nu, \Omega, t) \mathrm{d}\Omega}{\int\limits_{\Omega^{+}}  B^\mathrm{m}(\nu_\mathrm{c}, \Omega) T_\mathrm{fg}^\mathrm{m}(\nu, \Omega, t) \mathrm{d}\Omega} \ .
    \end{equation}

A correction of this form requires accurate knowledge of the spatially dependent spectral structure of the foregrounds in addition to the accurate knowledge of the beam in the frequency range being corrected and an accurate foreground model at reference frequency $\nu_\mathrm{c}$, required by \autoref{Eq:CCsnapshot}.

If the spectral structure of the foreground model is approximated as isotropic, the foreground spectrum can be factored out of the numerator and denominator of \autoref{Eq:CCsnapshot2}, and one recovers \autoref{Eq:CCsnapshot}. Thus, the corrections provided by $B_\mathrm{factor}$ and $B_\mathrm{factor}^{\prime}$ differ only when one models spatially dependent foreground spectral structure.

In this work, we focus on the effectiveness of Mozdzen BFCC as defined in \autoref{Eq:CCsnapshot}, both because it is the nominal form of beam-factor chromaticity correction used in B18 and because, by not requiring an \textit{a priori} fixed model of the spatially dependent spectral structure of the foregrounds to generate the beam-factor, Mozdzen BFCC eliminates the potential for introducing spurious spectral structure into the data through errors in this model. However, we return to the question of the relative merits of the Monsalve versus Mozdzen formulations of beam-factor chromaticity correction in Appendix \ref{Sec:MitigationWithMonsalveBFCC1}.

% %%%%%%%%%%%%%%%%%%%%%%%%%%%%%%%%%%%%%%%%%%%%%%%%%%
\section{Definitions of perfect BFCC}
\label{Sec:PerfectBFCCDefinitions}
% %%%%%%%%%%%%%%%%%%%%%%%%%%%%%%%%%%%%%%%%%%%%%%%%%%

When constructing models for BFCC data that assume that its spectral structure is well described by physical models for the intrinsic spectral structure of the emission components (see e.g. B18; \citealt{2018Natur.564E..32H}, for examples), one is implicitly assuming that the BFCC has perfectly corrected the data as follows:
\begin{itemize}
\item \textit{Perfect BFCC definition 1} - a correction that produces data that is proportional to the autocorrelation spectrum that would be measured if the spectrometer had an achromatic beam\footnote{In the limit of a beam that is achromatic and uniform across the sky,  in each channel of the data one recovers a uniformly weighted average of the sky temperature above the horizon, $T_{\rm data,uniform}(\nu, t)$, at time $t$. For an achromatic but spatially non-uniform beam and spatially non-uniform and spectrally isotropic brightness temperature distribution, the recovered spectrum $T_{\rm data,non-uniform}(\nu, t) \propto T_{\rm data,uniform}(\nu, t)$; thus the models for the two data sets are identical up to a scale factor.}, $T_\mathrm{sky}(\nu, t)$ (such that an accurate analytic model for the intrinsic spectral structure of the sky alone can be used to model the chromaticity corrected data).
\end{itemize}

To assess the validity of assuming \textit{perfect BFCC definition 1,} it is of interest to understand the requirements on the spectral structure of the sky and the instrument under which it is valid.

From \autoref{Eq:Tcorrected}, we will verify shortly that \textit{perfect BFCC according to definition 1} is achieved under the following conditions,
\begin{enumerate}
\item One has an error-free model for the instrument beam:
\begin{equation}
\label{Eq:PerfectBeamModel}
B^\mathrm{m}(\nu, \Omega) \equiv B(\nu, \Omega) \ ,
\end{equation}
for use in constructing $B_\mathrm{factor}(\nu, t)$.
\item One has an error-free model for the foreground brightness temperature distribution overhead as a function of time, at reference frequency $\nu_\mathrm{c}$:
\begin{equation}
\label{Eq:PerfectBaseMap}
T_\mathrm{fg}^\mathrm{m}(\nu_\mathrm{c}, \Omega, t) \equiv T_\mathrm{fg}(\nu_\mathrm{c}, \Omega, t) \ ,
\end{equation}
also for use in constructing $B_\mathrm{factor}(\nu, t)$.
\item The measured sky brightness distribution has spatially independent spectral structure:
\begin{equation}
\label{Eq:SpatiallyIndependentSpectrum}
T_\mathrm{sky}(\nu, \Omega, t) = T_\mathrm{sky}(\nu_\mathrm{c}, \Omega, t)f(\nu) \ .
\end{equation}
where $f(\nu)$ is spatially isotropic but can be an arbitrary function of frequency.
\item The sky brightness distribution is proportional to the foreground brightness distribution,
\begin{equation}
\label{Eq:SkyProptoForeground}
T_\mathrm{sky}(\nu, \Omega, t) \propto T_\mathrm{fg}(\nu, \Omega, t) \ .
\end{equation}
\end{enumerate}

That the above conditions are sufficient for deriving BFCC data which is proportional to the autocorrelation spectrum that would be measured if the spectrometer had a uniform achromatic beam, can be verified by substituting Equations \ref{Eq:PerfectBeamModel}--\ref{Eq:SkyProptoForeground} into \autoref{Eq:Tcorrected}. Defining $\gamma = T_\mathrm{sky}(\nu, \Omega, t) / T_\mathrm{fg}(\nu, \Omega, t)$ as the constant of proportionality between the sky and foreground brightness distributions, such that $T_\mathrm{sky}(\nu, \Omega, t) = \gamma T_\mathrm{fg}(\nu, \Omega, t) = \gamma T_\mathrm{fg}(\nu_\mathrm{c}, \Omega, t) f(\nu)$, and $T_\mathrm{m_{0}}(t) = \int B(\nu_\mathrm{c}, \Omega) T_\mathrm{fg}(\nu_\mathrm{c}, \Omega, t) \mathrm{d}\Omega$ as the beam weighted average foreground brightness at reference frequency $\nu_\mathrm{c}$, we have,
\begin{align}
\label{Eq:TcorrectedC3pt1}
T_{\rm corrected}(\nu, t) ={}& T_\mathrm{data} / B_\mathrm{factor}\\ \nonumber
={}&  \left[\int\limits_{\Omega^{+}} B(\nu, \Omega) \gamma T_\mathrm{fg}(\nu_\mathrm{c}, \Omega, t)f(\nu) \mathrm{d}\Omega + n \right] \\ \nonumber
&\times \frac{\int\limits_{\Omega^{+}} B(\nu_\mathrm{c}, \Omega) T_\mathrm{fg}(\nu_\mathrm{c}, \Omega, t) \mathrm{d}\Omega}{\int\limits_{\Omega^{+}} B(\nu, \Omega) T_\mathrm{fg}(\nu_\mathrm{c}, \Omega, t) \mathrm{d}\Omega} \\ \nonumber
={}& \gamma T_\mathrm{m_{0}}(t) f(\nu) + \frac{n}{B_\mathrm{factor}(\nu, t)}  \ .
\end{align}
Here, $\gamma$ and $T_\mathrm{m_{0}}$ are frequency-independent constants and, correspondingly, $T_{\rm corrected}(\nu) \propto f(\nu)$. Additionally, from \autoref{Eq:SpatiallyIndependentSpectrum}, $T_\mathrm{sky}(\nu, t) \propto f(\nu)$; as such, $T_{\rm corrected}(\nu) \propto T_\mathrm{sky}(\nu)$ and BFCC fulfils \textit{perfect BFCC definition 1}.

Nevertheless, conditions (iii) and (iv) limit the usefulness of \textit{perfect BFCC definition 1}. In practice, each of the conditions (i)--(iv) is violated at some level; however, conditions (i) and (ii) can, in principle, be approached with arbitrary precision given sufficiently accurate measurements of the beam and sky. This is not true of conditions (iii) and (iv). Rather, in reality, the spectral structure of the foreground brightness distribution on the sky is not isotropic, violating condition (iii) and the 21-cm and foreground brightness temperature distributions are uncorrelated, violating condition (iv).The impact of spatially dependent foreground spectral structure on modelling BFCC data is discussed in \autoref{Sec:RealisticModels}. For condition (iv) to hold would require that the foregrounds and signal have the same spectral structure. This latter point, in particular, limits the usefulness of \textit{perfect BFCC definition 1} in the context of 21-cm signal estimation, given that the spectral distinctiveness of the 21-cm signal from the foregrounds provides a primary means of separating the two signal components.

Given this, we consider the following redefinition of perfect BFCC, which, we will show in \autoref{Sec:PerfectBFCC}, allows condition (iv) to be eliminated:
\begin{itemize}
\item \textit{Perfect BFCC definition 2} - we take our second and final definition of perfect BFCC to be a correction for which, in the limit of an error-free model for $B^\mathrm{m}(\nu, \Omega)$ and $T_\mathrm{fg}^\mathrm{m}(\nu_\mathrm{c}, \Omega, t)$, a closed-form solution for $T_{\rm corrected}(\nu)$ can be derived in terms of the (assumed known) intrinsic spectral structure of the sky observed by the instrument and the calculated beam-factor.
\end{itemize}

While \textit{perfect BFCC definition 2} provides no guarantee that $T_{\rm corrected}(\nu)$ will be proportional to $T_\mathrm{sky}(\nu)$ (and in general it will not be), it nevertheless fulfils the more important attribute that under conditions (i)--(iii), and as long as one has an accurate model for $B_\mathrm{factor}(\nu, t)$ and the intrinsic spectral structure of the individual signal components comprising $T_\mathrm{sky}(\nu)$, one can fit in an unbiased manner $T_{\rm corrected}(\nu)$ derived from data violating condition (iv).

% %%%%%%%%%%%%%%%%%%%%%%%%%%%%%%%%%%%%%%%%%%%%%%%%%%
\section{Reducing the complexity of $m_\mathrm{pert}$ using Monsalve BFCC}
\label{Sec:MitigationWithMonsalveBFCC1}
% %%%%%%%%%%%%%%%%%%%%%%%%%%%%%%%%%%%%%%%%%%%%%%%%%%

In \autoref{Sec:RealisticModels}, we demonstrated that when writing the spectrum of the power law component of the foreground emission as a spatially varying power law and dividing the contribution of this emission to BFCC data into a subcomponent with spatially independent spectral structure and a smaller spatially dependent spectral perturbation, assuming an error-free model for the beam-factor, it is only the latter subcomponent that is not perfectly corrected for instrumental chromaticity using Mozdzen BFCC. We proceeded to derive a compact and constrained model for this subcomponent; however, it is of interest to consider whether an alternate form of BFCC could correct for both the spectrally isotropic and anisotropic contributions of the foreground to the data.

In the hypothetical limit that one has accurate knowledge of the spatially dependent spectral structure of the foregrounds in addition to accurate knowledge of the beam in the frequency range of the data and an accurate foreground model at reference frequency, $\nu_\mathrm{c}$, Monsalve BFCC has this potential. In fact, in the limit of perfect knowledge of these components\footnote{In the hypothetical limit of perfect knowledge of the spatially dependent spectral structure of the foregrounds, in addition to accurate knowledge of the beam in the frequency range of the data and an accurate foreground model at reference frequency $\nu_\mathrm{c}$, the numerator of the Monsalve beam-factor ($B_\mathrm{factor}^{\prime}(\nu, t)$; see \autoref{Eq:CCsnapshot2}) is equal to the foreground component of the data. Thus, it could simply be subtracted from $T_\mathrm{data}(\nu, t)$ to recover the underlying 21-cm signal, rendering the beam-factor chromaticity correction unnecessary.}, if one were to chromaticity correct the data using $B_\mathrm{factor}^{\prime}(\nu, t)$, the correction would be perfect by definition 2 in Appendix \ref{Sec:PerfectBFCCDefinitions}.

However, in contrast, in the realistic scenario of an imperfect model for the spatially dependent spectral structure of the foregrounds, the numerator of $B_\mathrm{factor}^{\prime}$ is no longer equal to the foreground component of the spectrometer data. In this case, dividing by $B_\mathrm{factor}^{\prime}$ will at-best provide a partial correction for true instrumental chromaticity while simultaneously introducing new foreground model-dependent spurious spectral structure into the corrected spectrum. The level of spurious spectral structure introduced into the corrected spectrum by $B_\mathrm{factor}^{\prime}$ in this scenario is a function of the accuracy of the model for the spatially dependent spectral structure of the foregrounds. As the accuracy of the model decreases, the correction for true instrumental chromaticity in the spectrum will deteriorate and the level of spurious spectral structure will increase, eventually becoming a dominant source of foreground systematic structure.

Between these two extremes, there will be a transition regime in which residual instrumental chromaticity when using $B_\mathrm{factor}$ or $B_\mathrm{factor}^{\prime}$ are comparable. We leave to future work the investigation of the level to which the spatially dependent spectral structure of the foregrounds must be known for chromaticity correction using $B_\mathrm{factor}^{\prime}$ to match or improve on chromaticity correction using $B_\mathrm{factor}$. Nevertheless, we note that, with either correction, some level of residual instrumental chromaticity in the spectrum is inevitable given realistic uncertainties on the beam and the sky, thus a BFCC model with an $m_\mathrm{pert}(\nu)$ component, such as \autoref{Eq:PLLogPolynomialLSTAverage}, will be necessary in either case.

% %%%%%%%%%%%%%%%%%%%%%%%%%%%%%%%%%%%%%%%%%%%%%%%%%%
\section{A parametric beam-factor extension to the BFCC formalism}
\label{Sec:ParametricBFCC}
% %%%%%%%%%%%%%%%%%%%%%%%%%%%%%%%%%%%%%%%%%%%%%%%%%%

In principle, one could extend BFCC, with either the Mozdzen or Monsalve formulations of the beam-factor, to use a parametric rather than fixed beam-factor model. In this case, one would fit the parameters of the beam-factor model jointly with those of the model for the BFCC data, such that uncertainties on the inputs to the beam-factor (the beam and base-map models for Mozdzen BFCC and the beam, base-map and spectral structure models for Monsalve BFCC) are directly accounted for. If the uncertainties associated with the components of the beam-factor are sufficiently low for residual spectral structure imparted to the BFCC data, due to errors in the beam-factor components, to be small relative to the noise in the data, such a joint fit is unnecessary. The computational expense of evaluating the integrals associated with calculating the beam-factor (e.g. \autoref{Eq:CCsnapshot2}) greatly exceeds (by $>6$ orders of magnitude for the beam-factor and BFCC model considered in this work) that of evaluating a closed-form model for BFCC data (\autoref{Eq:PLLogPolynomialLSTAverage}). Thus, in this regime, the computational efficiency of one's analysis is significantly enhanced. If residual spectral structure imparted to the BFCC data, due to errors in the beam-factor components, is statistically significant, jointly fitting for a model of the residual structure with the existing parameters of the BFCC model provides a computationally efficient alternative to jointly fitting for a parametric beam-factor model with the model for the BFCC data. In fact, noting that the Monsalve formulation of the beam-factor reduces to the Mozdzen formulation in the limit that one assumes that the foreground component of the beam-factor model has spatially independent spectral structure, one way to view $m_\mathrm{pert}(\nu)$ (see \autoref{Sec:BFCCDataModel}) is as exactly such a model component. In this case, $m_\mathrm{pert}(\nu)$ describes the residual spectral structure resulting from neglecting spatial fluctuations in the spectral structure of the foregrounds in the beam-factor model. As is shown in \autoref{Sec:Results}, the structure due to this approximation can be accurately absorbed by $m_\mathrm{pert}(\nu)$, preventing bias in the recovered 21-cm signal.

% %%%%%%%%%%%%%%%%%%%%%%%%%%%%%%%%%%%%%%%%%%%%%%%%%%
\section{Intrinsic foreground model derivation}
\label{Sec:IntrinsicForegroundModelDerivation}
% %%%%%%%%%%%%%%%%%%%%%%%%%%%%%%%%%%%%%%%%%%%%%%%%%%

If one describes the intrinsic sky brightness with \autoref{Eq:TskySDFgStationaryIonosphere} and assumes that it is observed with a hypothetical uniform achromatic antenna with beam $B_\mathrm{u}$ (or, similarly, data taken with a chromatic beam was perfectly corrected for instrumental chromaticity ex post facto)\footnote{In practice, a uniform achromatic antenna is not physically realisable and spectrometer data cannot be perfectly corrected for instrumental chromaticity ex post facto with BFCC (see \autoref{Sec:ComparisonBFCCDataModels}), so these assumptions are approximate and not necessarily sufficiently accurate for unbiased 21-cm signal recovery, as will be shown in \autoref{Sec:Results}.}, substituting these into \autoref{Eq:Tdata}, an unbiased model for the spectrum of the non-21-cm component of the resulting data set at a given time is given by,
\begin{multline}
\label{Eq:TskySDFgStationaryIonosphereUB}
T_\mathrm{sky}(\nu, t) = T_{\mathrm{e}}(1-e^{-\tau_\mathrm{ion}(\nu)}) + T_{\gamma}e^{-\tau_\mathrm{ion}(\nu)} + \\
\int\limits_{\Omega^{+}} B_\mathrm{u} \Bigg[ T_\mathrm{plfg}(\nu_\mathrm{c}, \Omega, t)\left(\frac{\nu}{\nu_\mathrm{c}}\right)^{-\beta_{\Omega,t}} e^{-\tau_\mathrm{ion}(\nu)} \Bigg] \mathrm{d}\Omega
\ ,
\end{multline}
where we have used the direction-independence of the first two terms to take them outside the integral. If we assume that $\beta_{\Omega,t}$ and $T_\mathrm{plfg}(\nu_\mathrm{c}, \Omega, t)$ are uncorrelated random fields with $\beta_{\Omega,t}$ drawn from a Gaussian distribution,
\begin{multline}
\label{Eq:GaussianSIdistribution}
\mathrm{Pr}(\beta) = \frac{1}{\sqrt{2\pi\sigma_{\beta}^{2}}} \exp\left[-\frac{1}{2}\left( \frac{\beta-\beta_{0}}{\sigma_{\beta}} \right)^2\right]
\ ,
\end{multline}
where $\beta_{0}$ and $\sigma_{\beta}$ are the mean and standard deviation of the distribution, respectively, and noting that the only spatially-dependent quantities on the RHS of \autoref{Eq:TskySDFgStationaryIonosphereUB} are $T_\mathrm{plfg}(\nu_\mathrm{c}, \Omega, t)$ and $\beta_{\Omega,t}$, we can rewrite the final term on the RHS of \autoref{Eq:TskySDFgStationaryIonosphereUB} as (e.g. \citealt{2012MNRAS.419.3491L}),
\begin{multline}
\label{Eq:TskySDFgStationaryIonosphereUB2}
\iint\limits_{-\infty}^{\infty} B_\mathrm{u} \Bigg[ \mathrm{Pr}(T) \mathrm{Pr}(\beta) T\left(\frac{\nu}{\nu_\mathrm{c}}\right)^{-\beta} e^{-\tau_\mathrm{ion}(\nu)} \Bigg] \mathrm{d}T \mathrm{d}\beta
\ .
\end{multline}
Here, $\mathrm{Pr}(T)$ describes the probability distribution from which the sky brightness temperature at frequency, $\nu_\mathrm{c}$, is drawn.
Evaluating the integral with respect to $T$ and substituting,
\begin{multline}
\label{Eq:GaussianxPL}
\mathrm{Pr}(\beta) \left(\frac{\nu}{\nu_\mathrm{c}}\right)^{-\beta} = \frac{1}{\sqrt{2\pi\sigma_{\beta}^{2}}} \exp\left[-\frac{1}{2}\left( \frac{\beta-\beta_{0}}{\sigma_{\beta}} \right)^2 - \beta \log\left(\frac{\nu}{\nu_\mathrm{c}} \right) \right]
\ ,
\end{multline}
we can rewrite \autoref{Eq:TskySDFgStationaryIonosphereUB2} as,
\begin{multline}
\label{Eq:TskySDFgStationaryIonosphereUB3}
\int\limits_{-\infty}^{\infty} \Bigg[  \frac{\bar{T}}{\sqrt{2\pi\sigma_{\beta}^{2}}} \exp\left[-\frac{1}{2}\left( \frac{\beta-\beta_{0}}{\sigma_{\beta}} \right)^2 - \beta \log\left(\frac{\nu}{\nu_\mathrm{c}} \right) \right]  e^{-\tau_\mathrm{ion}(\nu)} \Bigg] \mathrm{d}\beta
\ ,
\end{multline}
where $\bar{T}$ is the mean of $T_\mathrm{plfg}(\nu_\mathrm{c}, \Omega, t)$.

Evaluating \autoref{Eq:TskySDFgStationaryIonosphereUB3}, and substituting it back into \autoref{Eq:TskySDFgStationaryIonosphereUB}, we have,
\begin{multline}
\label{Eq:TskySDFgStationaryIonosphereUB4}
T_\mathrm{sky}(\nu, t) = T_{\mathrm{e}}(1-e^{-\tau_\mathrm{ion}(\nu)}) + T_{\gamma}e^{-\tau_\mathrm{ion}(\nu)} \\
+ \bar{T}e^{-\tau_\mathrm{ion}(\nu)} \left(\frac{\nu}{\nu_\mathrm{c}} \right)^{\frac{\sigma_{\beta}^{2}}{2}\log\left(\frac{\nu}{\nu_\mathrm{c}} \right) - \beta_{0}}
\ .
\end{multline}
Using \autoref{Eq:TauIon} for $\tau_\mathrm{ion}$, linearising the first term in \autoref{Eq:TskySDFgStationaryIonosphereUB4}, neglecting\footnote{This was found to have only a small effect on recovery of the 21-cm signal in \citealt{2018Natur.564E..32H} (producing an order of magnitude $1\%$ change in the amplitude).} $T_{\gamma}$, averaging over time, and defining,
\begin{align}
\label{Eq:IntrinsicCoeffDefinitions}
b_{0} &=  \bar{T}e^{-\tau_\mathrm{ion}(\nu)} \ ,\\ \nonumber
b_{1} &= 2.5 - \beta_{0} \ ,\\ \nonumber
b_{2} &= \frac{\sigma_{\beta}^{2}}{2} \ ,\\ \nonumber
b_{3} &= \tau_0 \ ,\\ \nonumber
b_{4} &= T_{\mathrm{e}}\tau_0 \ ,
\end{align}
we recover the Intrinsic foreground model considered in B18,
\begin{multline}
\label{Eq:B18IntrinsicForegroundModelAppendix}
T_\mathrm{Intrinsic,fg}^\mathrm{model}(\nu) = b_{0}\left(\frac{\nu}{\nu_\mathrm{c}} \right)^{-2.5 + b_{1} + b_{2}\log\left(\frac{\nu}{\nu_\mathrm{c}} \right)} \mathrm{e}^{-b_{3}\left(\frac{\nu}{\nu_\mathrm{c}} \right)^{-2}} + b_{4}\left(\frac{\nu}{\nu_\mathrm{c}} \right)^{-2}
\ .
\end{multline}
Here, $b_{i}$ with $i \in [0,\cdots,4]$ are foreground and ionospheric parameters to be determined in the fit of the model to the data.

From the definitions in \autoref{Eq:IntrinsicCoeffDefinitions}, for \autoref{Eq:B18IntrinsicForegroundModelAppendix} to provide a physical model for the emission components requires that $b_{i}$ with $i \in [1,2,3]$ are restricted to small values and $b_{i}$ with $i \in [2,3,4]$ are strictly positive, given that,
\begin{enumerate*}
\item $\beta_{0} \sim 2.5$ is a reasonable estimate for the mean spectral index in the frequency range relevant for CD (e.g. \citealt{2019MNRAS.483.4411M}),
\item $\sigma_{\beta} \ll 1$ (e.g. \citealt{2019MNRAS.483.4411M}), and,
\item $\tau_0 \ll 1$ (e.g. \citealt{2015RaSc...50..130R}).
\end{enumerate*}
We incorporate this information, when fitting \autoref{Eq:B18IntrinsicForegroundModelAppendix} to the simulated data in \autoref{Sec:Results}, using priors on the parameters of the model as listed in \autoref{Tab:DataModelIntrinsicPriors}.

When deriving \autoref{Eq:B18IntrinsicForegroundModelAppendix}, we assumed that $\beta_{\Omega,t}$ and $T_\mathrm{plfg}(\nu_\mathrm{c}, \Omega, t)$ are uncorrelated random fields. In practice, $T_\mathrm{plfg}(\nu_\mathrm{c}, \Omega, t)$ and $\beta_{\Omega,t}$ have a non-zero correlation coefficient, with the temperature spectral index steepening with decreasing sky-brightness temperature away from the Galactic plane. Additionally, $\beta_{\Omega,t}$ is only approximately Gaussian. In combination, this reduces the accuracy with which the foreground component of \autoref{Eq:B18IntrinsicForegroundModelAppendix} can model the foreground component of \autoref{Eq:TskySDFgStationaryIonosphereUB} (in simulations using an $N_\mathrm{side}=512$ resolution \textsc{healpix} sky model, from the $\mu\mathrm{K}$ to $\mathrm{mK}$ level). Nevertheless, in the absence of instrumental chromaticity, in terms of absolute error, the foreground component of \autoref{Eq:B18IntrinsicForegroundModelAppendix} remains an excellent model the foreground component of \autoref{Eq:TskySDFgStationaryIonosphereUB} relative to the order of magnitude $10~\mathrm{mK}$ noise level associated with the publicly available EDGES low data. Systematic errors above this level when fitting realistic simulated BFCC corrected EDGES data with \autoref{Eq:B18IntrinsicForegroundModelAppendix} can be attributed to the compounding effect of instrument-induced chromaticity imperfectly-corrected by BFCC.

%%%%%%%%%%%%%%%%%%%%%%%%%%%%%%%%%%%%%%%%%%%%%%%%%%

% Don't change these lines
\bsp	% typesetting comment
\label{lastpage}
\end{document}